\documentclass[useAMS,usenatbib]{mnras}
\usepackage{hyperref}
\usepackage{epsf}

\title[Validation of low-degree helioseismic solar-cycle changes]{Validation of solar-cycle changes in low-degree helioseismic parameters from the Birmingham Solar-Oscillations Network}
\author[R. Howe, G.R. Davies, W.J. Chaplin, Y.P. Elsworth, and S.J. Hale]{R. Howe$^{1}$\thanks{E-mail:
rhowe@nso.edu (RH)},  G.R. Davies$^{1}$, W.J. Chaplin$^{1}$, Y.P. Elsworth$^{1}$,
and S.J. Hale$^{1}$
\\
$^{1}$School of Physics and Astronomy, University of Birmingham, Edgbaston, Birmingham B15 2TT, United Kingdom}

\begin{document}
\maketitle

\label{firstpage}

\begin{abstract}
We present a new and up-to-date analysis  of the solar low-degree $p$-mode parameter shifts from the Birmingham Solar-Oscillations Network (BiSON) over the past 22 years, up to the end of 2014. We aim to demonstrate that they are not dominated by changes in the asymmetry of the resonant peak profiles of the modes and that the previously published results on the solar-cycle variations of mode parameters are reliable.
We compare the results obtained using
a conventional maximum likelihood estimation algorithm and a new one based on the Markov Chain Monte Carlo (MCMC) technique, both taking into account mode asymmetry.
We assess the reliability of
the solar-cycle trends seen in the data by applying the same analysis to
artificially generated spectra. We find that the two methods are in good agreement. Both methods accurately reproduce the input frequency shifts in the
artificial data and underestimate the amplitude and width changes by a small amount, around 10 per cent.
We confirm earlier findings that the frequency and line width are positively correlated, and the mode amplitude anticorrelated, with the level of solar activity, with the energy supplied to the modes remaining essentially unchanged. For the mode asymmetry the correlation with activity is marginal, but the
MCMC algorithm gives more robust results than the MLE.
The magnitude of the parameter shifts is consistent with earlier work.
There is no evidence that the frequency changes we see arise from changes in the asymmetry, which would need to be much larger than those observed in order to give the observed frequency shift.

\end{abstract}

\section{Introduction}

The variation of solar acoustic mode ($p$-mode) parameters with solar activity is one of the best-known results in helioseismology, and has been studied at a wide range of spatial and temporal scales over the last few decades. In the light of recent developments in mode parameter estimation techniques it seems timely to revisit these findings. In this work we re-analyse the Sun-as-a-star data from the Birmingham Solar-Oscillations Network (BiSON) over the last two solar cycles and verify our ability to detect subtle variations using two different algorithms, one that uses conventional maximum-likelihood estimation and a new one based on Bayesian principles

\subsection{Historical Background}
\subsubsection{Frequency Variation}
Small changes in the frequencies of the low-degree modes, positively correlated with proxies for the solar activity such as sunspot number and 10.7\,cm radio flux, were reported for example by \citet{1985Natur.318..449W} using data from the Active Cavity Radiation Monitor (ACRIM), and later confirmed in ground-based observations by \citet{1989A&A...224..253P} and \citet{1990Natur.345..322E}, while \citet{1990Natur.345..779L} first reported solar-cycle changes in the frequencies of medium-degree modes from resolved-Sun observations at the Big Bear Solar Observatory. \citet{1990Natur.345..779L} also pointed out that the variation increased with frequency, following the inverse of the so-called `mode inertia.' This indicates that the
mechanism responsible for the shifts is located close to the solar surface. The frequency dependence of the variation is harder to detect in low-degree data due to the smaller number of modes available, but within a few years it was reported by \citet{1994ApJ...434..801E}. 

In resolved-Sun data the frequency shifts follow the
activity variation in location as well as in time  \citep*[e.g.][and references therein]{2002ApJ...580.1172H}; modes of the same radial order $n$ and degree $l$ but different azimuthal order $m$ have different distributions of power with latitude and so are differently affected as the magnetic activity belts move in latitude over the solar cycle. 
Even for low-degree modes, \citet{2004MNRAS.352.1102C,2004ApJ...610L..65J,2007ApJ...659.1749C,2015arXiv150207607S} have reported  marginally significant differences in the response of the frequency to activity for modes of different degree, consistent with the different latitudinal distribution of power for the different spherical harmonics.


\subsubsection{Power and width variations}

Magnetic activity influences not only the frequency of the
acoustic modes but also their power and lifetime. \citet{1993MNRAS.265..888E}
reported an anticorrelation between solar activity and the amplitude of low-degree modes, with about a 30 per cent change between solar minimum and solar maximum. The
positive correlation between mode width and activity took longer to establish, due to the difficulties introduced by systematic effects such as the broadening of the modes when the duty cycle is low. However, \citet{2000MNRAS.313...32C}, after a careful analysis of BiSON data, reported a change of around 24 per cent in the line width and a similar decrease in the mode power (amplitude squared) between solar minimum and maximum. The combination of these two results suggested that the
solar-cycle change affects the mode damping rather than the excitation, with the rate of energy supplied to the modes remaining constant.
\citet*{2000ApJ...531.1094K,2000ApJ...543..472K} found activity-correlated changes in the mode width for medium-degree Global Oscillation Network Group (GONG) data, which could also be
localized to the latitudes where magnetic activity is present. 
\citep*{2002ApJ...572..663K}. Again, the energy supply rate appears to remain constant.

\citet{2007A&A...463.1181S} studied the line width variation in 9.5 years of BiSON data and found marginal evidence for a degree-dependence of the sensitivity, with the $l=0$ shifts about half the size of those for $l\ge 1$, consistent with what would be expected from the latitudinal distribution of the modes in relation to the latitudinal migration of the activity belts during the solar cycle. They  concluded that previous estimates of Sun-as-a-star line width changes may have been overestimated by about 50 per cent because the shifts were averaged over all the modes.



\subsubsection{Asymmetry}
As discussed below in Section~\ref{sec:noise}, the mode peaks show a small asymmetry due to the correlation of mode excitation with the noise that drives it \citep{1993ApJ...410..829D}. This term is negative for velocity observations and generally positive for measurements made in intensity; it is smallest where the modes are strongest and for low-degree observations can reach a value of a few per cent at the extremes of the $p$-mode band and a fraction of one per cent at 3\,mHz.  
\cite{2007ApJ...654.1135J} reported a fractional change of about 15 per cent between solar maximum and solar minimum in the asymmetry of low-degree modes in data from BiSON and the Global Oscillations at Low Frequencies (GOLF) instrument on the Solar and Heliospheric Observatory (SOHO) spacecraft, with the strongest (most negative) asymmetry at solar maximum. As the
uncertainty in the asymmetry depends strongly on the signal-to-noise ratio of the modes and also on the duty cycle, the result was most clearly seen in the GOLF data and was described as `marginally significant' for BiSON.



%


\subsection{Mechanisms}
There is still some uncertainty about the precise mechanisms responsible
for the parameter shifts.
Attempts have been made to model the frequency shifts by invoking the direct
effects of magnetic field at the tachocline \citep{1986Natur.323..603R}, the sunspot anchoring zone around 50\,Mm below the surface \citep{2005A&A...439..713F}, the photosphere \citep{1985ApJ...298..867B}, or the chromosphere \citep{1989ApJ...338..538C,1994SoPh..152..261J}, but none of these have predicted
shifts of the observed magnitude. The shifts have also been considered as an indirect effect of temperature changes associated with the activity belts \citep{1988ApJ...331L.131K} and as an effect of a change in acoustic cavity size \citep{2005ApJ...625..548D}. None of these models is universally accepted. In any case, it is possible to consider the frequency shifts as a solar-cycle proxy in themselves, and one of the few that has some sensitivity to layers below the photosphere.

\subsection{Motivation}
Two relatively recent developments have prompted us to revisit the mode parameter variations over the last two decades of BiSON observations.

First, although there is broad consensus on the size and sign of the different solar-cycle effects on mode parameters, \citet{2013ASPC..478..137K} has reported results from a re-analysis of GONG, Michelson Doppler Imager (MDI), and Helioseismic and Magnetic Imager (HMI) medium-degree data showing frequency changes smaller by about a factor of two than have been seen in the standard analysis of these data sets. The author attributes the discrepancy to the use of asymmetric peak profiles in the fitting, and also reports variations in the peak width and asymmetry, with the asymmetry changes possibly accounting for the `missing' frequency shifts. These results make it timely to re-investigate the
frequency variations over the past two solar cycles of BiSON observations and assess the validity of our mode-parameter variations using fits that take into account the asymmetry of the modes.


Secondly, there is growing evidence of, and interest in,
temporal variations in the mode frequency that cannot be described simply
as a linear function of the global activity level. These may take the form of fluctuations superimposed on the linear trend with activity proxy or of changes in the strength of the trend over time; the two are not necessarily clearly distinguished, as the presence of an additional short-term variation may reduce the strength of the correlation with activity. 
Following the long, anomalous solar minimum after Solar Cycle 23 and the relatively weak Cycle 24, there has been particular interest in any changes in 
frequency shifts 
that might indicate changes in the structure of the outer solar layers. 
\citet{2012ApJ...758...43B} found that the shifts in Cycle 23 BiSON data 
did not follow the trend with activity extrapolated from the Cycle 22 data, with the mismatch showing up first in the low-frequency shifts and only later
at higher frequencies.
On the other hand, \citet{2015arXiv150207607S}, using GOLF data, reported 
a frequency increase in Cycle 24, particularly for $l=1$ modes, that was larger than that expected from the increase in activity level when compared to the 
results from Cycle 23.
It seems that the interpretation of such results is quite sensitive to which modes are averaged and what period is taken as the reference.

There have also been reports of variations in the mode frequencies on time scales that do not correspond to the 11-year solar cycle. \cite{2006MNRAS.369..933H} reported aperiodic fluctuations in individual low-degree  mode freqencies after the subtraction of the activity effect, which were correlated between different instruments but not with any activity index and not correlated between different modes. These variations were attributed to stochastic fluctuations in the solar spectrum. \citet{2010ApJ...718L..19F,2011JPhCS.271a2025B,2012A&A...539A.135S,2013ApJ...765..100S} later reported a quasi-biennial variation in the mode frequencies, superimposed on the 11-year variation, 
which they interpreted as evidence of a `second solar dynamo' located in the near-surface layers. This variation appears to be correlated with other activity proxies in high-activity periods but persists during solar minima when the proxies are flat. Along similar lines, \citet{2011ApJ...739....6J} found that the GONG medium-degree frequencies showed hints of a `double minimum' between Solar Cycles 23 and 24, with the frequencies first reaching their lowest value ahead of the activity minimum; this may reflect a similar phenomenon to that seen in the low-degree data.


All of these results make it timely to re-examine the parameter variations and assess their reliability.



\subsection{Arrangement}
The arrangement of the paper is as follows:
in Section~\ref{sec:noise} we discuss the issues related to correlated and uncorrelated noise and
peak asymmetry and go on to describe the specifics of the data simulation. In Section~\ref{sec:fitting} we describe the algorithms used for extracting the mode parameters. In Section~\ref{sec:artdata} we describe the tests carried out on the artificial data and discuss the results and their implications for the analysis of BiSON data. In Section~\ref{sec:fitresults} we present the results from fitting the BiSON data, and in Section~\ref{sec:conclusion} we discuss our conclusions.


\section{Generation of artificial data}
\label{sec:noise}

In order to test for possible biases introduced into the mode parameters and their variations we have carried out a series of tests on artificial data. The generation of the artificial data was based on the solarFLAG simulator, which was used to make
data for the hare-and-hounds study on rotational frequency splittings
discussed by 
\citet*{2006MNRAS.369..985C} and also to test peak-bagging of Sun-as-a-Star helioseismic data in \citet{2007ApJ...654.1135J}. 
The original solarFLAG simulator
did not include effects of
correlated excitation, or of correlations of the excitation with
background noise.
An important
consequence was that the artificial mode peaks in the frequency power
spectrum showed no asymmetry, unlike their real solar
counterparts. Since any analysis which seeks to extract accurate
estimates of the frequencies must cope with this asymmetry it was felt
we needed a simulator that could provide such a test.
The departure of the mode shape from the pure Lorenztian predicted by a simple harmonic oscillator has three main causes.
They are the localization of the noise source, the impact of correlated noise on mode excitation, and finally the more subtle effect of
the presence of the wings of nearby correlated modes.
We have used a simple but very powerful method to introduce in the
time domain the effects of asymmetry, which is based on the framework
proposed by
\citet*{2006MNRAS.371.1731T}.
The influence of source localization \citep{1999MNRAS.309..761C} does not need to be explicitly considered as it produces a line shape that is the
same as from the correlated background. We include it implicitly by our choice of correlation coefficient.
Thus, there are just two
factors in our method that contribute to the asymmetry of the
artificial mode peaks. First, background noise is correlated with the
excitation of the modes, and second, overtones of the same angular
degree and azimuthal order have excitation functions that are
correlated in time. This correlated mode excitation is based on the
description given in \citet*{2008AN....329..440C} 
and the method is described in detail in the following sections.
Further details, and analytical descriptions of
the examples in Sections~\ref{sec:corrnoi} and~\ref{sec:correx} below,
may be found in \citet{2008AN....329..440C}.

\subsection{Correlated noise and mode excitation}
\label{sec:corrtut}
In \cite{2006MNRAS.371.1731T}
it was hypothesized that the
excitation function of a mode of angular degree $l$, azimuthal degree
$m$, and frequency $\nu$, is the same as that component of the solar
background (granulation) noise that has the same spherical harmonic
projection, $Y_{lm}$, in the corresponding range in frequency in the
Fourier domain. Let us start by considering the impact on the line shape of this
correlated background noise.
We do this in the context of how the solarFLAG
simulator generates artificial data on single $p$-modes.

 \label{sec:corrnoi}

The basis of the solarFLAG simulator is the method discussed by
\citet{1997MNRAS.287...51C}
 for generating time series of individual $p$
modes. The method uses the Laplace transform solution of the equation
of a forced, damped harmonic oscillator to make the output velocity of
each artificial mode. Oscillators are re-excited at each time sample
-- the chosen cadence is typically 40\,sec or 60\,sec for
Sun-as-a-star data -- with small `kicks' from a time series of random
noise. This procedure mimics the stochastic excitation of the solar $p$-modes. For the moment we shall assume the random noise data are drawn
from a normal (i.e., white) distribution having zero mean. We shall
see later that in the final version of the simulator we used
`granulation-like' noise -- to mimic the spectrum of solar
granulation -- which has a spectrum that looks white only locally in
the vicinity of the oscillator resonance.

\begin{figure*}
 \centerline{\epsfxsize=5.5cm\epsfbox{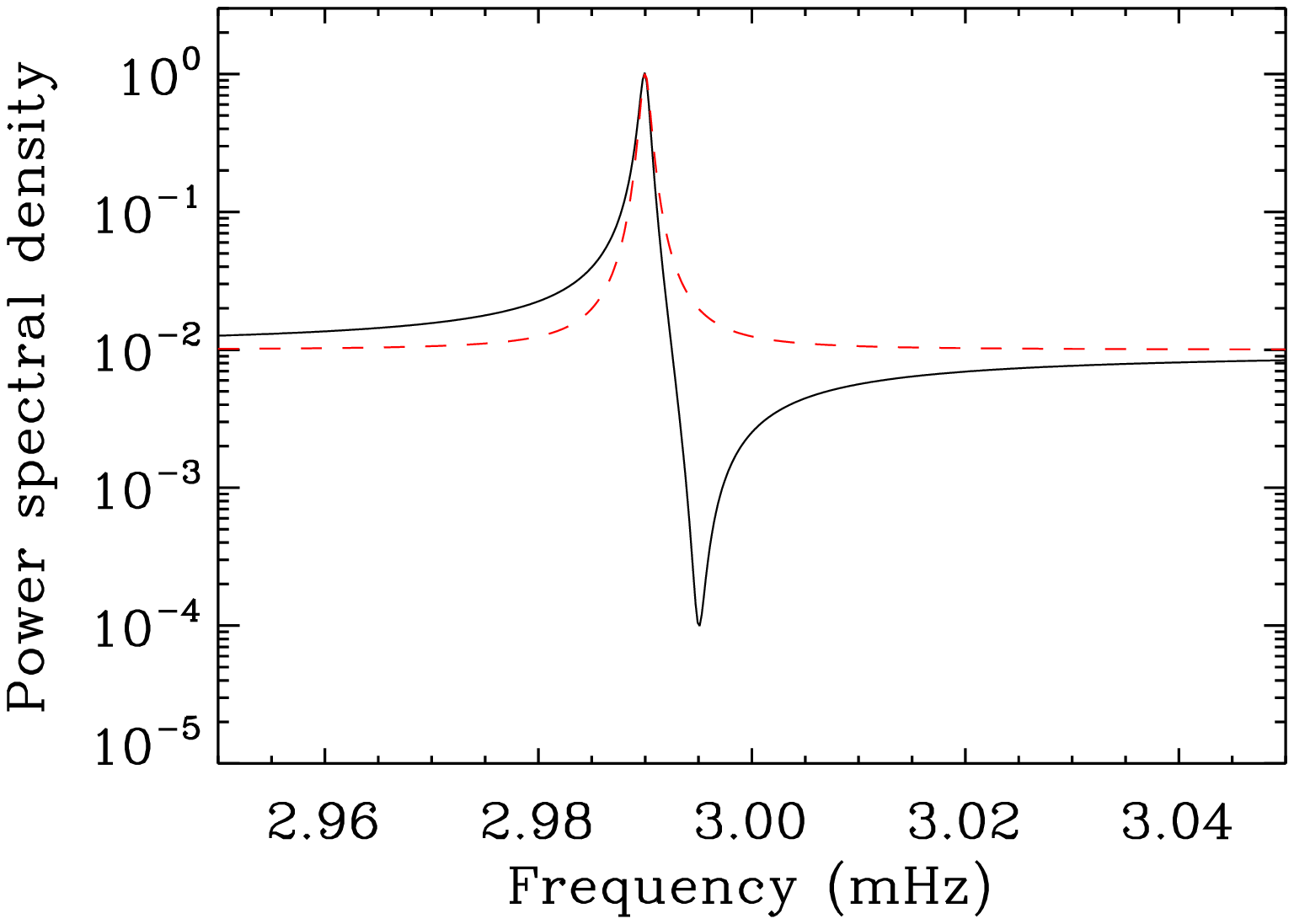}
               \epsfxsize=5.5cm\epsfbox{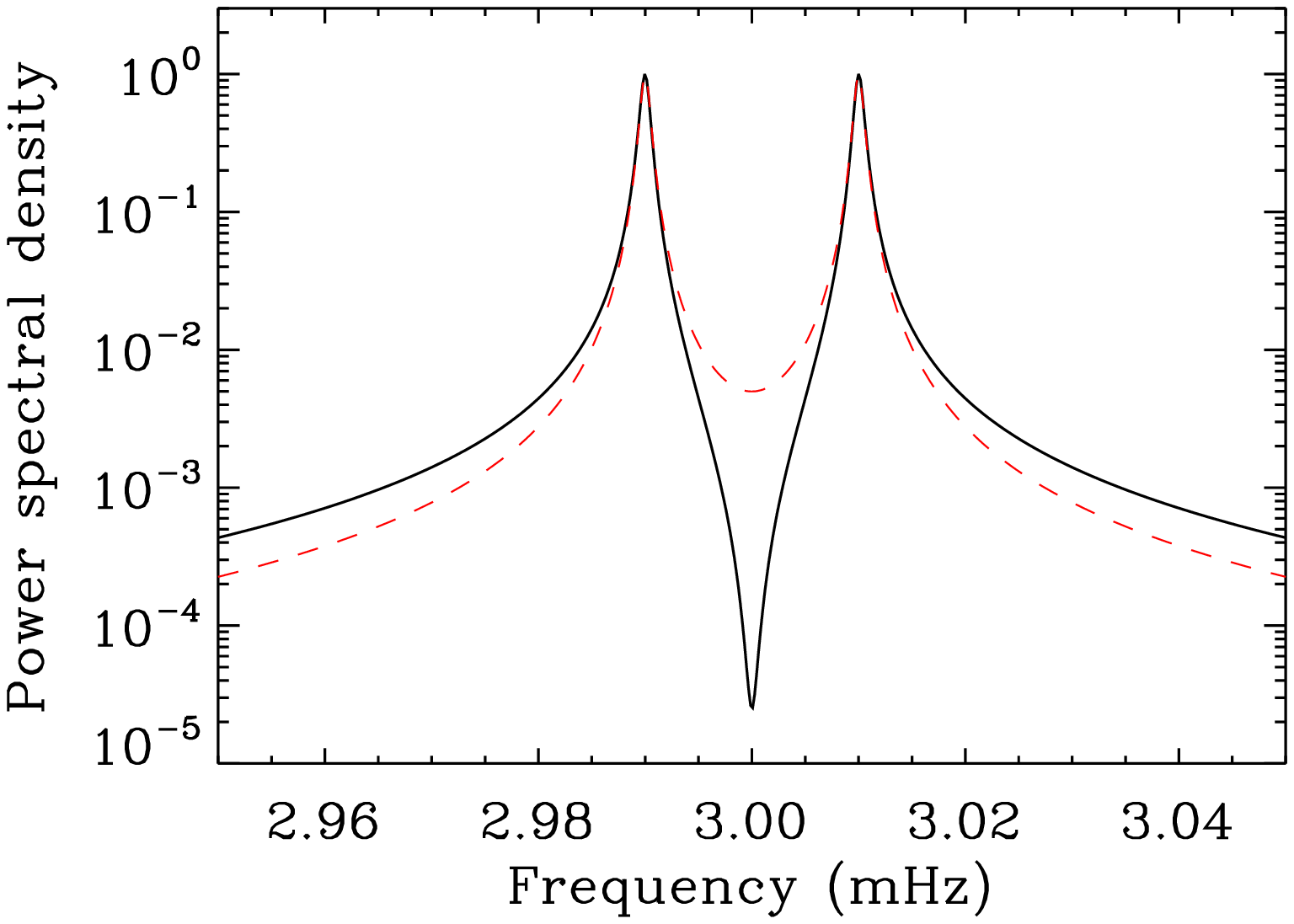}
               \epsfxsize=5.5cm\epsfbox{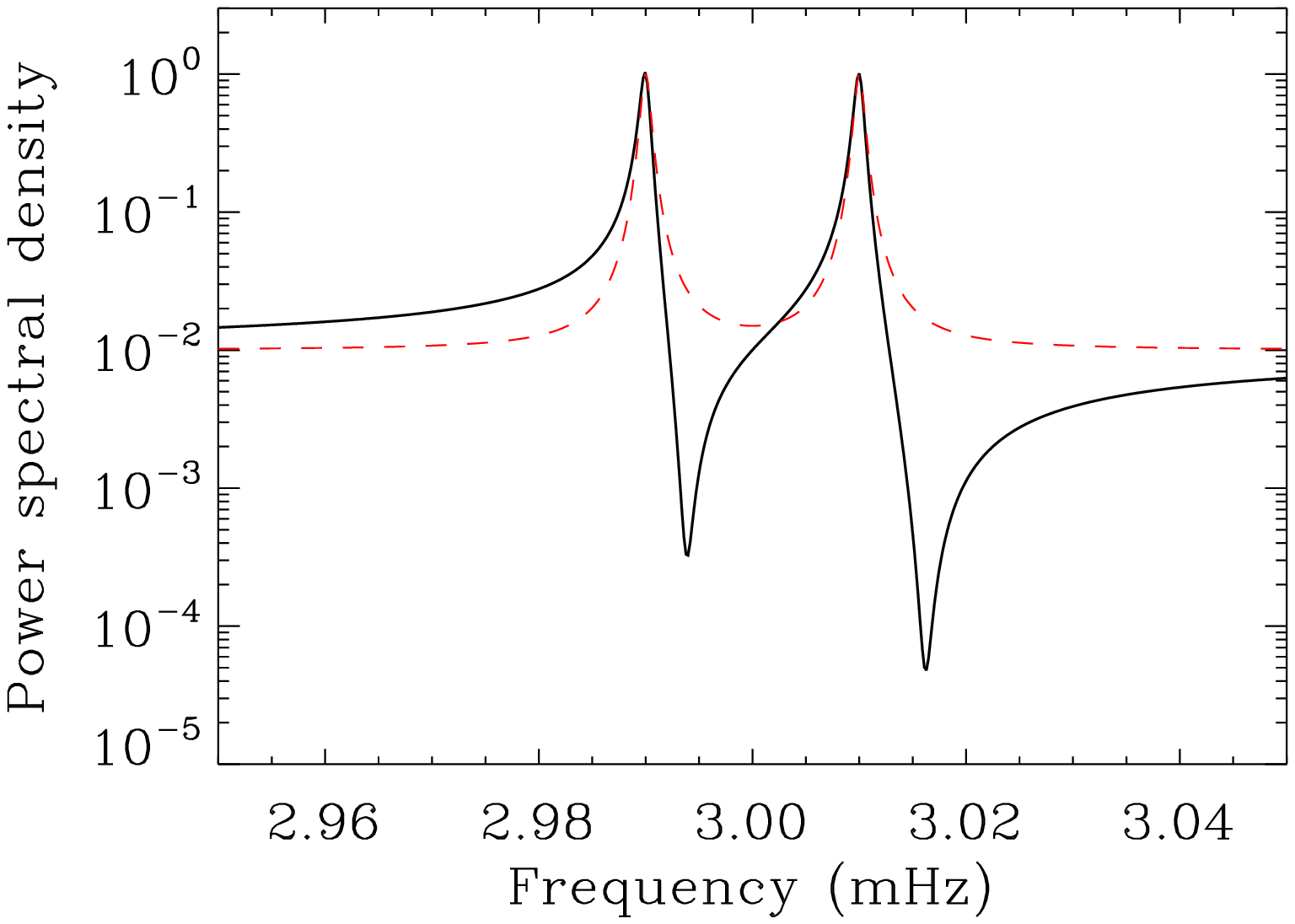}}

 \caption{
Limit frequency power spectra for scenarios with one or two
 modes. Left-hand panel: a single mode, with background noise. The dashed red
 line shows the spectrum for no correlation; the solid black line for
 when the mode is correlated with the background noise. Centre panel:
 two modes. The dashed red line shows the limit spectrum when the excitation
 of the modes is uncorrelated; the solid black line when the excitation
 is correlated. Right-hand panel: Two modes, with background
 noise. The dashed red line shows the limit spectrum when there is no
 correlation of the excitation, or with the background noise; the
 solid black line when the excitation is correlated, and there is
 correlation with the background noise.}

 \label{fig:corr3}
 \end{figure*}

The left-hand panel in Fig.~\ref{fig:corr3} shows, on a logarithmic scale, the limit power spectral
density of two scenarios, which illustrate the impact of correlated
noise. In both scenarios the power spectral density has contributions
from a mode -- of frequency $\nu=2990\,\rm \mu Hz$, and line width
$\Gamma=1\,\rm \mu Hz$ -- and from white background noise. The thin
dashed red line shows the result for a scenario where the excitation of the
mode is \emph{uncorrelated} with the background noise. The solarFLAG
simulator would generate the data for this scenario by using one
time series of random noise to excite the mode; to the velocity output
of the mode would then be added another, completely uncorrelated
time series of appropriately scaled random noise, to represent the
background (here the height-to-background ratio is 100). Since there
is no correlation, the limit frequency power spectrum is given simply
by the incoherent addition of the limit spectrum of the mode (a
Lorentzian) and the limit spectrum of the noise (here, a flat offset
for white noise).

Now, what if we use the excitation time series as the background noise?
The excitation and the background will now be 100-per-cent correlated,
and the power spectral density will have a peak that is asymmetric. 
This situation, for a large  (about ten per cent) value of the asymmetry, is shown with the thick
black line in the left-hand panel of Fig.~\ref{fig:corr3}.
If only the effect of the background correlation  had been considered the peak asymmetry would have been positive with an excess of power on the high-frequency side of the peak.
However, the asymmetry seen in low-$l$ solar p modes observed in
Doppler velocity is negative and hence we show the effect of negative correlation with excess
power on the low-frequency side of the resonance. The observed asymmetry is much lower than illustrated here.


Next, we consider the impact of correlated mode excitation.

 \label{sec:correx}

An important implication of the framework proposed by
\cite{2006MNRAS.371.1731T}
is that overtones with the same ($l$,\,$m$)
should have excitation functions that are correlated in time. (Note
that the $Y_{lm}$ for ($l$,\,$m$) and ($l$,\,$-m$) are orthogonal, and are
therefore assumed to have independent, i.e., uncorrelated,
excitation.) To illustrate the impact of this correlation, we look at
the simplest possible scenarios, which have just two modes in the
frequency power spectrum. Consecutive overtones of the low-$l$ solar p
modes are separated in frequency by $\sim 135\,\rm \mu Hz$. Here, we
consider two modes separated in frequency by $20\,\rm \mu Hz$. This
smaller frequency spacing exaggerates the impact of the correlated
excitation on the observed power spectral density, and therefore
allows us to show more clearly the effect of the correlation. We again
assume each mode has a line width of $1\,\rm \mu Hz$. The impact of the
actual $\sim 135$-$\rm \mu Hz$ spacing is considered in
Subsection~\ref{sec:simu} below.

The thin dashed red line in the middle panel of Fig.~\ref{fig:corr3} shows
the limit spectrum for two modes whose excitation is uncorrelated in
time. There is no background noise. The solarFLAG simulator would
generate the data for this scenario by using independent time series of
random noise to excite each oscillator. Now, what happens if the
simulator excites both oscillators with the same time series of random
noise? The excitation is now 100-per-cent correlated in time, and the
power spectral density shows clearly that the peaks are asymmetric
(thick dark line). This asymmetry comes from the interaction of the
tails of the mode peaks. We draw an important conclusion from this
example: correlated  excitation of  modes will give a contribution to the
observed asymmetry that is dependent on how the tails of the individual modes overlap. 
It turns out that this effect is important even for quite well separated modes. 

In our final example, we consider both sources of asymmetry together and we add background noise to each two-mode
scenario above. The thin dashed red line in the  right-hand panel of
Fig.~\ref{fig:corr3} is for a scenario where the background noise and
the mode excitation are all uncorrelated. The thick dark line instead
shows what happens if we add correlated background noise to the two
correlated modes. This means the excitation and background noise are
all 100 per cent correlated, and we see that addition of the
background further modifies the shape of the power spectral density,
relative to the correlated-mode example with no background noise. As
such, there are now two factors which contribute to the peak
asymmetry: there is a contribution from the correlated excitation of individual modes (and also potentially from source localization); and
a contribution from the correlated background.

One other important point to note is that correlation of the
excitation in time does \emph{not} imply correlation of the mode
amplitudes in time. This can be understood by considering the analogy
of damped, stochastically forced oscillators. Modes of different
frequencies will be `kicked' by the common excitation at different
phases in their oscillation cycles, and provided the frequencies
differ by more than a few line widths
\citep[see][]{2008AN....329..440C}
-- a condition easily met by consecutive overtones of the
low-$l$ modes, which are separated by $\sim 135\,\rm \mu Hz$ -- there
will be significant differences in how the amplitudes vary in time,
due to the excitation.


 \subsection{The solarFLAG simulator}
 \label{sec:simu}

 \subsubsection{General overview}
 \label{sec:gensimu}

The solarFLAG data sets simulate full-disc `Sun-as-a-star' Doppler
velocity observations of the Sun, such as those made by the
BiSON and GOLF.
The solarFLAG data sets are made with a full cohort of simulated low-$l$
modes, covering the ranges $0 \le l \le 5$. The frequencies of the
modes come from a standard solar model of the user's choice. A surface
term is also added to these frequencies, based on polynomial fits to
the differences between the standard model frequencies and frequencies
from analysis of BiSON and GOLF data. For data presented in this
paper 
the
frequencies came from model BS05(OP) of \citet{2005ApJ...621L..85B}.

A database of $p$-mode power and line width and asymmetry estimates,
obtained from analyses of GOLF and BiSON data, was used to guide the
choice of the other mode input parameters.  The hypothetical solarFLAG
instrument was assumed to make its observations from a location in, or
close to, the ecliptic plane.  This is the perspective from which
BiSON (ground-based network) and GOLF view the Sun. The rotation axis
of the star is then always nearly perpendicular to the line-of-sight
direction, and only a subset of the $2l+1$ components of the
non-radial modes are clearly visible: those having even $l+m$. These
components are represented explicitly in the solarFLAG time
series. The visibility for given ($l$,\,$m$) also depends, although to a
lesser extent, on the spatial filter of the instrument
\citep[e.g.][]{1989MNRAS.239..977C,2009MNRAS.397..793B}.
Here, we adopted BiSON-like visibility
ratios.

\begin{figure*}
\centerline{\epsfxsize=16.0cm\epsfbox{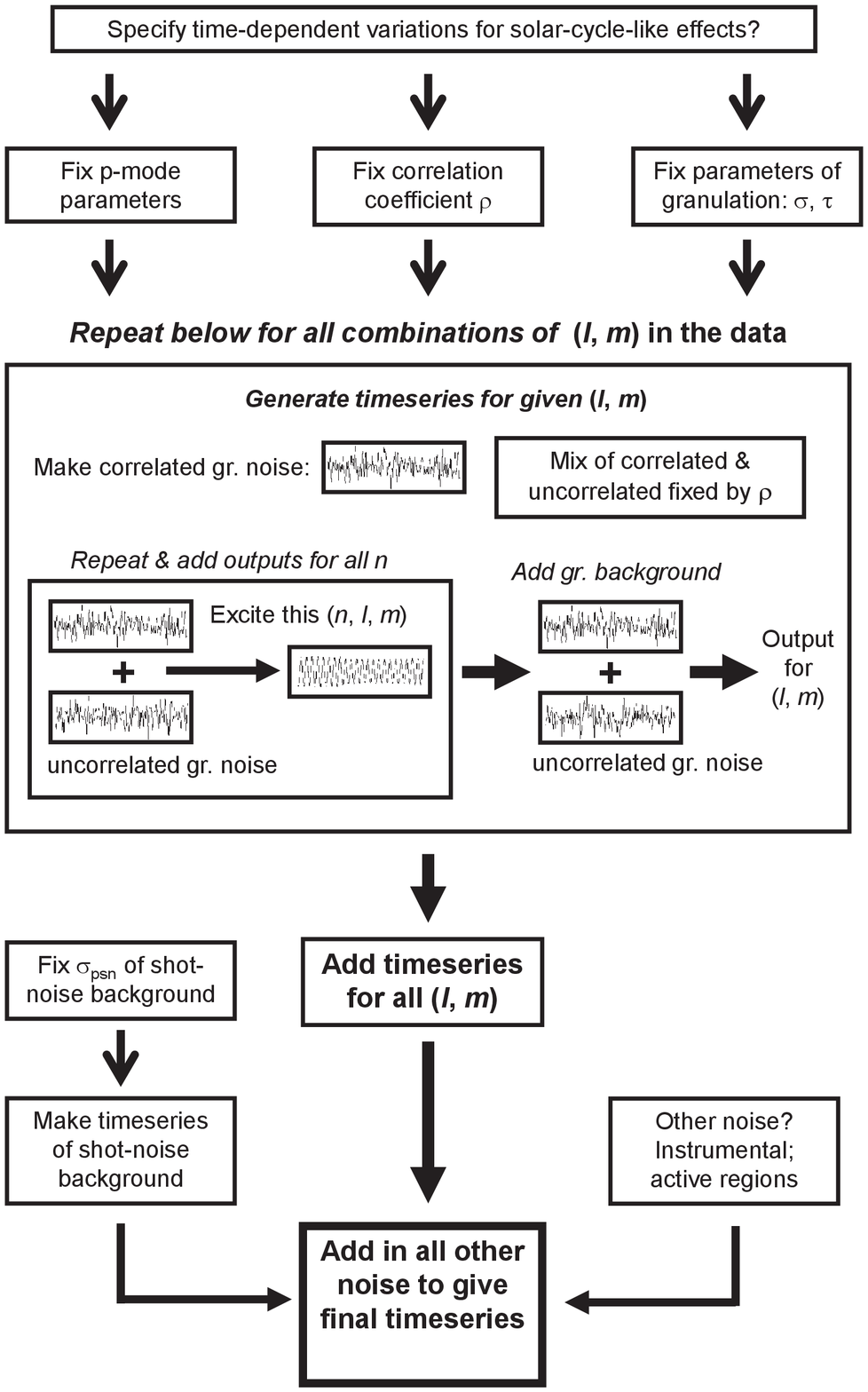}}

 \caption{A schematic representation of the solarFLAG simulator.
}

 \label{fig:sim}
 \end{figure*}

Fig.~\ref{fig:sim} shows a schematic representation of the solarFLAG
simulator, the details of which we discuss next.  



Before the simulator can begin its calculations, all the input
parameters must be specified. The $p$-mode input parameters are held in
a control file (with mode powers having been suitably scaled to
reflect the visibility filter of the Sun-as-a-star observations). The
other input parameters relate to the granulation-like noise, and other
background noise, and are specified at run time by command-line
inputs. Time series of granulation-like kicks are used to excite the modes,
and are also used to give correlated background noise. The timescale
and standard deviation of this noise must be chosen. A single
constant, $\rho$, is also set and this fixes the coefficient of
correlation for the correlated excitation, and the correlation with
the background noise. This gives the user the flexibility to `tune'
the asymmetry of the mode peaks (as well as other effects arising from
the correlations), 
since higher
correlation gives larger peak asymmetry. The input parameters of other
sources of noise, such as the standard deviation, $\sigma_{\mathrm{psn}}$, of the photon-shot noise,
are also specified at this point.

The main block of the solarFLAG code generates output time series for
each ($l$,\,$m$). It makes the time series of all the overtones, $n$,
for that ($l$,\,$m$). The modes are excited by time series of
granulation-like noise. For each ($l$,\,$m$) there is a noise
time series that gets used again and again to excite the overtones:
this is the \emph{correlated gr. noise} time series. 
When the user fixes the correlation at less than 100\,per
cent, a suitable fraction of \emph{uncorrelated gr. noise} (see
schematic in Fig.~\ref{fig:sim}) must be mixed in when the modes are excited. These
uncorrelated noise time series must be seeded with completely
independent random noise and made afresh for each mode. After all
overtones have been made, a suitably scaled mixture of correlated and
uncorrelated noise is added to the time series. This gives the
cumulative time series for each ($l$,\,$m$). Execution of the main code
block is then repeated for a total of 14 different ($l$,\,$m$)
combinations\footnote{Data on modes cover the range $0 \le l \le 5$;
and there are $l+1$ $m$-components visible in the Sun-as-a-star data
at each $l$. This actually sums to give 21 combinations of
($l$,\,$m$). However, components at $l=4$ and 5 that do not have
$l=\pm m$ are so weak that they are not included in the time series
generation. Discounting these modes reduces the number of combinations
to 14.}.

After time series for all the ($l$,\,$m$) have been made they are added
together. 
The final stage of processing adds in the
simulated time series of other uncorrelated noise sources. The
schematic includes a contribution from photon-shot noise. Other
sources may also be included at this point (e.g., active-region noise 
and instrumental noise) but this was not done for the data used in this work.

A complete description of the different elements that make up the
complex frequency amplitude (and frequency power) spectrum of a
solarFLAG time series is presented in Appendix~\ref{sec:appfull}.

We now turn to specifics on how the data are made, and how the effects
of correlations are quantified.

 \subsubsection{Achieving correlated excitation of overtones}
 \label{sec:corrsimu}

As noted above, we use time series of granulation-like noise to excite
the overtones, $n$, of the same $l$ and $m$. Later we will refer to this  as `correlated noise'. The granulation-like noise is made using white-noise
input to a low-order, autoregressive model of the AR[1] type 
\citep[e.g., see][]{2005MNRAS.361..887K},
i.e.,
 \begin{equation}
 u(t)=u(t-\Delta t)\exp{-\Delta t / \tau} + \delta(t).
 \label{eq:ar1}
 \end{equation}
Here, $u(t)$ is the output time series of granulation-like noise;
$\tau$ is the time constant of the model, which should be given a
value to mimic the lifetime of the solar granulation (see below);
$\Delta t$ is the cadence at which samples are generated by the
autoregressive process; and finally the $\delta(t)$ are random numbers
drawn from a normal distribution having zero mean and sample standard
deviation $\sqrt{\sigma^2 \Delta t / \tau}$, where $\sigma$ fixes the
amplitude of the granulation-like noise. The power spectral density,
$n(\nu)$, of this granulation-like noise follows the
approximate but useful \citet{1985ESASP.235..199H}
power-law model 
i.e.,
 \begin{equation}
 n(\nu) = \frac{4 \sigma^2 \tau}{1+(2 \pi \nu \tau)^2}.
 \label{eq:harbasic}
 \end{equation}
The default option would be to excite all overtones of the same
($l$,\,$m$) with the same time series $u(t)$, meaning the excitation of
the overtones would be 100-per-cent correlated (as in
Section~\ref{sec:correx}). 
As indicated earlier, we decided to add to the solarFLAG
simulator the option to let the user choose a coefficient of
correlation, $\rho$, 
common to 
all modes.  We describe in
Section~\ref{sec:assmag} below how the asymmetry of a mode is fixed by
the combination of $\rho$, the noise background, and the input mode
paraeters.

Overtones of the same ($l$,\,$m$) are excited by a
composite time series of kicks, $u(t)$, where:
 \begin{equation}
 u(t)=|\rho|^{1/2} u_{\rm c}(t) + \left( 1-|\rho| \right)^{1/2} u_{\rm u}(t).
 \label{eq:kick}
 \end{equation}
In the above, the $u_{\rm c}(t)$ and $u_{\rm u}(t)$ are both
time series of granulation-like noise. However, the $u_{\rm c}(t)$ are
kicks that are \emph{common} to all the overtones (the
\emph{correlated gr. noise} time series shown in red in
Fig.~\ref{fig:sim}); while the $u_{\rm u}(t)$ are completely
\emph{independent} (shown as \emph{uncorrelated gr. noise} in
Fig.~\ref{fig:sim}).

So far in this section we have not said anything about tuning the
correlation with the background noise. Correlated background noise is
provided by the correlated part of the $u(t)$. After all overtones of
the same ($l$,\,$m$) have been made, the $u(t)$ is added to the
time series to give background noise (having first been appropriately
scaled
). In the formulation
presented by
\citet{2008AN....329..440C}, 
another 
coefficient, $\alpha$, was used to describe the correlation with the
background noise. Here, we assume this correlation has the same size
as the correlation between the excitation of different modes, so that
we set $\alpha = \rho$. 
The \emph{a priori} choice of $\rho$ therefore also
fixes the correlation of the overtones with the background noise.

How do we juggle all the choices implied by the descriptions above to
give a time series where both the S/N in the modes \emph{and} the asymmetry
of the modes are realistic? That is the question we turn to next.
\label{sec:assmag}
We use two main factors to contribute
to the peak asymmetry: correlation of the modes with the background
noise; and correlated excitation of the overtones. We will now consider each of these in turn as it relates to controlling the asymmetry of the artificially generated modes.

 \subsubsection{Asymmetry due to correlated noise}
 \label{sec:asscorr}

The contribution to the asymmetry from correlations with the noise
background may be dealt with analytically in a fairly straightforward
manner. Let $b_{nlm}$ be the asymmetry, due to correlated noise, of
the $n$th overtone of a given ($l$,\,$m$). The frequency of this mode
is $\nu_{nlm}$. The asymmetry (as it appears in the mode parametrization function in Equation~\ref{eq:rh1}) is given by
\citep{2006MNRAS.371.1731T}:
 \begin{equation}
 b_{nlm} = \rho \sqrt{\frac{n_{lm}(\nu_{nlm})}{H_{nlm}}},
 \label{eq:ass}
 \end{equation}
where $H_{nlm}$ is the height of the mode peak in the frequency power
spectrum. Notice the presence of the factor $\rho$ in Equation~\ref{eq:ass}: it
is needed because it is only the \emph{correlated} part of
$n_{lm}(\nu_{nlm})$ that contributes to the asymmetry. $n_{lm}(\nu_{nlm})$ is the granulation-noise background
at the frequency of the resonance, which is just
(cf. Equation~\ref{eq:harbasic}):
 \begin{equation}
 n_{lm}(\nu_{nlm}) = \frac{4 \sigma_{lm}^2 \tau}{1+(2 \pi \nu_{nlm} \tau)^2}.
 \label{eq:har}
 \end{equation}
We give $\sigma_{lm}$ in Equation~\ref{eq:har} an explicit
dependence on the combination ($l$,\,$m$). This gives us the ability to
choose different $\sigma_{lm}$ for different ($l$,\,$m$), to take proper account of the different mode visibilities. This plays a
key part in tuning the asymmetry, as we now go on to explain.

We have assumed that at a given frequency the relative  sizes of the
granulation noise and the mode amplitudes on the Sun are independent
of degrees $l$ and $m$. An important consequence of this assumption is
that the asymmetries from the correlated noise will be the same for
all ($l$,\,$m$). In the solarFLAG simulator, after making overtones of
a given ($l$,\,$m$), the time series of mode amplitudes is multiplied
by a visibility factor, ${\cal S}_{lm}$, to mimic the visibility
filter of the Sun-as-a-star observations. To preserve the independence
of asymmetry on ($l$,\,$m$) we must therefore also modify the
$\sigma_{lm}$ of the granulation noise, according to:
 \begin{equation}
 \sigma_{lm} = \sigma
               \frac{{\cal S}_{lm}}
               {\displaystyle\sum_{{\rm all}~(l,m)} {\cal S}_{lm}}.
 \label{eq:siglm}
 \end{equation}
Here, $\sigma$ is the equivalent standard deviation of the cumulative
granulation noise background, $n(\nu)$, so that:
 \begin{equation}
 n(\nu) = \frac{4 \sigma^2 \tau}{1+(2 \pi \nu \tau)^2} =
          \displaystyle\sum_{{\rm all}~(l,m)}
          \frac{4 \sigma_{lm}^2 \tau}{1+(2 \pi \nu \tau)^2}.
 \label{eq:harcum}
 \end{equation}
We have now covered all the steps needed to see how the asymmetry is
tuned.

The $p$-mode parameters (i.e., frequencies, heights, line widths) are
fully specified on input. The asymmetry is then tuned by three free
parameters: the coefficient of correlation $\rho$; the equivalent
standard deviation of the cumulative granulation noise background,
$\sigma$; and the timescale of the granulation, $\tau$. Once $\sigma$
and $\tau$ have been chosen, we use Equation~\ref{eq:siglm} to
determine the $\sigma_{lm}$ for each combination of ($l$,\,$m$). This
in turn determines the asymmetry, due to the correlated noise,
according to Equations~\ref{eq:ass} and~\ref{eq:har}.

\begin{figure*}


\centerline{\epsfxsize=5.5cm\epsfbox{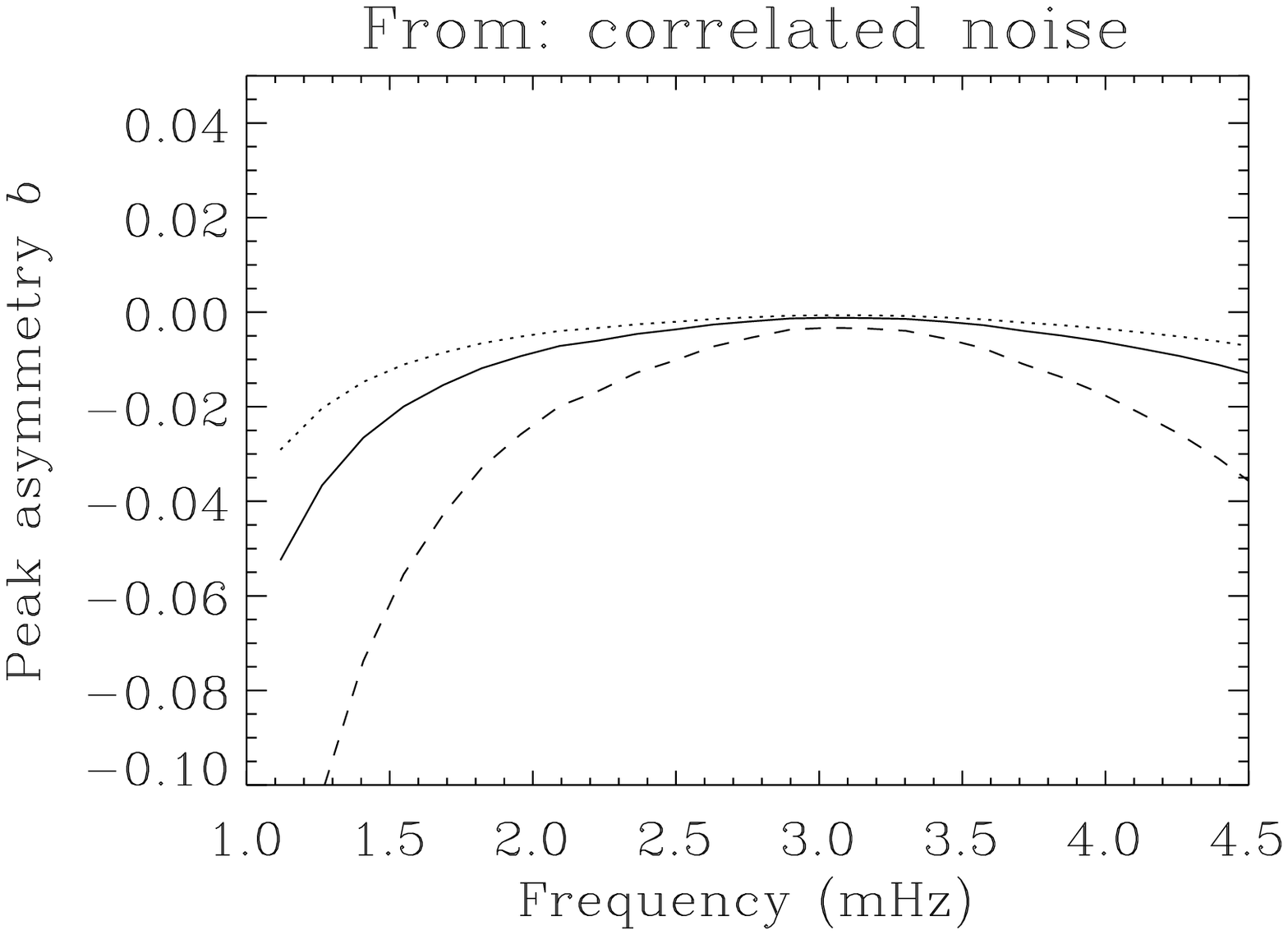}
               \epsfxsize=5.5cm\epsfbox{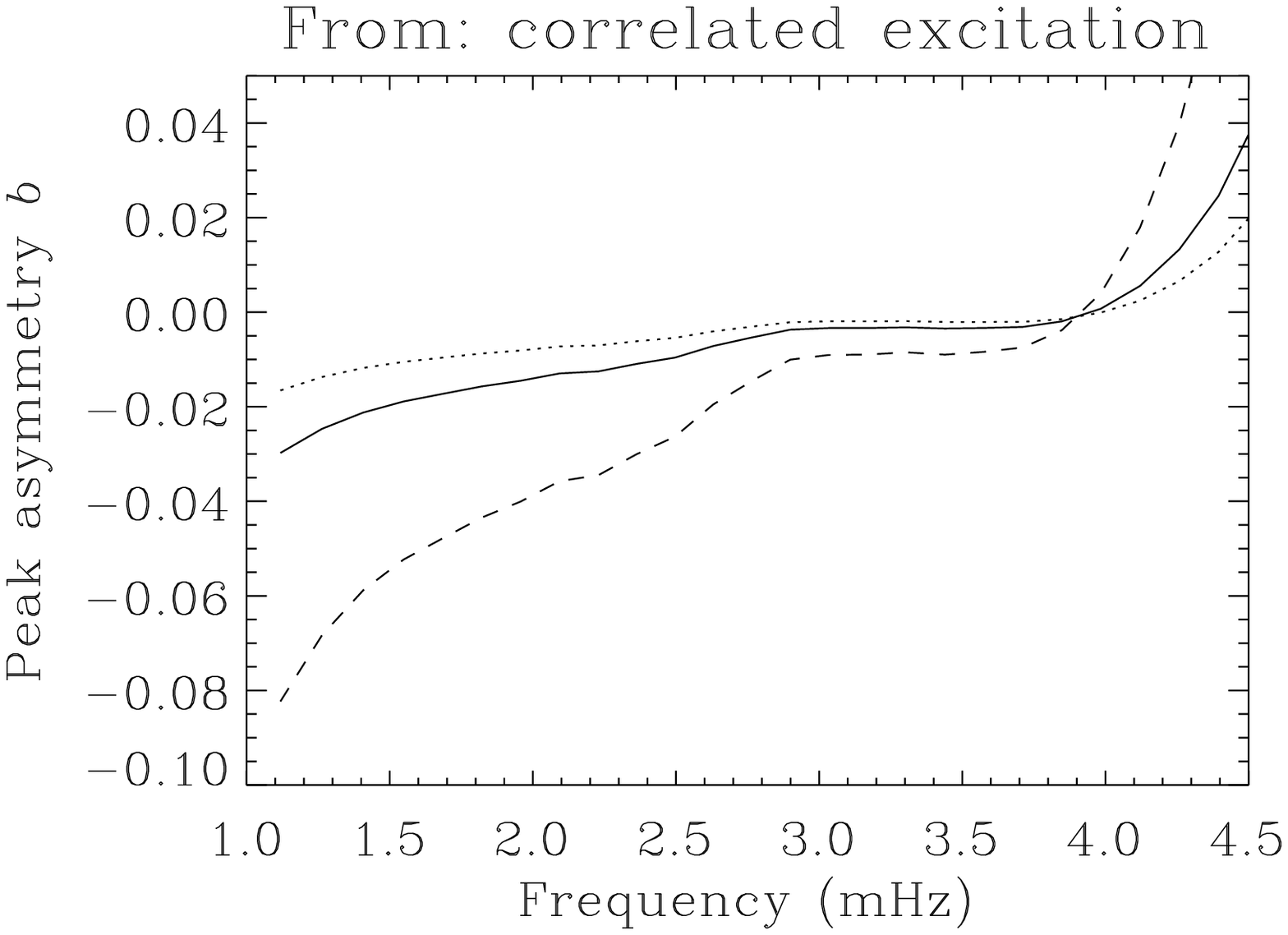}
               \epsfxsize=5.5cm\epsfbox{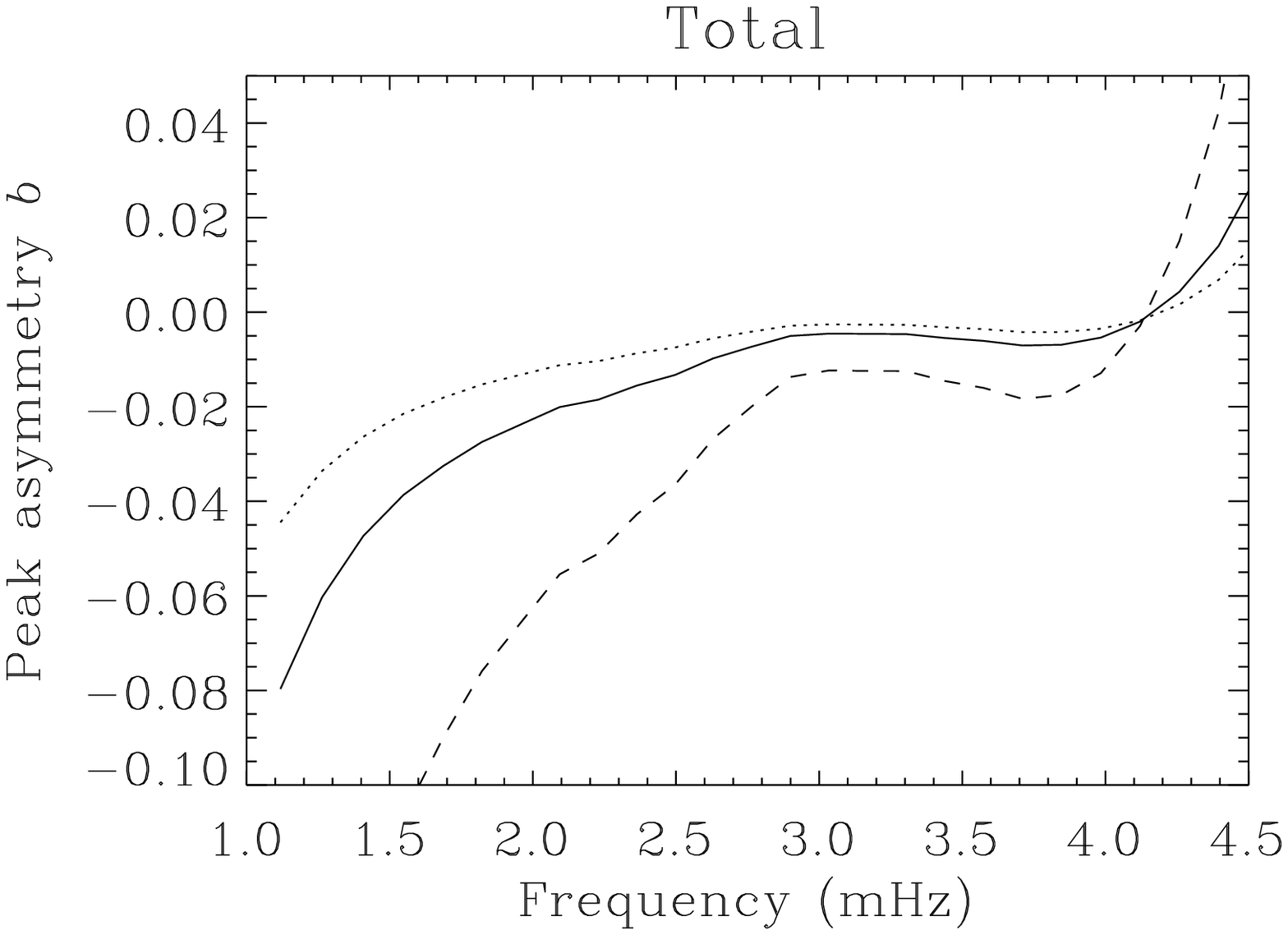}}

 \caption{Impact of choice of $\rho$ on mode peak asymmetry. The
 left-hand panel shows the asymmetry due to correlation with
 background noise (Section~\ref{sec:asscorr}); the centre panel the
 asymmetry due to correlated mode excitation
 (Section~\ref{sec:assex}); and the right-hand panel the total
 asymmetry, due to the complex interaction of the two
 contributions. All panels show results for $\sigma=0.2\,\rm
 m\,s^{-1}$ and $\tau=260\,\rm sec$. Different linestyles show results
 for different $\rho$: dotted for $\rho=-0.20$; solid for $\rho=-0.36$
 and dashed for $\rho=-1.00$.}

 \label{fig:assrho}
 \end{figure*}

The left-hand panel of Fig.~\ref{fig:assrho} shows resulting
asymmetries given by the parameters $\sigma=0.2\,\rm m\,s^{-1}$ and
$\tau=260\,\rm sec$. The dotted line shows results for $\rho=-0.20$,
the solid line for $\rho=-0.36$, and the dashed line for
$\rho=-1.00$. Here, we used negative $\rho$ to give the negative
asymmetries observed in Doppler velocity data. The magnitude of the
asymmetry increases in direct proportion to the magnitude of $\rho$ is
increased (this follows trivially from inspection of
Equation~\ref{eq:ass}). When $\rho=-0.36$ we get asymmetries that
resemble quite closely those seen in the real observations. The
asymmetries are largest at low and high frequency, because the ratio
of the granulation background to mode amplitude is highest there. Very
similar looking plots are given by varying $\sigma$; inspection of
Equations~\ref{eq:ass} and~\ref{eq:har} indicates that the magnitude
of the asymmetry is proportional to $\sigma$.

\begin{figure*}


\centerline{\epsfxsize=5.5cm\epsfbox{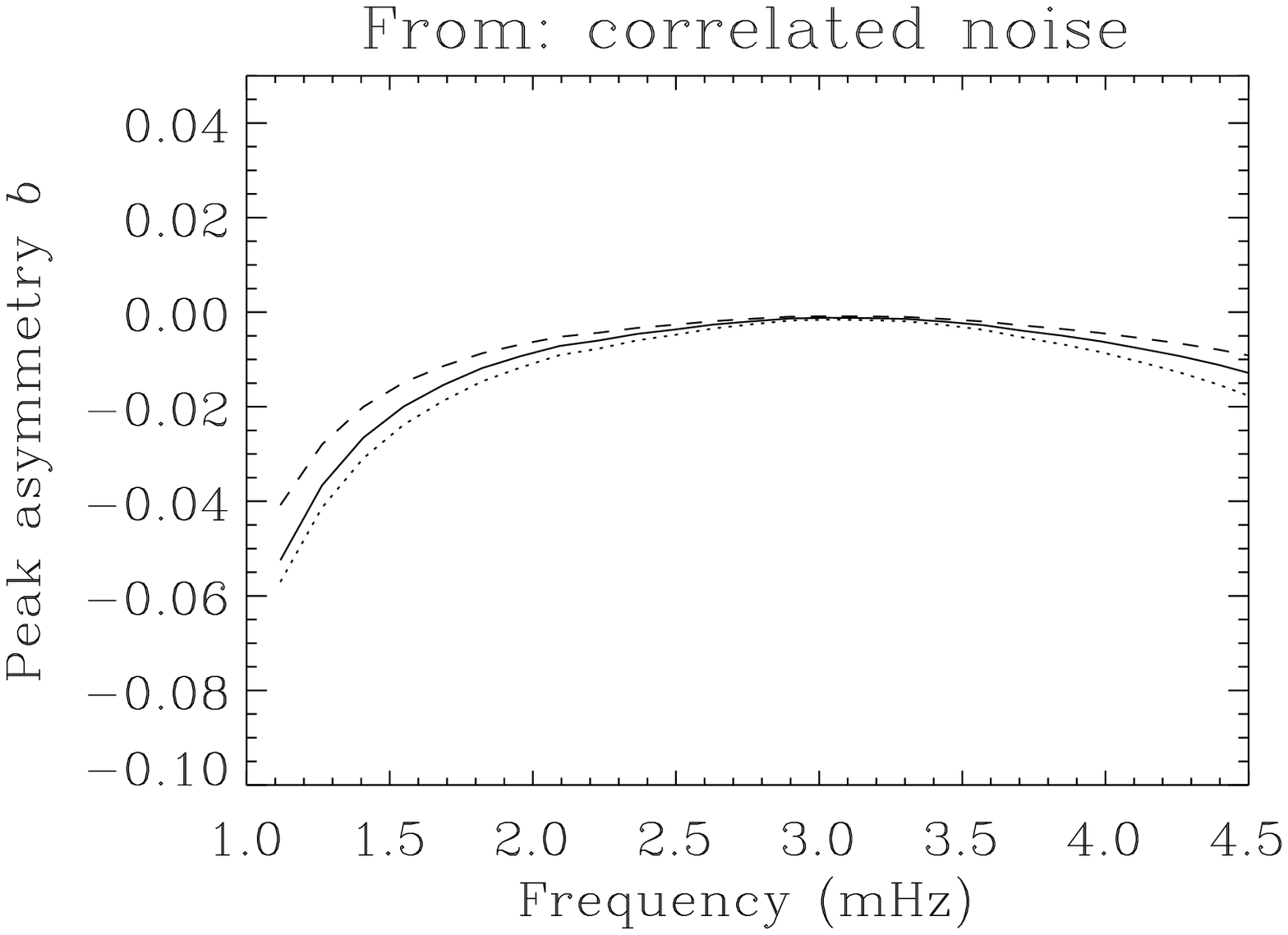}
               \epsfxsize=5.5cm\epsfbox{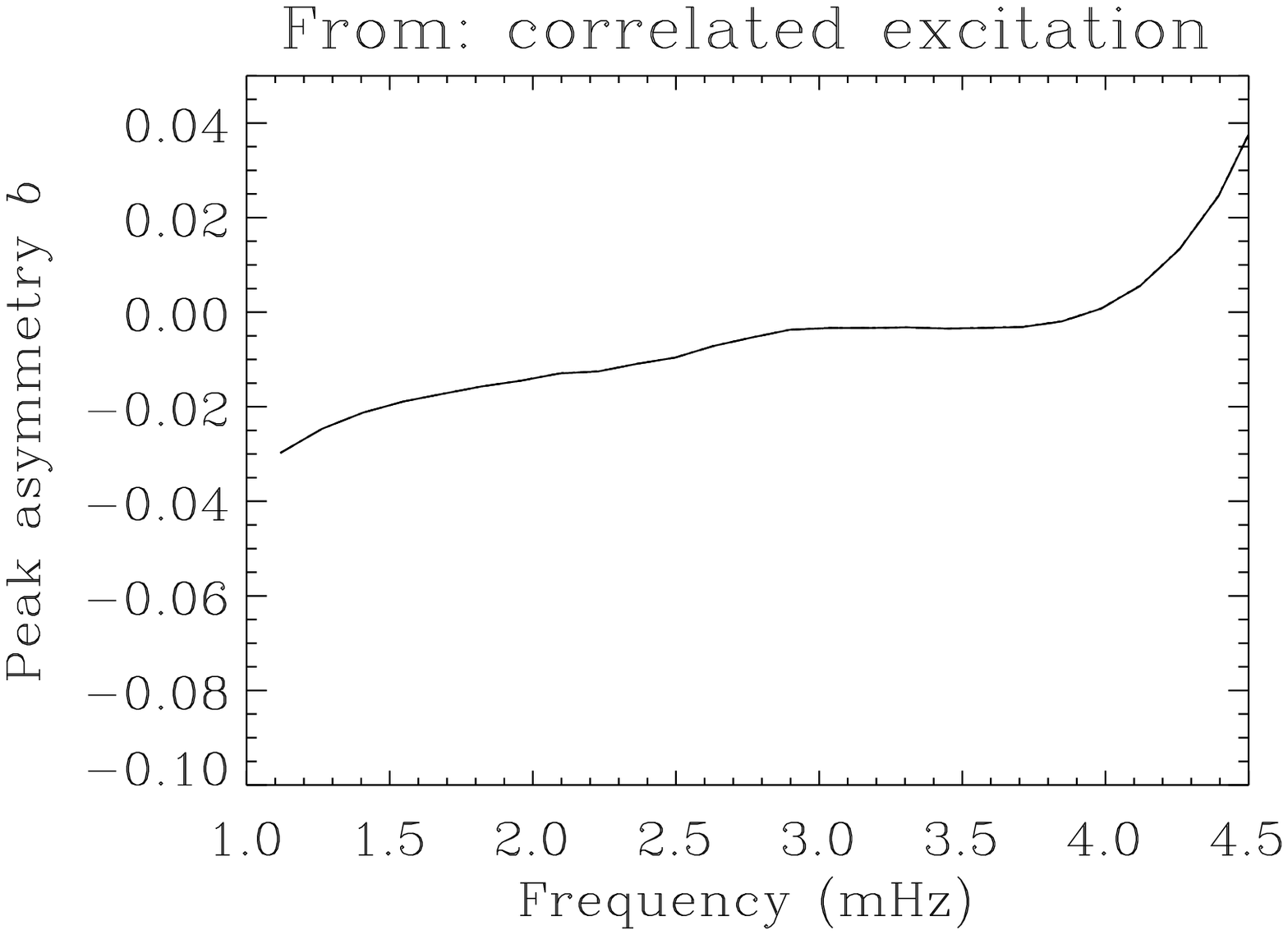}
               \epsfxsize=5.5cm\epsfbox{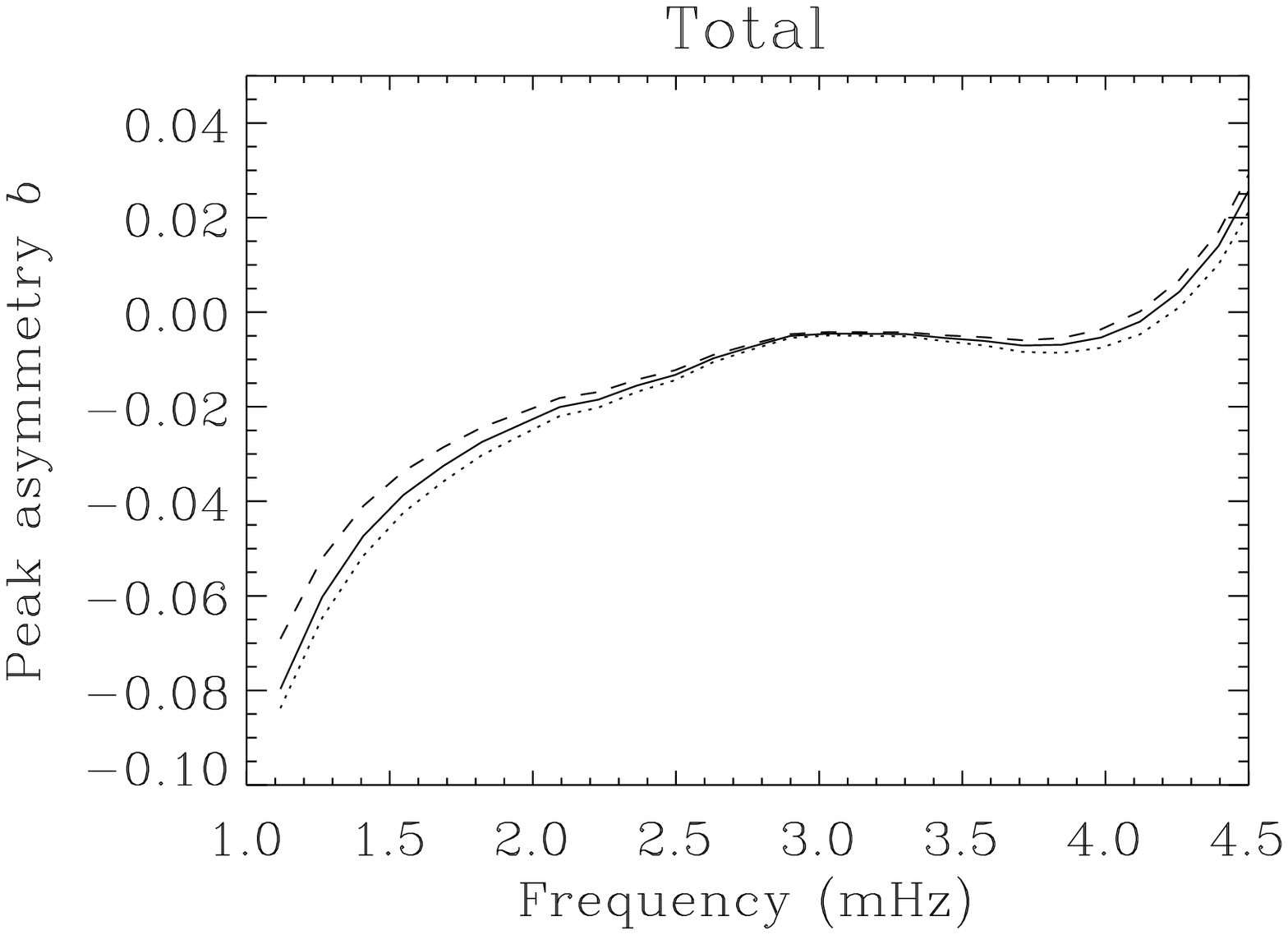}}

 \caption{Impact of choice of $\tau$ on mode peak asymmetry.  The
 left-hand panel shows the asymmetry due to correlation with
 background noise (Section~\ref{sec:asscorr}); the centre panel the
 asymmetry due to correlated mode excitation
 (Section~\ref{sec:assex}); and the right-hand panel the total
 asymmetry, due to the complex interaction of the two
 contributions. All panels show results for $\sigma=0.2\,\rm
 m\,s^{-1}$ and $\rho=-0.36$. Different linestyles show results for
 different $\tau$: dotted for $\tau=130\,\rm sec$; solid for
 $\tau=260\,\rm sec$ and dashed for $\tau=520\,\rm sec$.}

 \label{fig:asstau}
 \end{figure*}

The left-hand panel of Fig.~\ref{fig:asstau} shows instead the effect
of varying $\tau$ (for fixed $\sigma=0.2\,\rm m\,s^{-1}$ and
$\rho=-0.36$). The dotted line shows results for $\tau=130\,\rm sec$,
the solid line for $\tau=260\,\rm sec$ and the dashed line for
$\tau=520\,\rm sec$. Changes in $\tau$ clearly have a less pronounced
effect on the asymmetries than do similar relative changes in $\rho$.


 \subsubsection{Asymmetry due to correlated excitation}
 \label{sec:assex}

The contribution to the asymmetry 
overtones of the same ($l$,\,$m$) 
from the overlapping wings of  the overtones of the same ($l,\,m$) that have been excited with correlated noise 
is fixed by the frequency
separations, line widths and relative heights of those overtones, and
the choice of $\rho$. We later refer to this as `correlated excitation.' This contribution is far less amenable to a neat
analytical description than the correlated noise contribution
described above. For a given overtone it must describe, in the complex
plane, the contribution of power from all other overtones to which it
is correlated.  A full description of the complex frequency amplitude
spectrum, and frequency power spectrum, is given in
Appendix~\ref{sec:appfull}.
We may use the analytical descriptions
there to calculate numerically the asymmetries given to the mode peaks
by the correlated excitation. The resulting estimates are shown in the
central panels of Fig.~\ref{fig:assrho} and~\ref{fig:asstau}. Note
that the correlated excitation contribution does not depend on the
sign of $\rho$ (only the magnitude), and is independent of the choice
of $\tau$ (and $\sigma$).

At lower frequencies the asymmetry given to modes by the correlated
excitation is negative. Here, there are a larger number of other
overtones at frequencies above a mode than there are below it. The
reverse is true at higher frequencies, where the asymmetry is
positive. To understand the sign of the asymmetry in each region, we
refer back to the central panel of Fig.~\ref{fig:corr3}. There, we had
two correlated modes. The lower-frequency mode had negative asymmetry,
due to the correlated impact of its higher-frequency counterpart; 
the higher-frequency mode showed asymmetry of opposite sign, due
to the lower-frequency mode. We see this pattern repeated in the full
spectrum of overtones with the cross-over in behaviour located where
$p$-mode power is a maximum.

The right-hand panels of Fig.~\ref{fig:assrho} and~\ref{fig:asstau}
show the full asymmetry given to the modes.  It is worth pointing out
that this is not simply the sum of the asymmetries given by correlated
noise (left-hand panels) and correlated excitation (central
panels). Matters are a bit more complicated, because the correlated
noise and correlated excitation act together in a non-trivial manner
in the complex plane.

One final thing to mention with regard to the asymmetries is that
there is actually a third contribution, which comes from the fact that
the frequency response of the granulation-like noise used to excite
modes is not white (i.e., flat). The response in the vicinity of each
resonance of course rises with decreasing frequency, meaning there
will be a small negative asymmetry contribution. The effect is,
however, negligible. (The impact of the non-white response of the
excitation is modelled in Appendix~\ref{sec:appfull}.)

We have seen that the input frequencies, powers and line widths of
overtones, and the correlation coefficient $\rho$, fix the asymmetry
contribution from the correlated excitation. When in addition the
$\sigma$ and $\tau$ of the granulation-like noise are chosen the
asymmetry contribution from the correlated noise is also completely
specified.  Once any additional uncorrelated background has been
specified, these choices in principle also fix the observed S/N ratio
of the mode peaks. For the data analysed below, we chose {$\tau=260\,{\rm sec}$} -- the characteristic solar timescale --  and {$\rho=0.36$}, chosen to give a good match to the asymmetry measured in previous work. These values correspond to the solid curves in {Figures~\ref{fig:assrho} and \ref{fig:asstau}}. We also chose {$\sigma=0.2\,{\mathrm m\,s^{-1}}$} for correlated and {$\sigma=0.25\,{\mathrm m\,s^{-1}}$} for uncorrelated noise.  
We now go on to describe the S/N, in terms of the
commonly used background-to-height ratio, $\beta$.

 \subsubsection{Background-to-height ratio in mode peaks}
 \label{sec:beta}

The intrinsic background-to-height ratio, $\beta_{nlm}$, in a mode
peak is given by:
 \begin{equation}
 \beta_{nlm} = \frac{N(\nu_{nlm})}{H_{nlm}},
 \label{eq:beta}
 \end{equation}
where $N(\nu_{nlm})$ is the total background at the resonance. It has
a part due to the cumulative granulation background of all ($l$,\,$m$)
that appear in the time series. There will also be parts due to other
sources of uncorrelated noise. In what follows we specify just one of
the possible components: photon shot noise, $N_{\rm psn}$. This shot
noise is specified by its variance, $\sigma_{\rm psn}^2$. The power
spectral density due to this noise term is then:
 \begin{equation}
 N_{\rm psn} = 2 \sigma_{\rm psn}^2\, \Delta\,t.
 \label{eq:psn}
 \end{equation}
The total background may therefore be written:
 \begin{equation}
 N(\nu_{nlm}) = N_{\rm psn}+ \sum_{{\rm all}~(l,m)} n_{lm}(\nu_{nlm}),
 \label{eq:bg1}
 \end{equation}
or explicitly:
 \begin{equation}
 N(\nu_{nlm}) = 2 \sigma_{\rm psn}^2\, \Delta\,t +
                  \frac{4 \sigma^2 \tau}{1+(2 \pi \nu_{nlm} \tau)^2}.
 \label{eq:bg2}
 \end{equation}
The above implies that the intrinsic background-to-height ratio is in
principle specified by the choice of four parameters (when no
instrumental or other sources of noise are specified): the height of
the mode, $H_{nlm}$; the $\sigma$ and $\tau$ of the granulation-like
noise background; and the $\sigma_{\rm psn}$ of the uncorrelated
shot-noise background.

\subsubsection{Solar-cycle effects}

It is a fairly straightforward matter to include solar-cycle-like variations in the artificial timeseries data. Changes in frequency may be introduced by varying in time the natural frequency of the damped, driven oscillator used to simulate each mode component; whilst changes in amplitude and damping may be introduced by varying in time the damping constant for the oscillator. In all cases we require changes
that mimic those seen in the real data.  We base these changes on an artificial proxy of the real 10.7-cm radio flux variations observed over solar activity cycle 23.  Variations in the frequencies and damping rates are programmed to follow the artificial proxy, with appropriate calibration constants used to fix the absolute scale of the variation for each parameter.

Here, we used a high-order polynomial fit to the 10.7-cm radio flux versus time -- for the 11-year period beginning 1997 February 3 -- to provide the proxy activity, which we refer to as $X$ in later sections of the paper. The purpose of the fit was to provide a smooth proxy that captures the main 11-year cycle while removing shorter-term variations. Calibration constants for the frequencies and damping rates of every simulated mode were based on results from previous analyses of real BiSON data. The calibration constants for the frequencies depend not only on frequency -- the higher the frequency, the larger is the solar-cycle frequency shift -- but also the angular degree, $l$, and the azimuthal order, $m$, of each simulated component (e.g. Chaplin et al. 2004). The dependence on ($l$,\,$m$) is due to the strong latitudinal dependence of the real near-surface activity, which drives the changes in the mode parameters.

As noted earlier, due to the relatively large uncertainties in the measurements little information is available on the frequency and ($l$,\,$m$) dependence of detected solar-cycle changes in the mode amplitudes and damping rates for low-degree modes, but we know from medium- and high-degree studies \citep[e.g.][]{2000ApJ...531.1094K,2004ApJ...608..562H} that the changes are strongest in the middle of the five-minute band. However, the relative sizes of the variations are consistent with the hypothesis that both are the result of net changes to the damping only \citep[e.g.][]{2000MNRAS.313...32C}. Here, we have introduced solar-cycle variations that adhere to this finding, so that in our simulations we change only the damping constant. 
Further information on the introduced variations is included later in Section~\ref{sec:artdata}.

Finally, we note that variations in amplitude lead to solar-cycle-like changes in the mode peak asymmetries of our artificial data. This is because the height-to-background ratios of the mode peaks changes with the artificial proxy, due to variations in the intrinsic amplitudes (the simulated granulation and shot-noise components were stationary in time). Again, information on the programmed variations is in Section~\ref{sec:artdata}.

\section{Fitting methods}
\label{sec:fitting}

\begin{figure*}

\centerline{\epsfxsize=0.4\linewidth\epsfbox{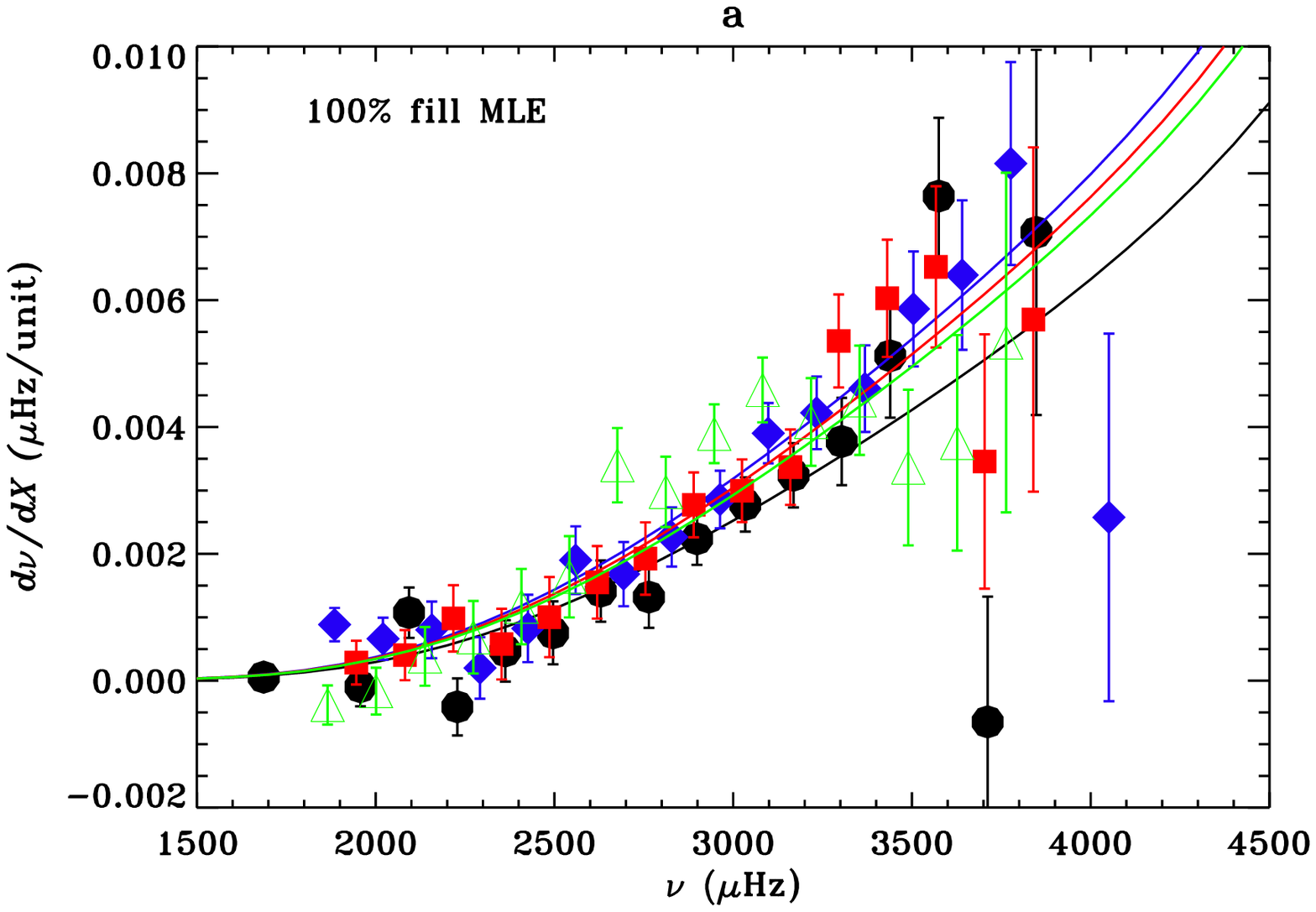}
\epsfxsize=0.4\linewidth\epsfbox{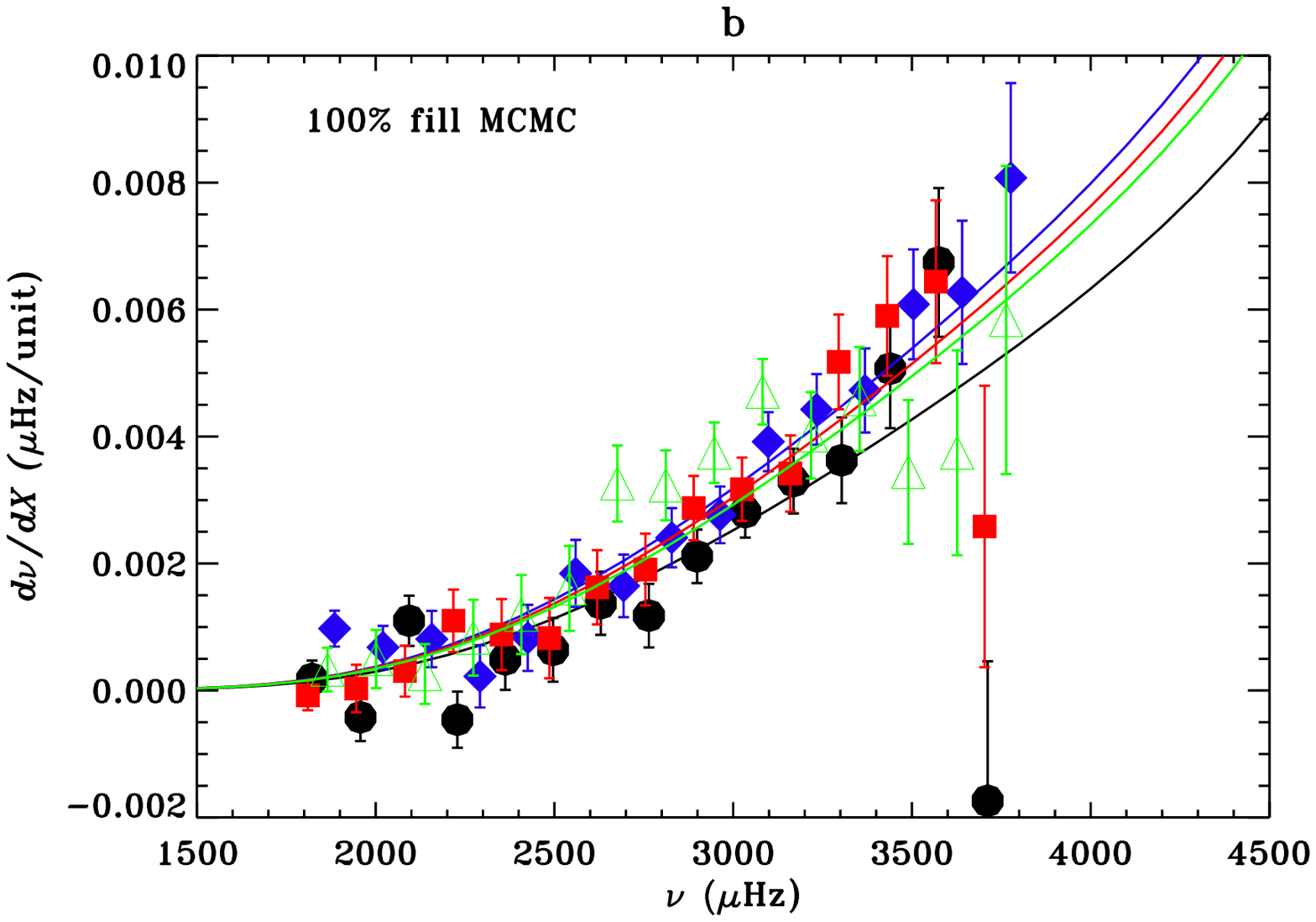}}
\centerline{\epsfxsize=0.4\linewidth\epsfbox{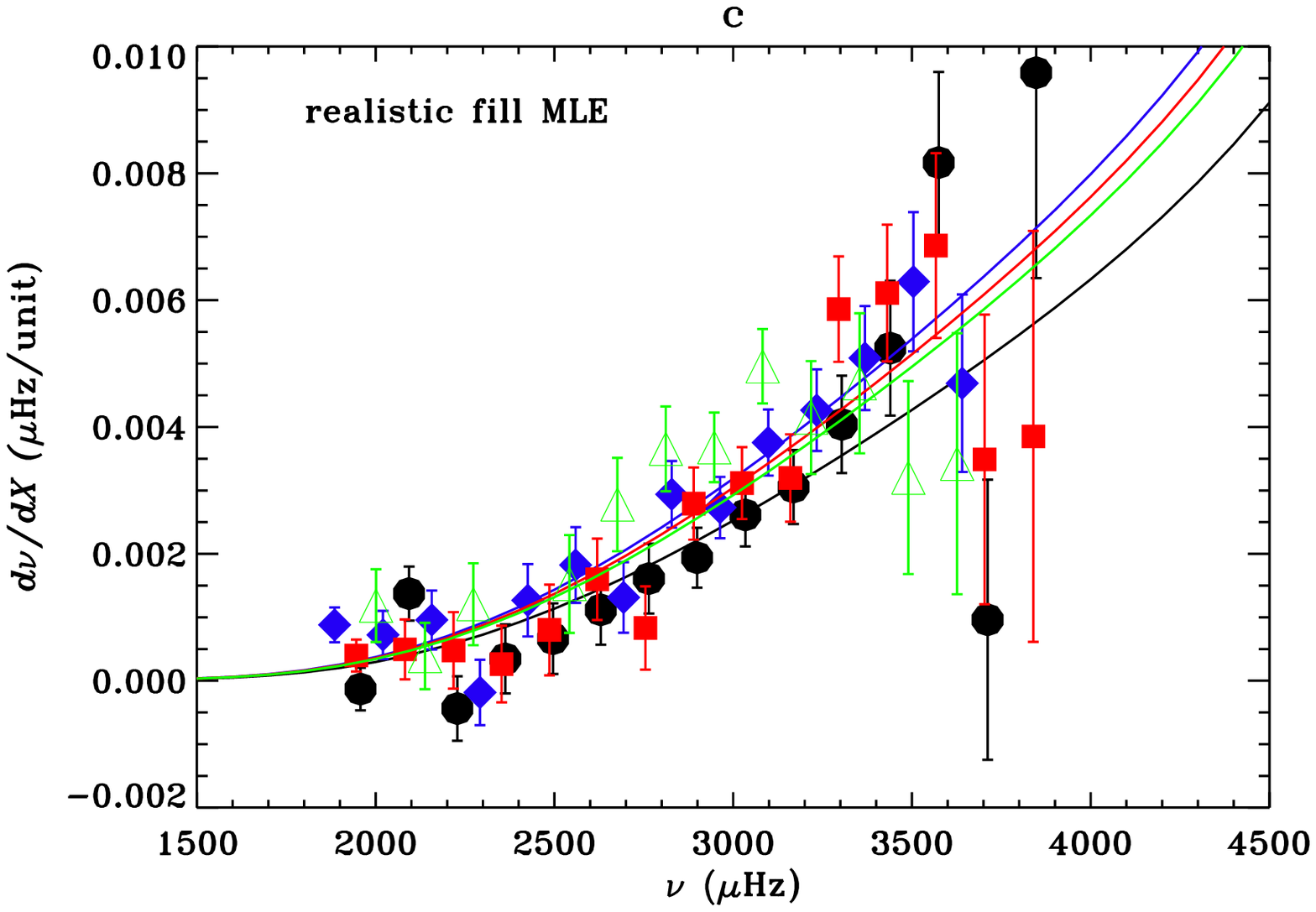}
\epsfxsize=0.4\linewidth\epsfbox{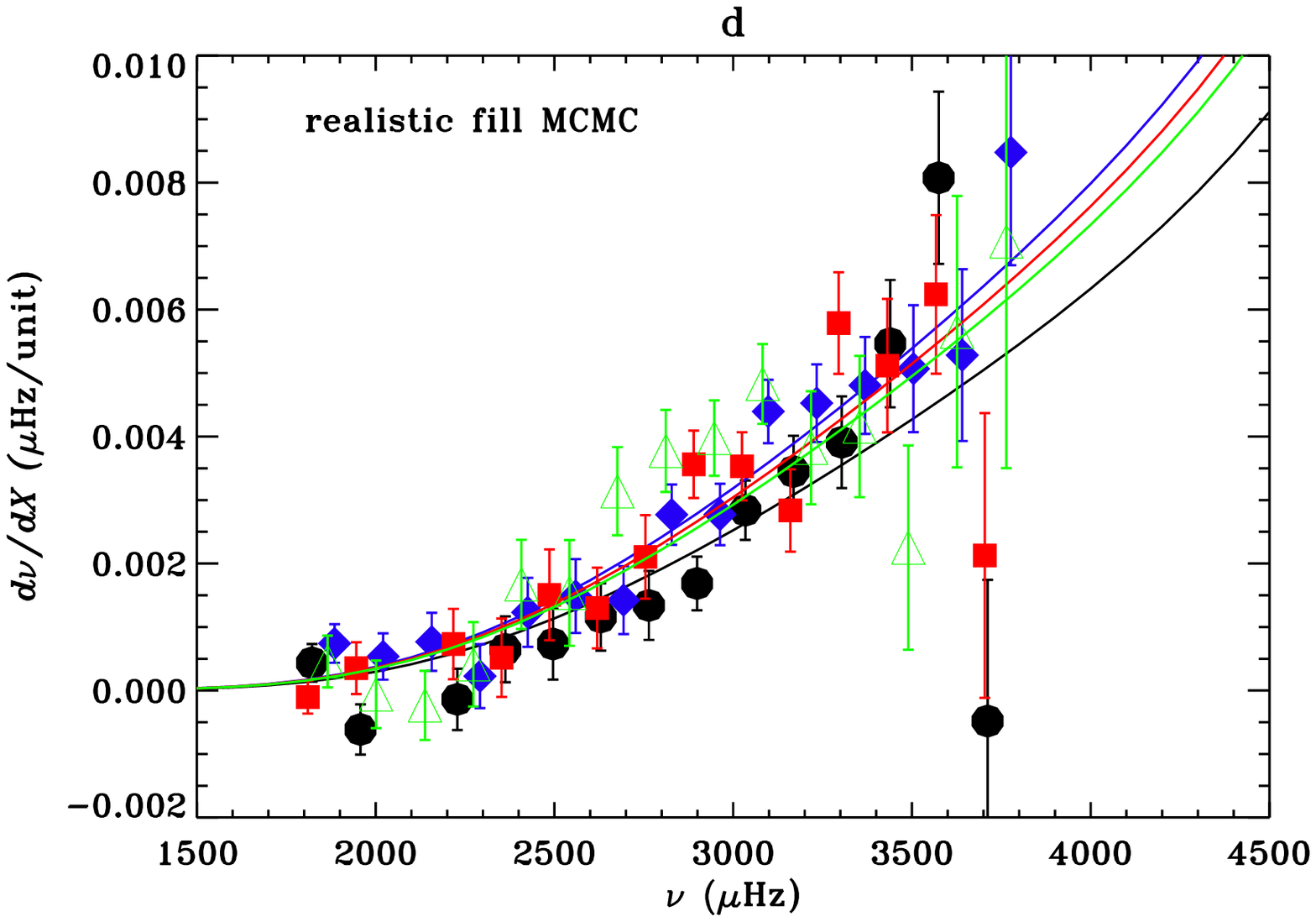}}

\epsfxsize=0.4\linewidth\epsfbox{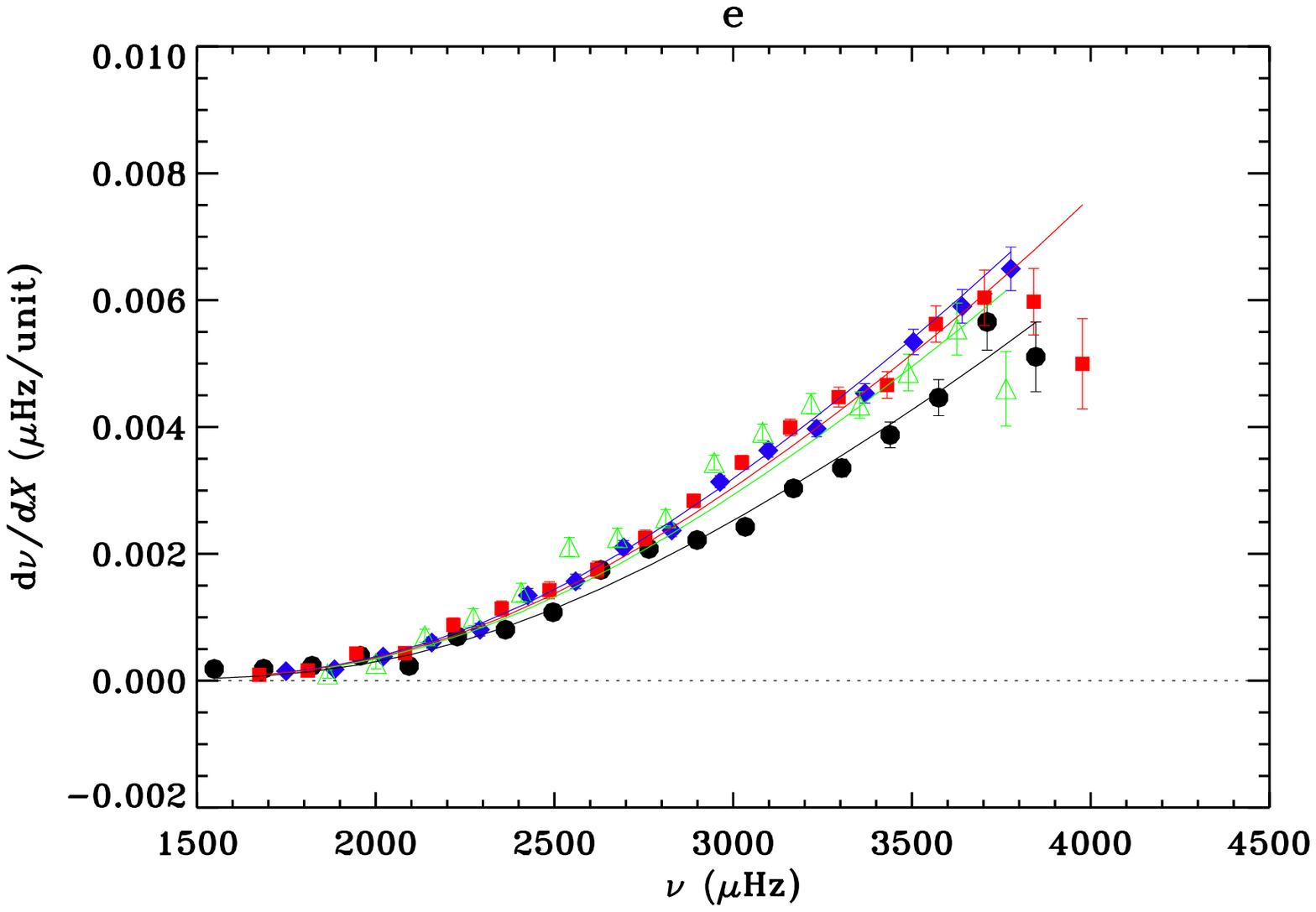}
\epsfxsize=0.4\linewidth\epsfbox{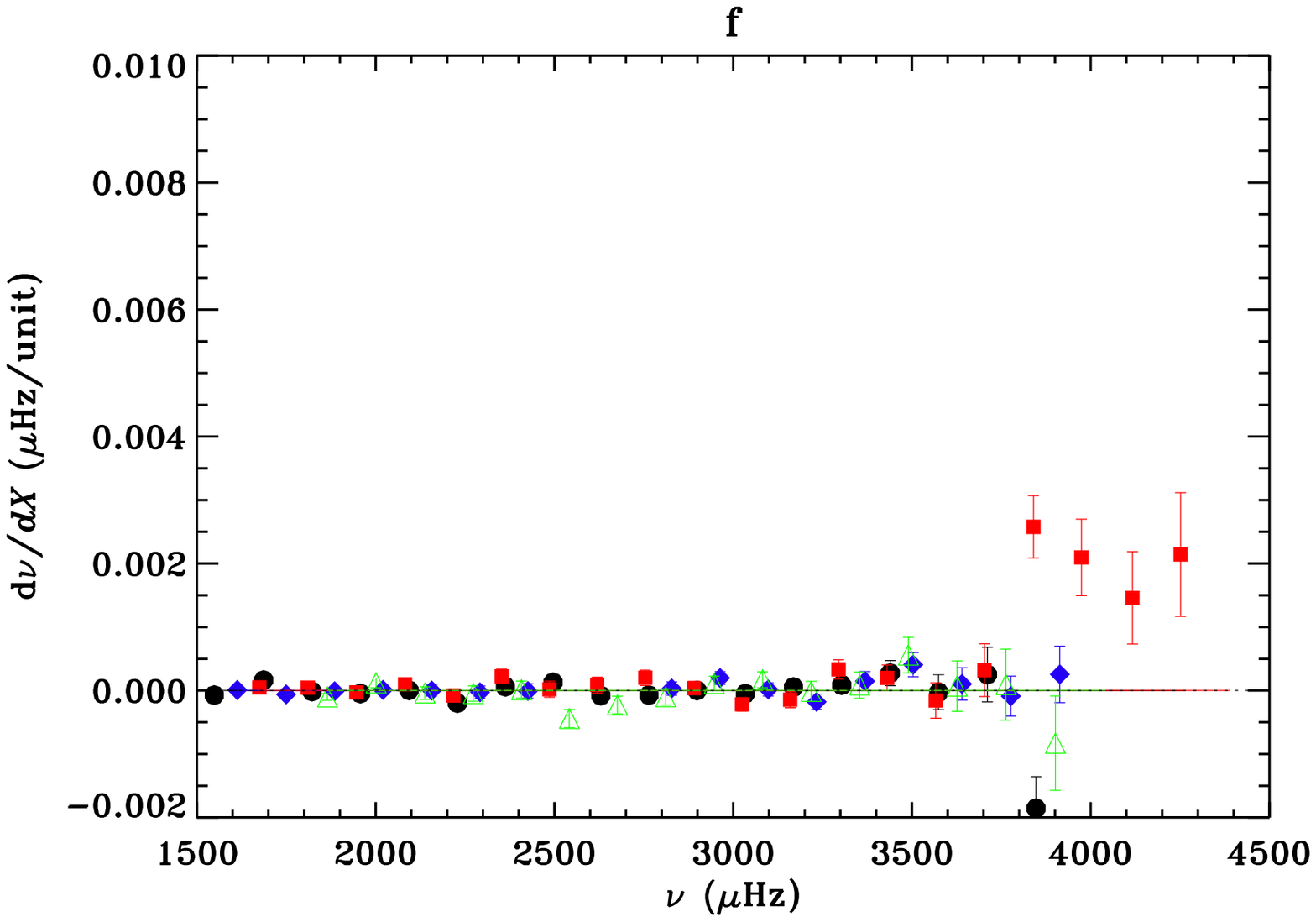}

\epsfxsize=0.4\linewidth\epsfbox{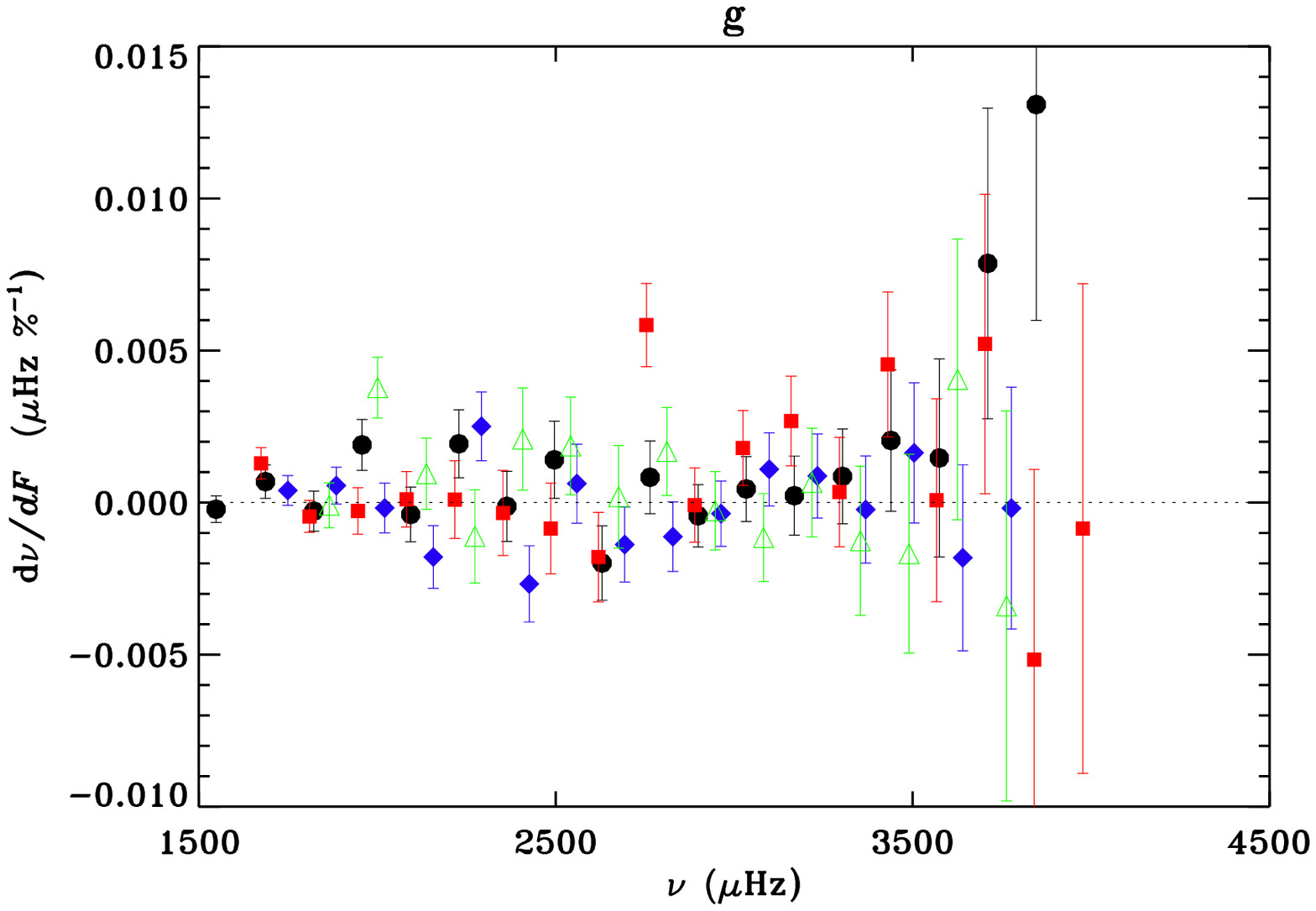}
\epsfxsize=0.4\linewidth\epsfbox{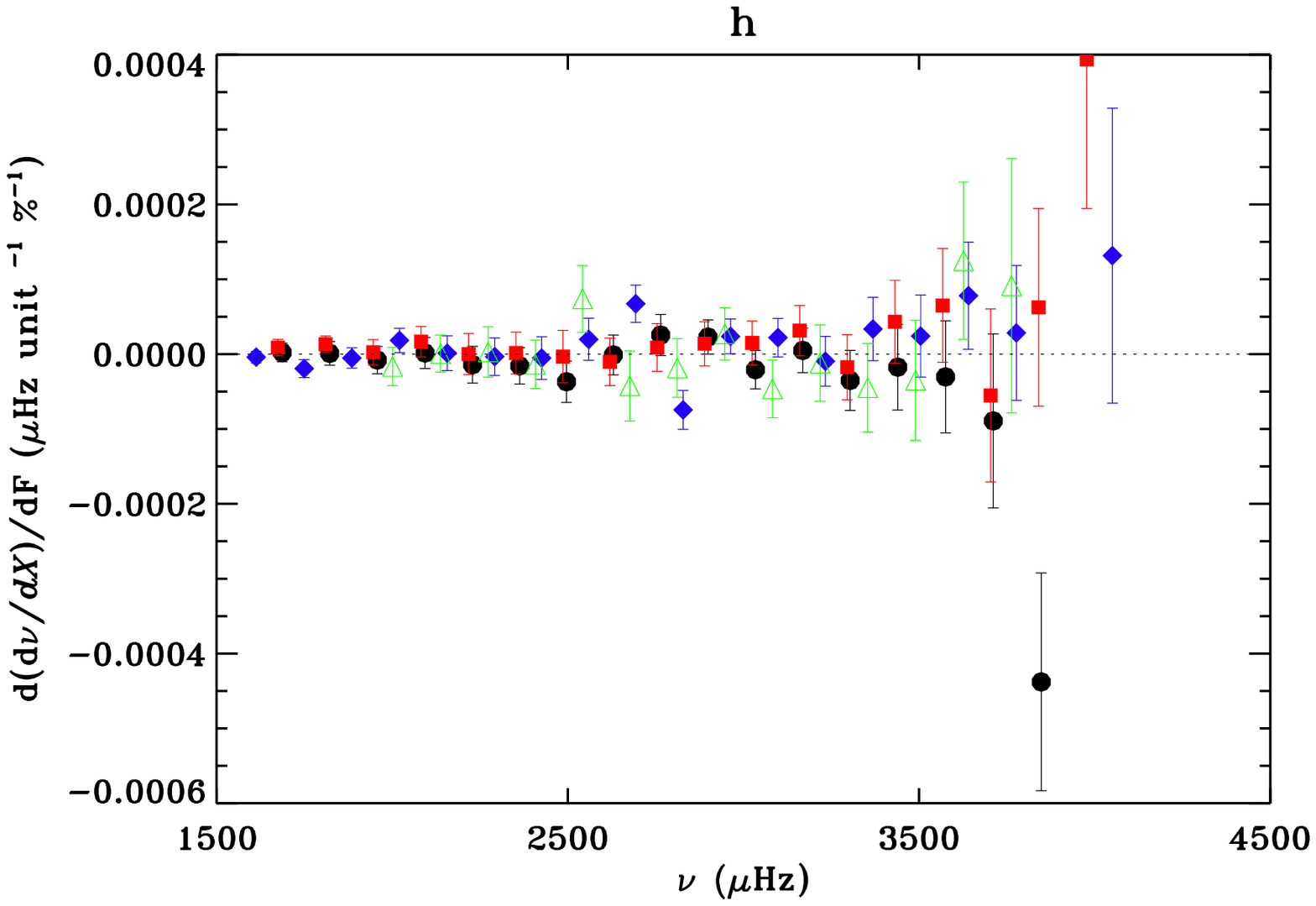}

\caption{Sensitivity of fitted frequency to activity level and fill for artificial data. Panels (a)--(d) show fitted (symbols) and input (lines) $d\nu_{nl}/dX$
for MLE (a, c) and MCMC (b,d) fits to 11 one-year sets of artificial data at different activity levels, mimicking the solar cycle.  The results are for 100 per cent (a, b) and realistic (c, d) duty cycle. The lower two rows, for MLE only, show the results of the tests with multiple realisations of the artificial data. Panel (e) shows the sensitivity of the frequency to activity index, $X$; panel (f) is similar to panel (e) but for a set of artificial time series where there is no input frequency variation. Panel (g) shows the sensitivity to duty cycle, $F$, for the test with 25 realisations of the artificial data. Panel (h) shows the rate of change with duty cycle $F$ of the sensitivity $d\nu/dX$ of mode frequency to activity index, for the  test where each of the 21 one-year BiSON window functions was applied to each of the 11 years of the SolarFLAG time series.  
Black circles represent $l=0$, blue diamonds $l=1$, red squares $l=2$, and
green open triangles $l=3$.}
\label{fig:flag2}
\end{figure*}

Part of the object of the current work is to compare the results from two different fitting algorithms. Both of these are conventional in the sense that they
use the ``pairwise'' fitting approach in which each $l=0/2$ and $l=1/3$ pair of peaks is fitted independently with its own background and parameter set.

The two methods we use are  labelled as MLE (Maximum Likelihood Estimate) and MCMC (Markov Chain Monte Carlo) but are not solely described by the method of optimization.  There are differences in the power spectrum model that we detail below.

\subsection{Mode Parametrization}
Common to both fitting methods, each mode component of degree $l$, radial order $n$ and azimuthal order $m$ is represented by a peak profile described by the equation
\begin{equation}
\label{eq:rh1}
P(\xi)=\left({h\over{1+\xi^2}}\right)\times[(1+b\xi)^2],
\end{equation}
where,
\begin{equation}
\xi=2(\nu-\nu_0)/\Gamma,
\end{equation}
$\nu_0$ is the frequency of the Lorentzian component, $\Gamma$ its
width,
$h$ its height, and $b$ a fractional parameter characterizing the
asymmetry. This expression simplifies to the  normal Lorentzian for $b=0$.
This is the profile of \citet{1998ApJ...505L..51N}, with the quadratic term in $b$ suppressed for greater stability \citep{2009ApJ...694..144F}. For this work, we have chosen to parametrize the peak in terms of the amplitude, $A$, where
\begin{equation}
\label{eq:rh2}
h=A^2/(\pi\Gamma).
\end{equation}
Neglecting the small asymmetry term, $A^2 \equiv \pi\Gamma h$ is proportional to the integrated energy of the mode 
\citep[see, for example,][but note that their $A$ is our $h$]{2000ApJ...543..472K}, and the energy supply rate $dE/dt$ is given by

\begin{equation}
{dE/dt} \propto A^2\Gamma  \propto \Gamma^2 h.
\label{eq:rh3}
\end{equation}

The likelihood spaces, (or the posterior probability distributions) are log-normal for the mode width and mode amplitude parameters, so we follow the common practice of varying the logarithms of the amplitude and width parameters rather than their raw values.

The background term is a constant for each
$l=0/2$ or $l=1/3$ pair, 
and in both algorithms the
asymmetry term $\alpha$ is also common to all of the peaks in each pair. In the code used for the MLE algorithm there is a separate $\Gamma$ for each $n$ and $l$, while the MCMC code uses the same $\Gamma$ for every peak within an $l=0/2$ or $l=1/3$ group. There is a separate amplitude parameter for each $n$ and $l$, but  the relative heights of the rotationally split components of different $m$ within an $n,l$ group are fixed in the MLE code at 1:0.41 for the $m=2:m=0$ component ratio for $l=2$ and $1:0.19$ for the $m=3:m=1$ ratio for $l=3$. In the MCMC code they are allowed to vary; the priors are set at a normal distribution of width 0.2 centered on 0.4 for the $l=2,m=0:l=2,m=2$ ratio and the $l=3,m=1:l=3,m=3$ ratio while for the $l=3,m=1:l=3,m=3$ ratio
we use a flat-topped Gaussian centered on 0.2 that is flat between 0.1 and 0.3 and has Gaussian sides that drop off with width 0.1. As we will see, the differences have no obvious impact on the final results.

The rotational splitting parameter was fixed at $0.4\,\mu{\mathrm Hz}$ for the MLE algorithm, while for the MCMC it was left as a free parameter with a Gaussian prior with centre $0.4$ and width $0.05\mu{\mathrm Hz}$. 

\subsection{Optimization}

The Fourier power spectrum has the statistics of $\chi^2$ with two degrees of freedom
rather than Gaussian statistics. The quantity to be mimimized is the negative log of the likelihood function \citep{1990ApJ...364..699A},
\begin{equation}
S \equiv -\ln{L}=\sum_i{\left\{{\ln{M_i} + {{O_i}\over{M_i}}}\right\}},
\end{equation}
where $M$ is the model, $O$ is the observed spectrum, and the sum is over the data points (frequency bins) in the fitting window.

In the MLE approach we simply find the values of the model parameters that give
the minimum value of $S$ and estimate the errors by inverting the Hessian matrix. For MCMC, on the other hand, we take the standard deviation of the posterior distribution for each parameter.

The Maximum Likelihood Estimation (MLE) code used here was a slightly modified version of the one developed for use with BiSON data and described by \citet{1999MNRAS.308..424C}. 
The MCMC code is a vanilla Markov Chain Monte Carlo approach \citep[see][for more details]{Gelman03,2014MNRAS.439.2025D}.

\subsection{Treatment of window function}
It is a well-known problem in ground-based helioseismology that the Earth's rotation causes the observations to be interrupted with a 24-hour periodicity, which gives rise to `sidelobe' peaks at multiples of $11.57\,\mu{\rm Hz}$ from each mode. At low degrees, where $l=0/2$ and $l=1/3$ pairs are separated in frequency by a similar amount, this is particularly problematic. The effect is greatly reduced by the use of multi-site observations but cannot be entirely eliminated, and it must therefore be taken into account in the data analysis.

As is conventional, we describe the modulation of the time series by a `window function' -- a time series corresponding to the observations in which each time sample is represented by a one for an observation and a zero for missing data. The observed spectrum can be considered as the convolution of the power spectrum of this window function with that of the uninterrupted observations.

In the MLE code used here, the sidelobe peaks are represented in the fitting model by peaks of the same width and asymmetry as each main peak, at $\pm 11.57\,\mu{\rm Hz}$ from its central frequency. The initial guess for the relative amplitude of this sidelobe peak was estimated from a spectrum in which a single sine wave was convolved with the window function of the time series. For modes in the main five-minute band  (between 2.0 and 3.5 mHz) this relative height was then allowed to vary in the first pass of fitting and held fixed for modes outside this frequency range. The mean value of the fitted sidelobe after the first pass was then used as the fixed value for all modes in the final pass. This allowed a larger number of successful fits to be recovered than if the sidelobe height had been left as a free parameter throughout.

The MCMC algorithm uses a different, more accurate but more computationally expensive approach in which the model spectrum is convolved with the window function spectrum at each optimization step. This method can also account for the broadening of the peak as the duty cycle of the observations is decreased; however, neither this approach nor the one used with the MLE algorithm take into account the correlations between frequency bins introduced by the missing data. Note that the treatment of the sidelobes is related to the specifics of the codes used here and is not intrinsic to the MLE or MCMC approach; in principle it would be possible to perform MLE with the convolution, as was indeed done for example by  \cite{2007ApJ...654.1135J,2009ApJ...694..144F}.

\section{Artificial data tests}
\label{sec:artdata}

\begin{figure*}
\centerline{\epsfxsize=0.4\linewidth\epsfbox{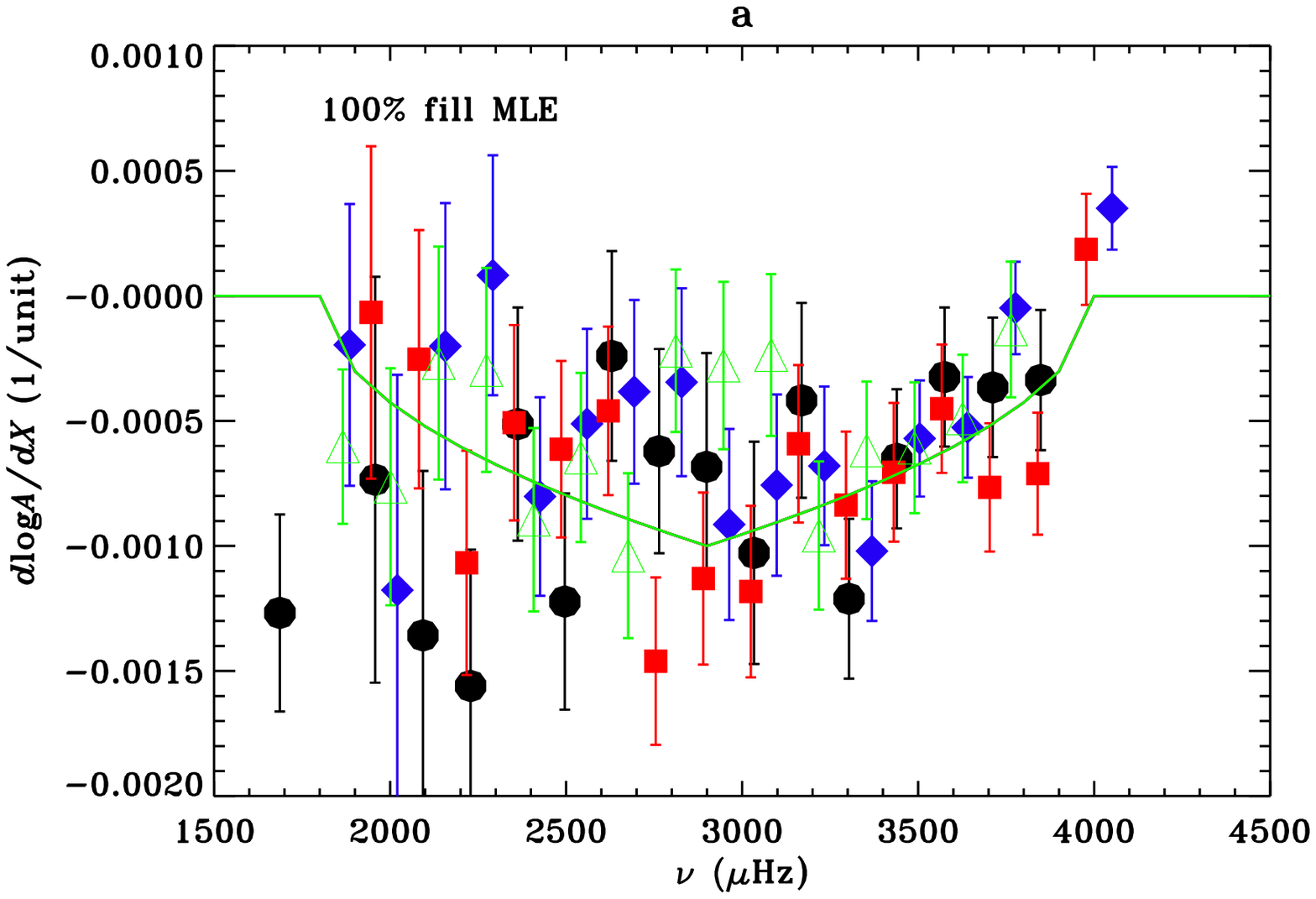}
\epsfxsize=0.4\linewidth\epsfbox{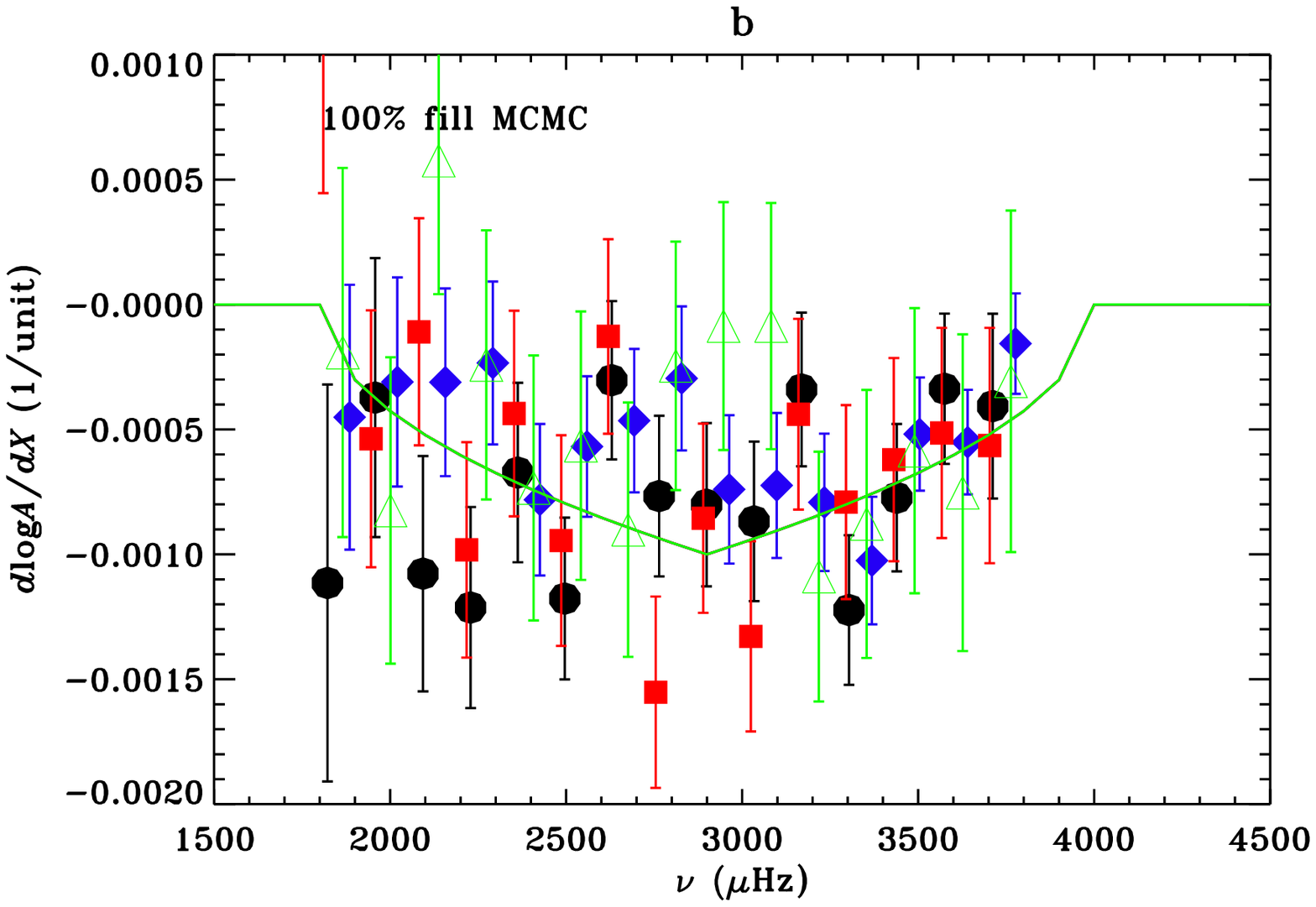}}
\centerline{
\epsfxsize=0.4\linewidth\epsfbox{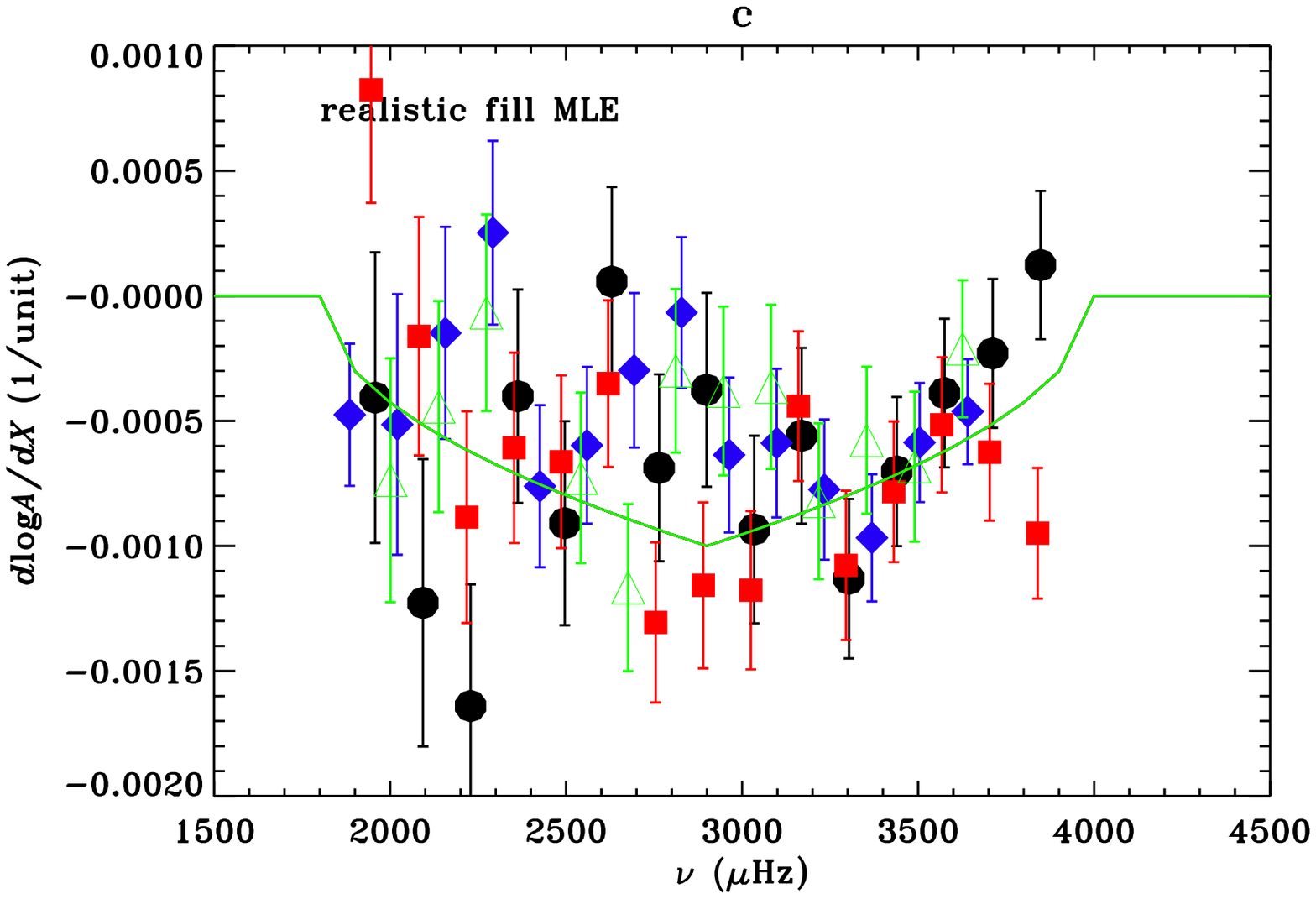}
\epsfxsize=0.4\linewidth\epsfbox{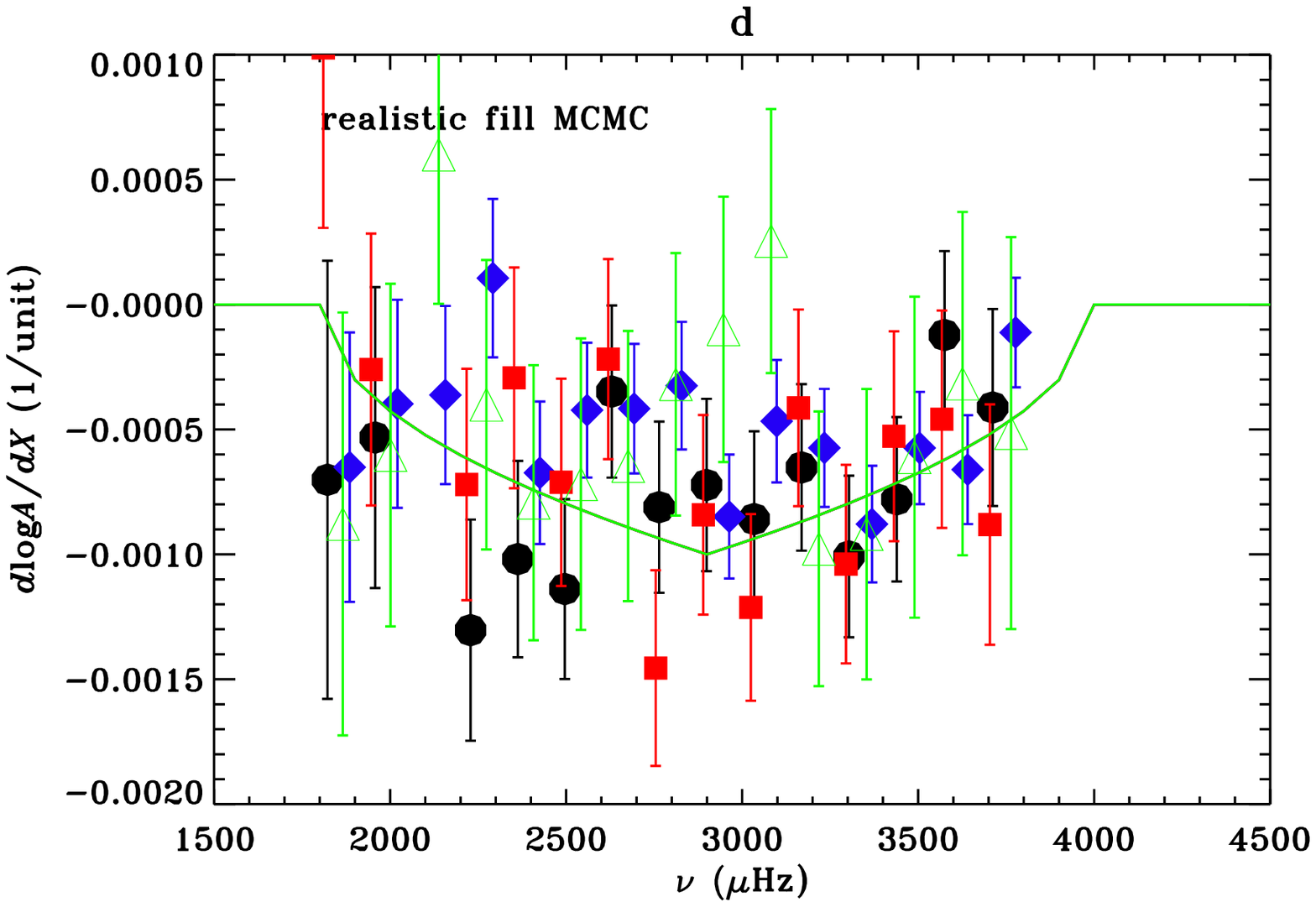}}

\epsfxsize=0.4\linewidth\epsfbox{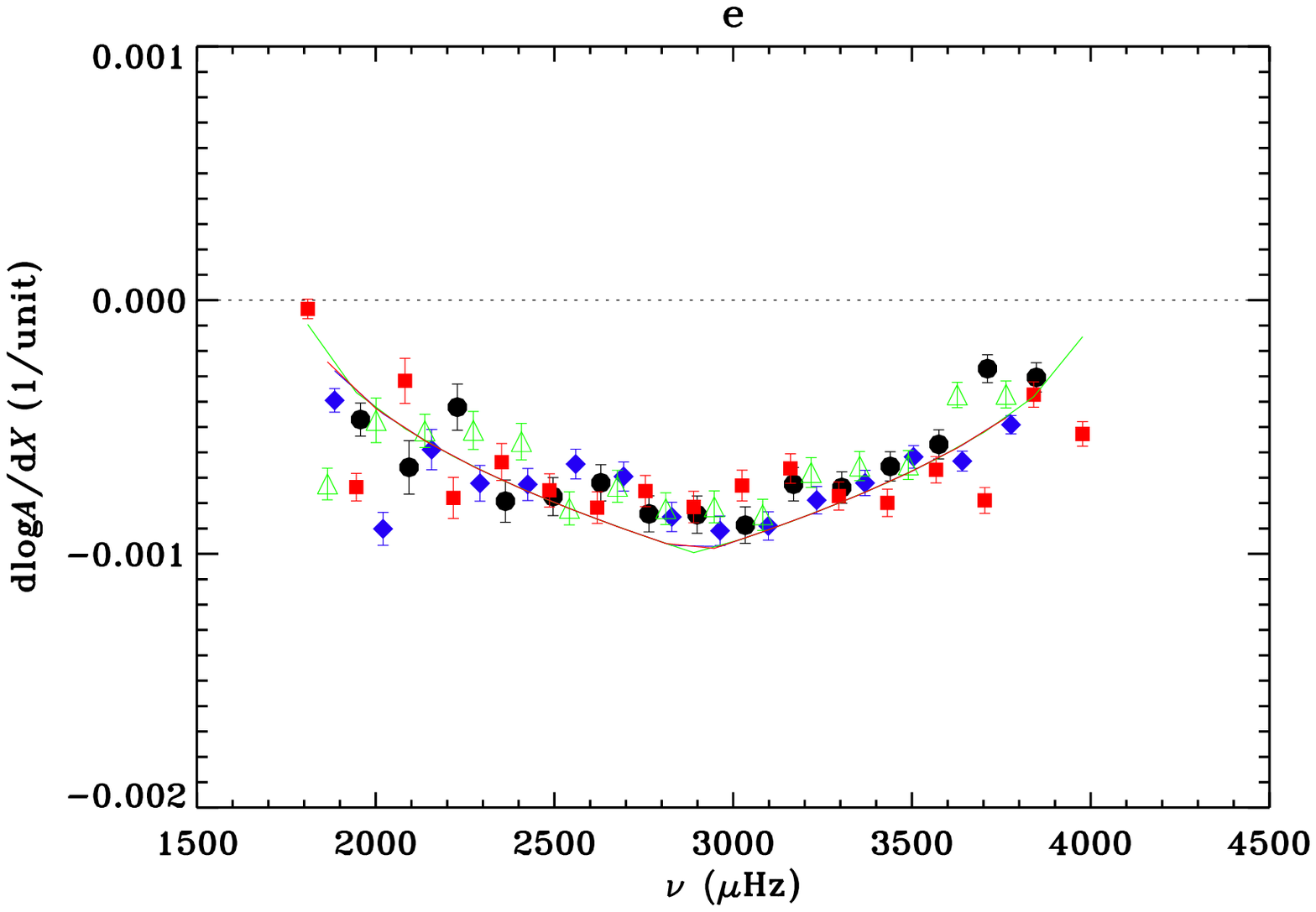}
\epsfxsize=0.4\linewidth\epsfbox{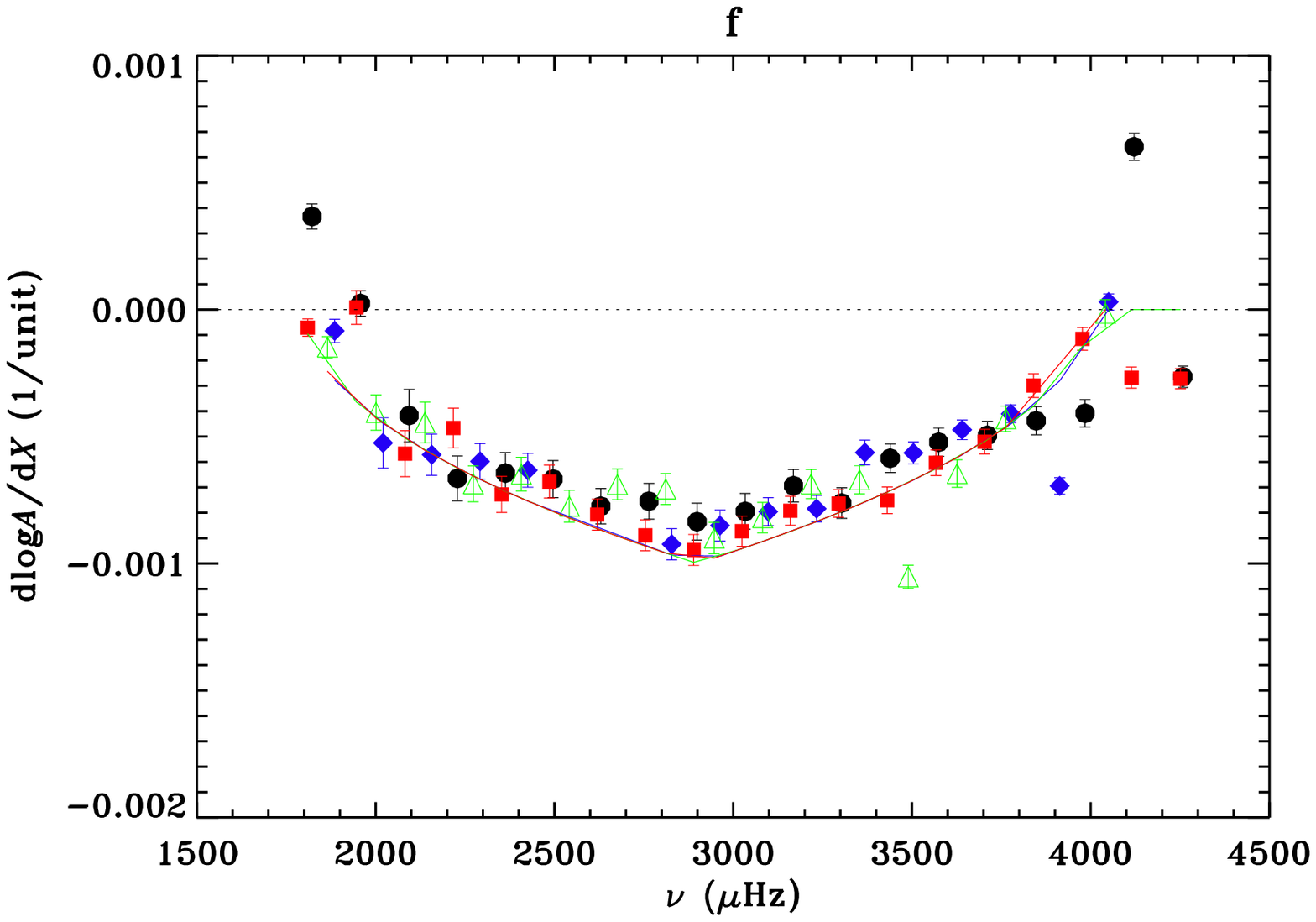}

\epsfxsize=0.4\linewidth\epsfbox{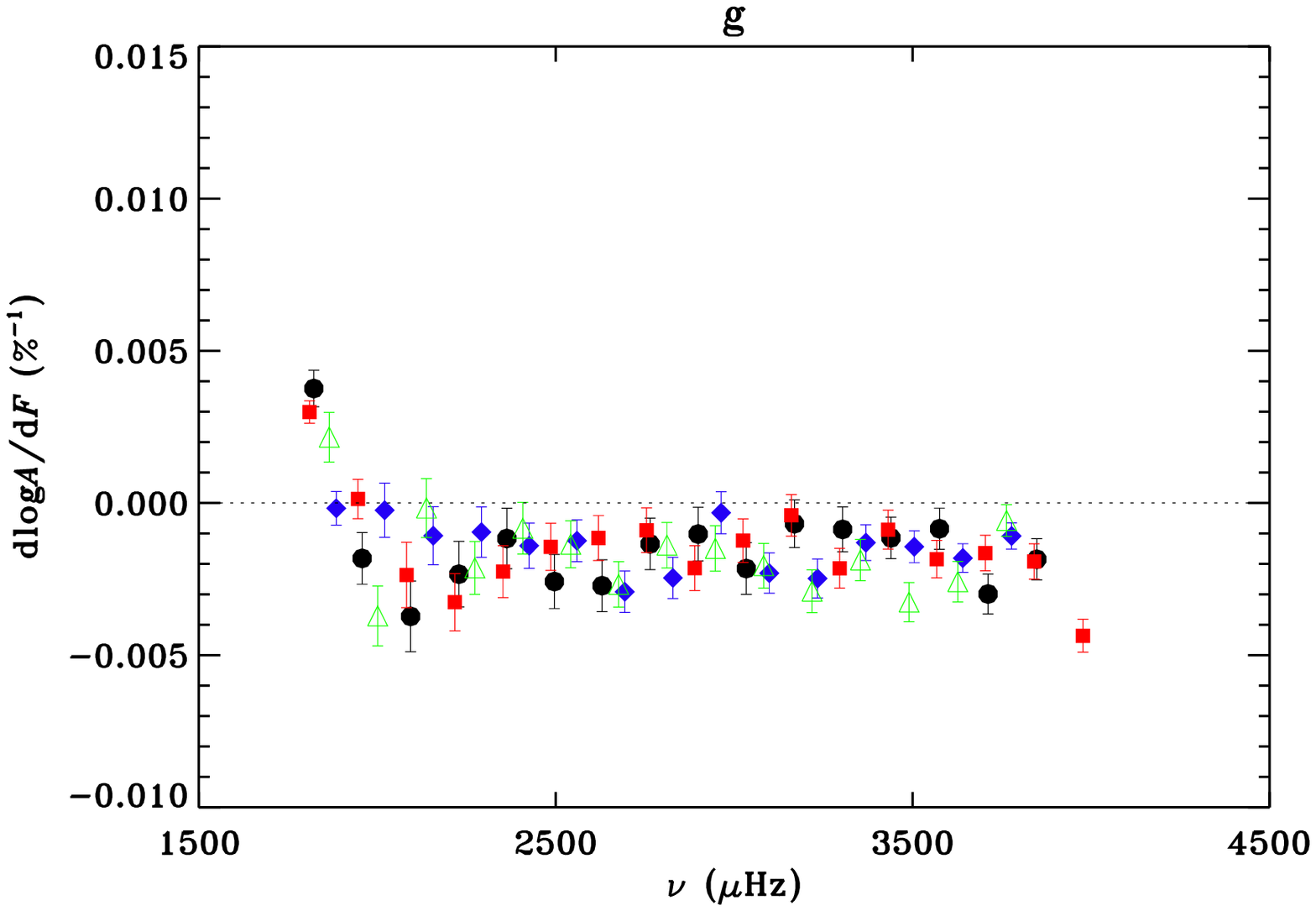}
\epsfxsize=0.4\linewidth\epsfbox{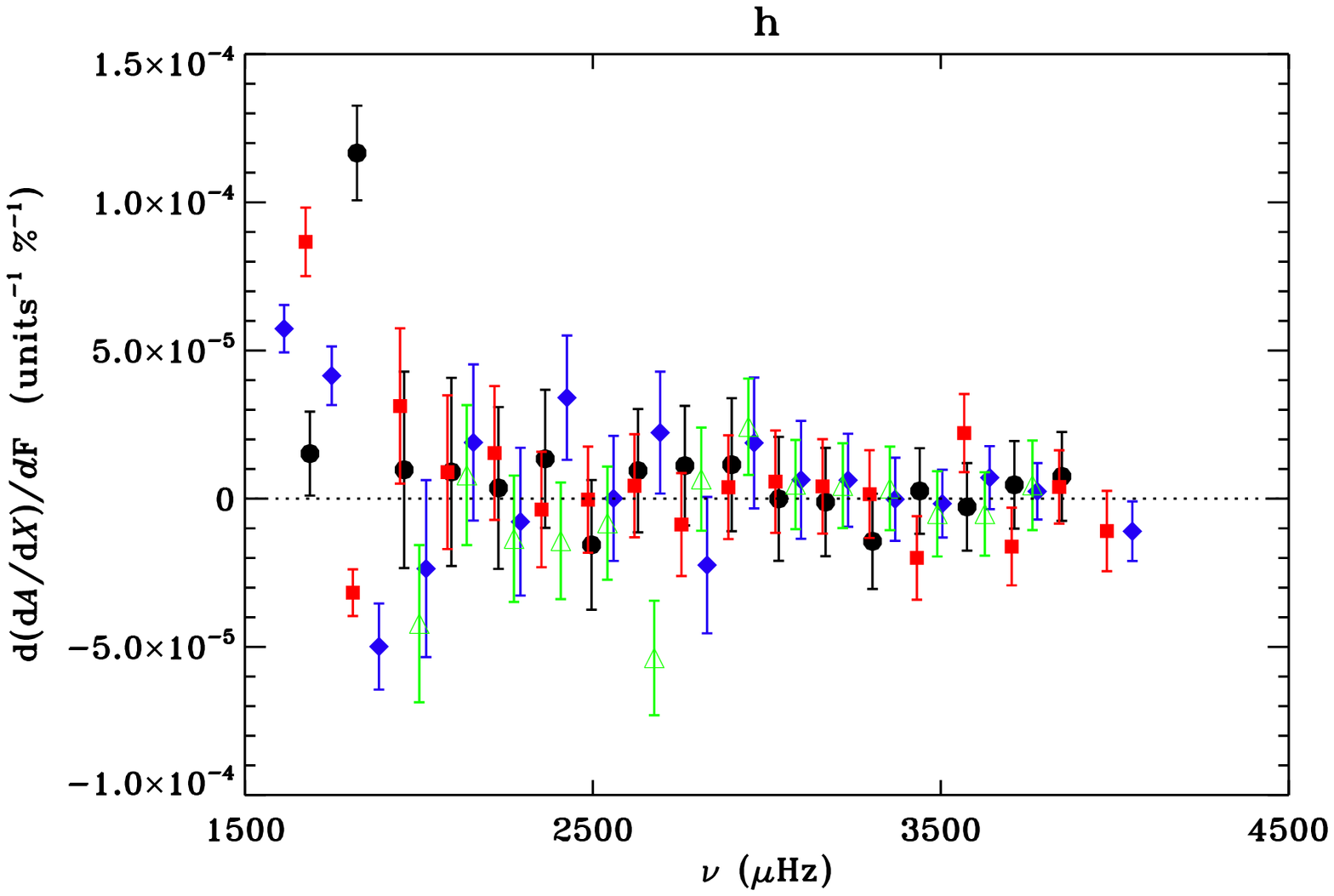}

\caption
{Sensitivity of mode amplitude to activity index and duty cycle. Panels (a)--(d) show fitted (symbols) and input (lines) $d\log A_{nl}/dX$
for MLE (a, c) and MCMC (b,d) fits to 11 one-year sets of artificial data at different activity levels, mimicking the solar cycle.  The results are for 100 per cent (a, b) and realistic (c, d) duty cycle. The bottom two rows, for MLE only, show the results of the tests with multiple realizations of the artificial data. Panel (e) and (f) show the sensitivity of the amplitude to activity index, $X$, 
with panel (f) showing the results for the test with no input frequency variation.
Panel (g) shows the sensitivity to duty cycle, $F$, for the test with 25 realizations of the artificial data. Panel (h) shows the rate of change with duty cycle $F$ of the sensitivity $d\log A_{nl}/dX$ of mode amplitude to activity index, for the  test where each of 21 one-year BiSON window functions was applied to each of the 11 years of the SolarFLAG time series. 
Black circles represent $l=0$, blue diamonds $l=1$, red squares $l=2$, and
green open triangles $l=3$.}
\label{fig:flag4}
\end{figure*}

\begin{figure*}

\epsfxsize=0.45\linewidth\epsfbox{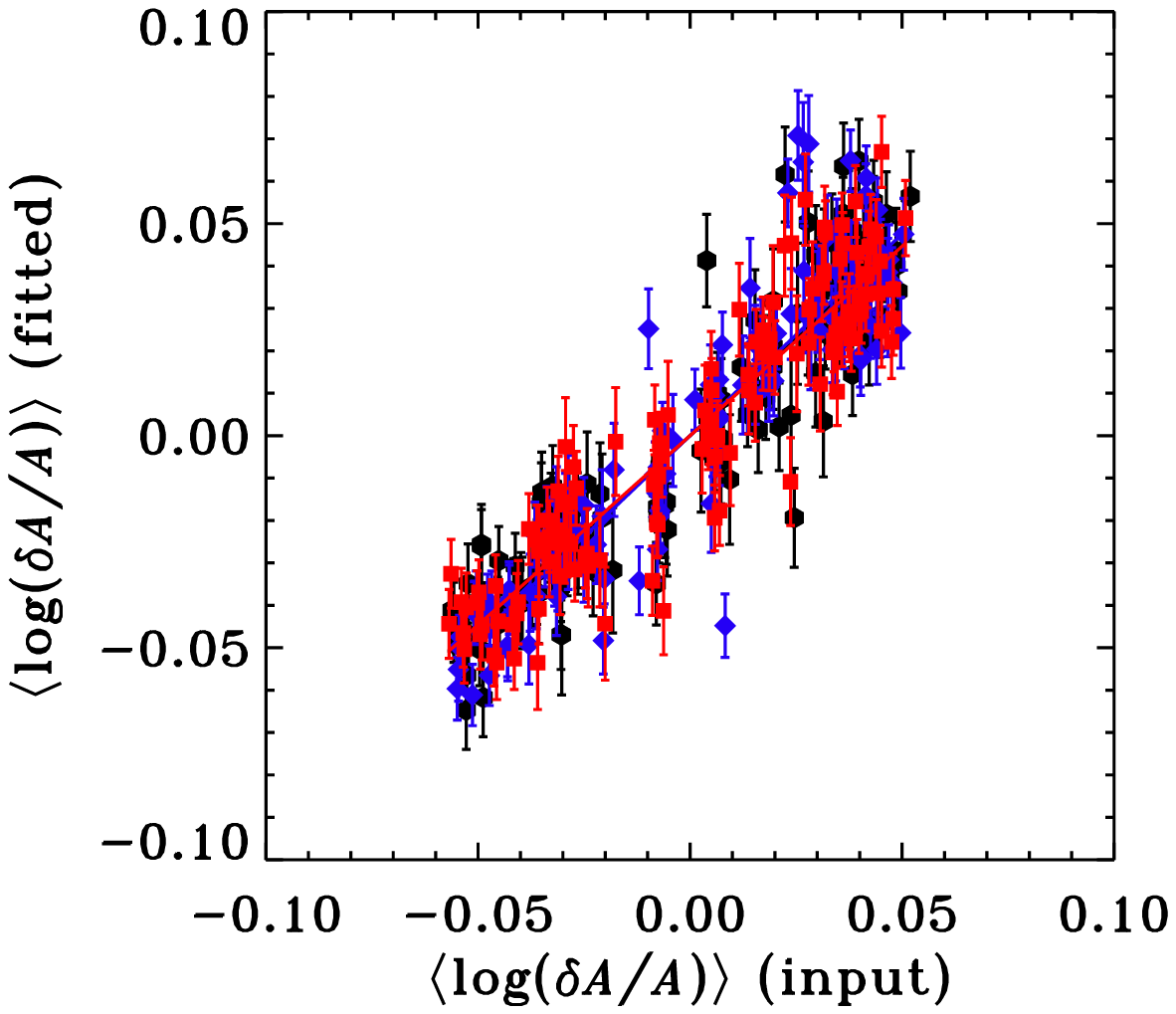}
\epsfxsize=0.45\linewidth\epsfbox{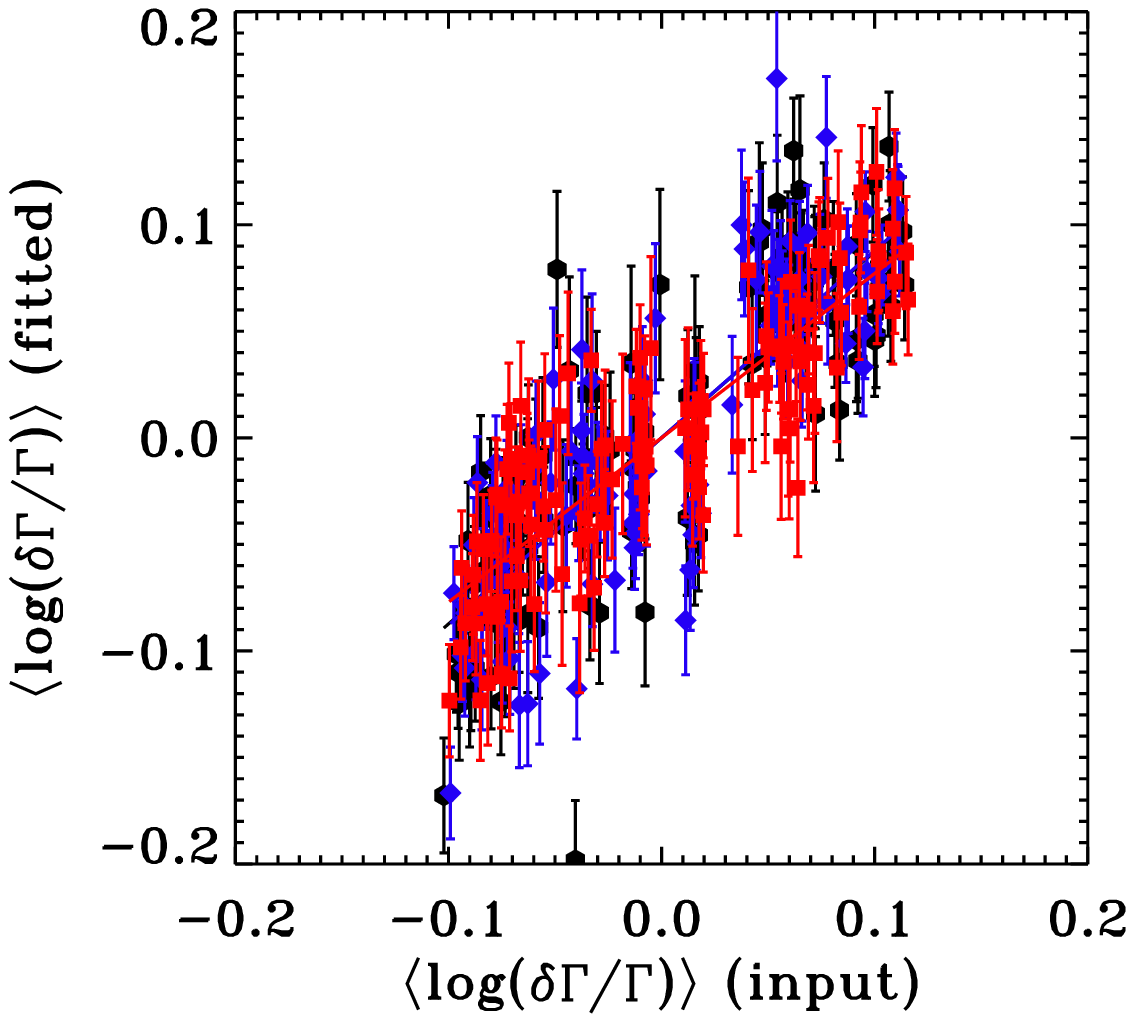}
\caption{Fitted amplitude (left) and linewidth (right) shifts averaged over
25 realizations of the artificial data with realistic duty cycle for each mode between 2.0 and 3.5 mHz, plotted against the input value. Black circles represent $l=0$, blue diamonds $l=1$, and red squares $l=2$. The correspondingly coloured lines represent least-squares fits for each $l$.}
\label{fig:flagx}
\end{figure*}

The object of the tests described here is to check that the MLE and MCMC algorithms give consistent and reliable results for the variation of the different mode parameters with the activity index.

The 11-year time series generated as described in Section~\ref{sec:noise} above was divided into contiguous 365-day segments and each segment was fitted using both the MLE and MCMC algorithms. The series were fitted both with 100 per centduty cycle and with a `realistic' duty cycle based on that of the BiSON time series for the 11 years starting with the notional start date of the artificial series, 1997 February 3. For each mode we then take an error-weighted mean over all the data sets and subtract this from the values to obtain the shifts, finally using a linear least-squares fit to the activity index to derive the sensitivity for each mode for comparison with the input values.

For the MLE method only, three additional, more extensive tests were carried out. In one test, 25 additional realizations of the 11-year artificial data series were fitted in one-year segments, while in another each one-year segment of a single realization of the artificial data was fitted with the duty cycle corresponding to each of the 21 one-year BiSON time series. The term `realization' here refers to a realization of the noise exciting the simulated modes. The extra fits make it easier to distinguish the systematic errors from the random noise;  by using 25 realizations we reduce the random errors by a factor of five, which is sufficient to clarify the trends. In the real observations we have 22 years of data, so the random errors on the mean parameters and sensitivities are a factor of $\sqrt{2}$ smaller than for a single realization of the simulated data, or approximately three times larger than for 25 realizations. Finally, to help rule out the possibility that the frequency shifts are caused by asymmetry changes, we prepared another set of artificial data realizations where the input amplitude and width changes -- and hence the asymmetry changes -- were as before but the input frequency change was set to zero.  We note that to generate frequency shifts as large as those observed, the asymmetry shift would need to be extremely and unrealistically large, as discussed below in Section~\ref{sec:testdisc}.

\subsection{Results}

\begin{figure*}

\epsfxsize=0.4\linewidth\epsfbox{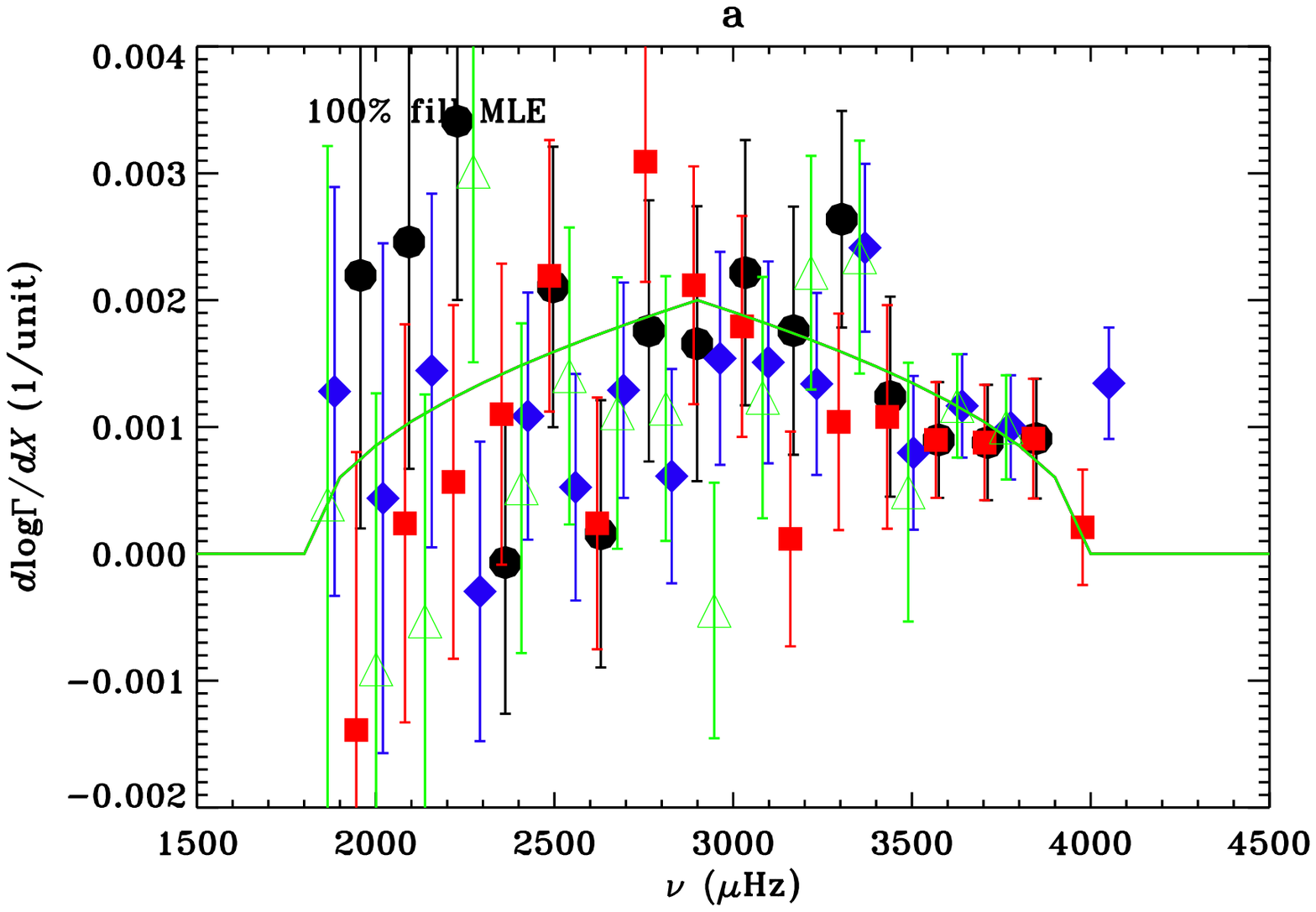}
\epsfxsize=0.4\linewidth\epsfbox{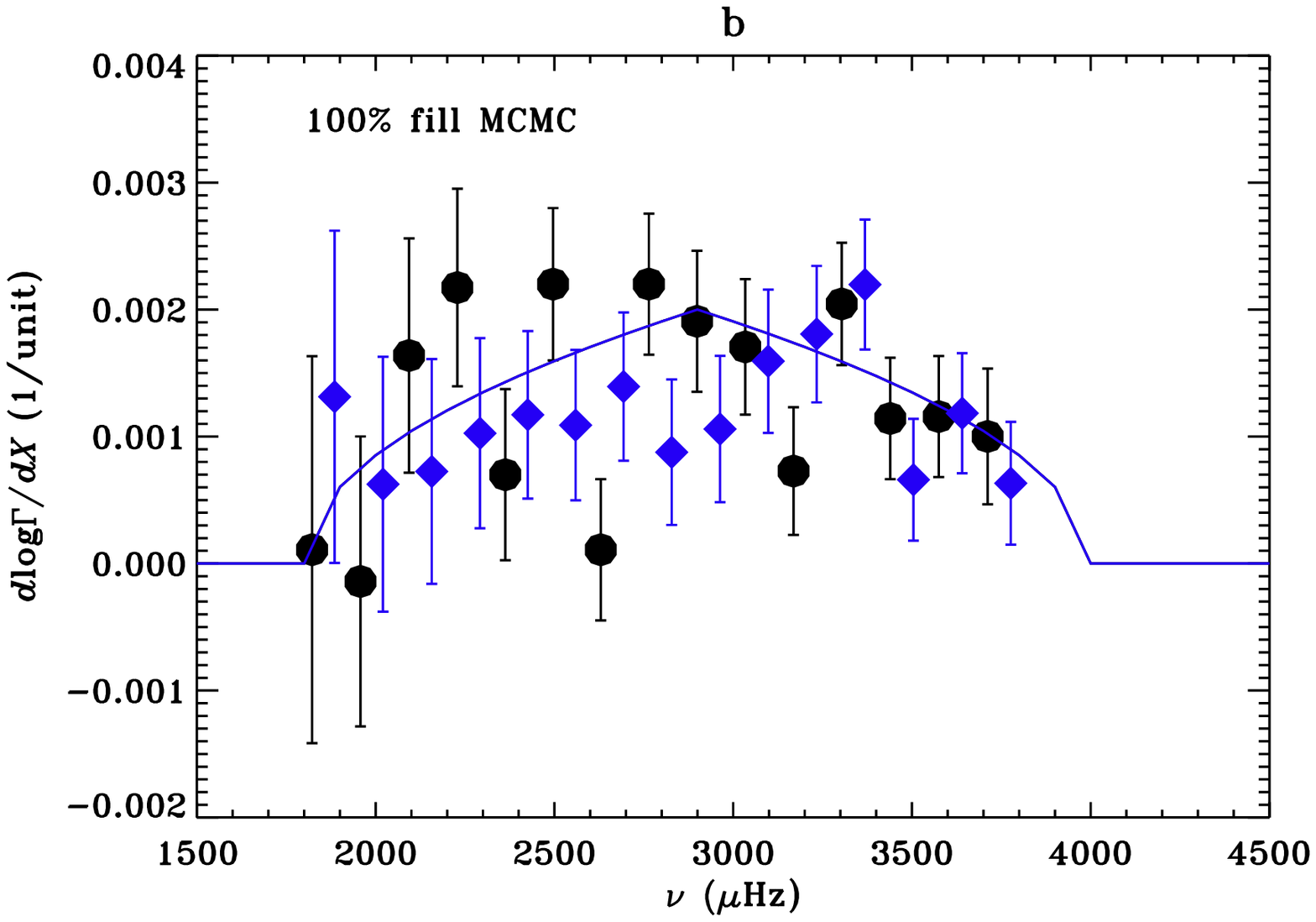}
\epsfxsize=0.4\linewidth\epsfbox{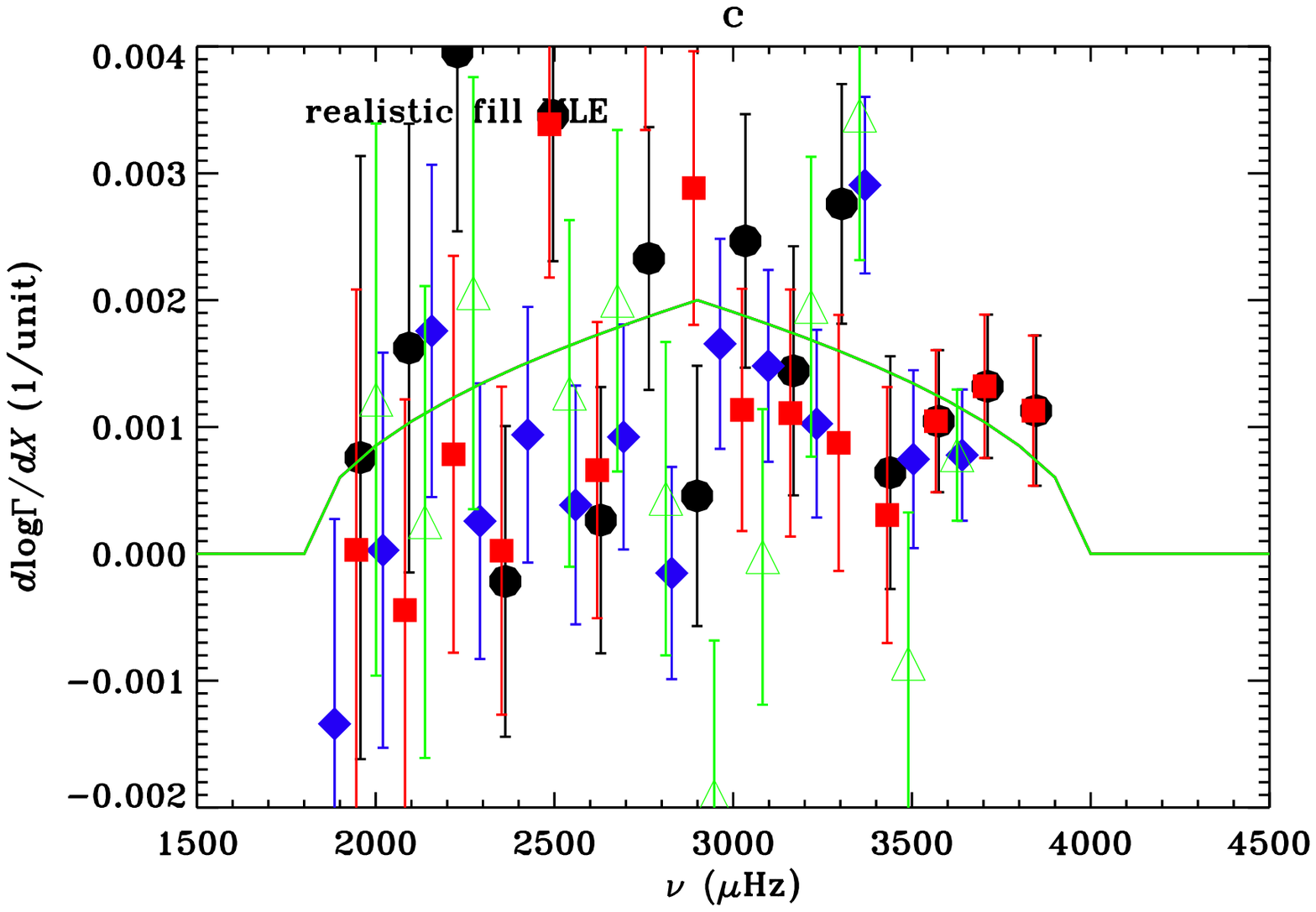}
\epsfxsize=0.4\linewidth\epsfbox{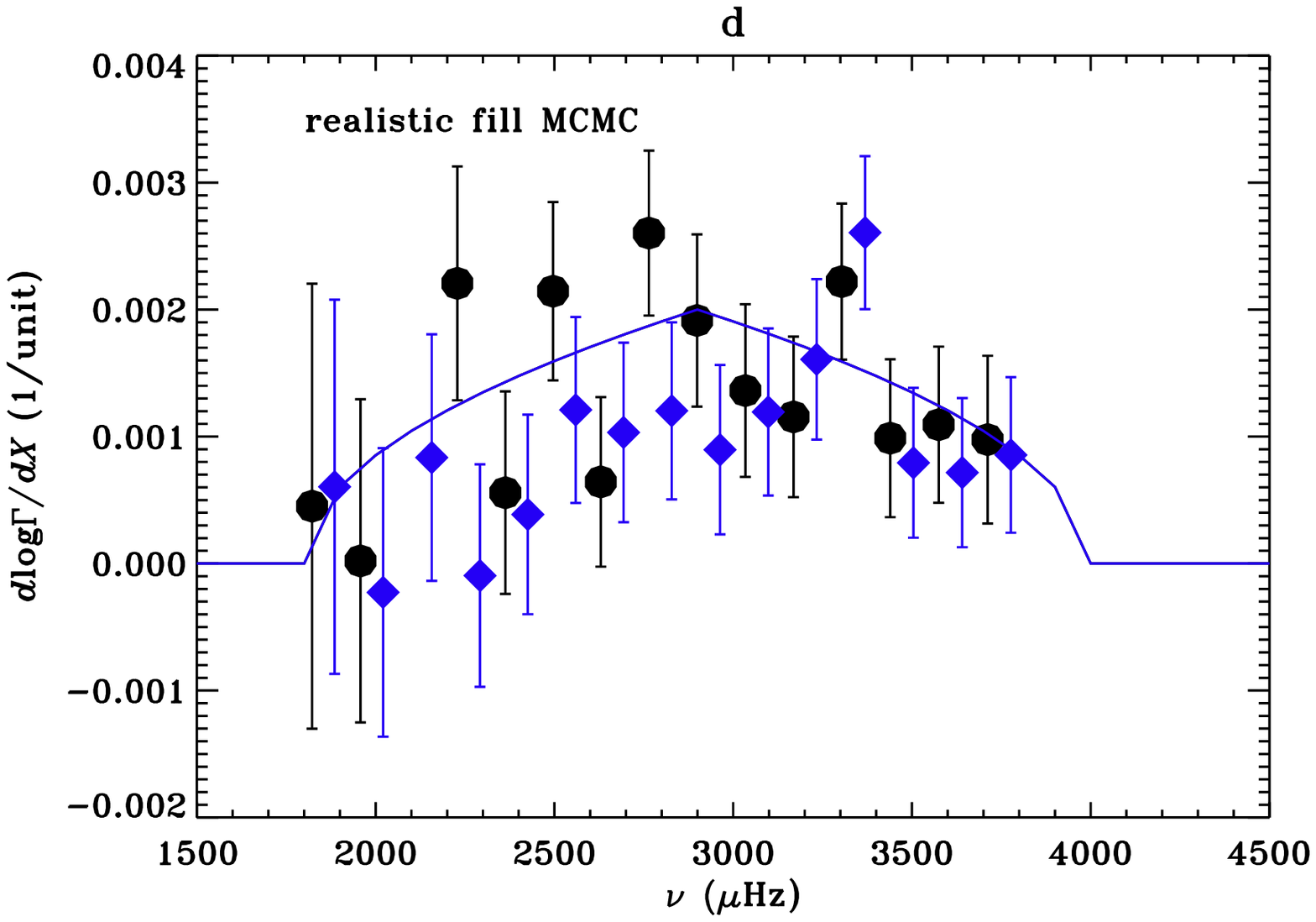}

\epsfxsize=0.4\linewidth\epsfbox{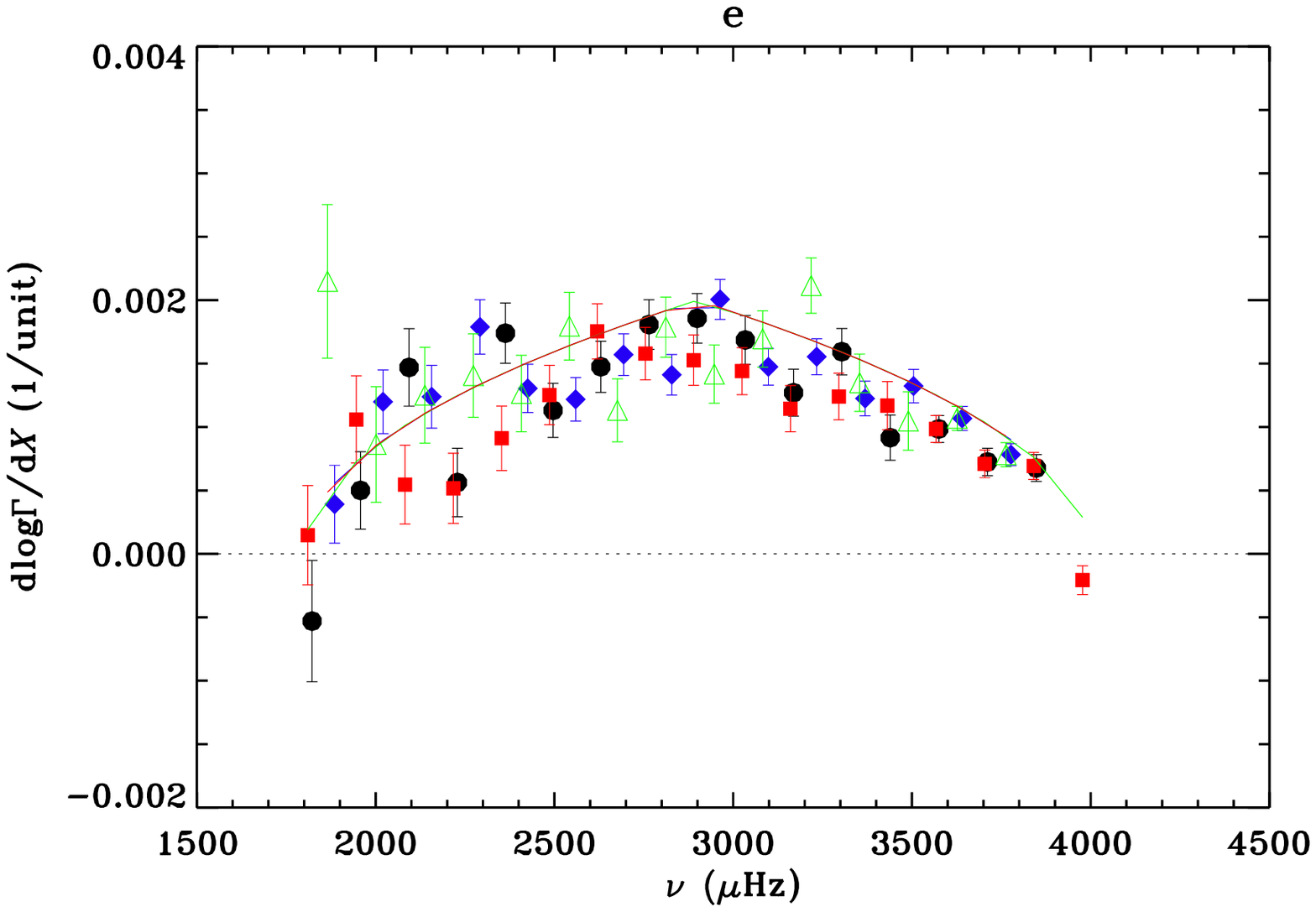}
\epsfxsize=0.4\linewidth\epsfbox{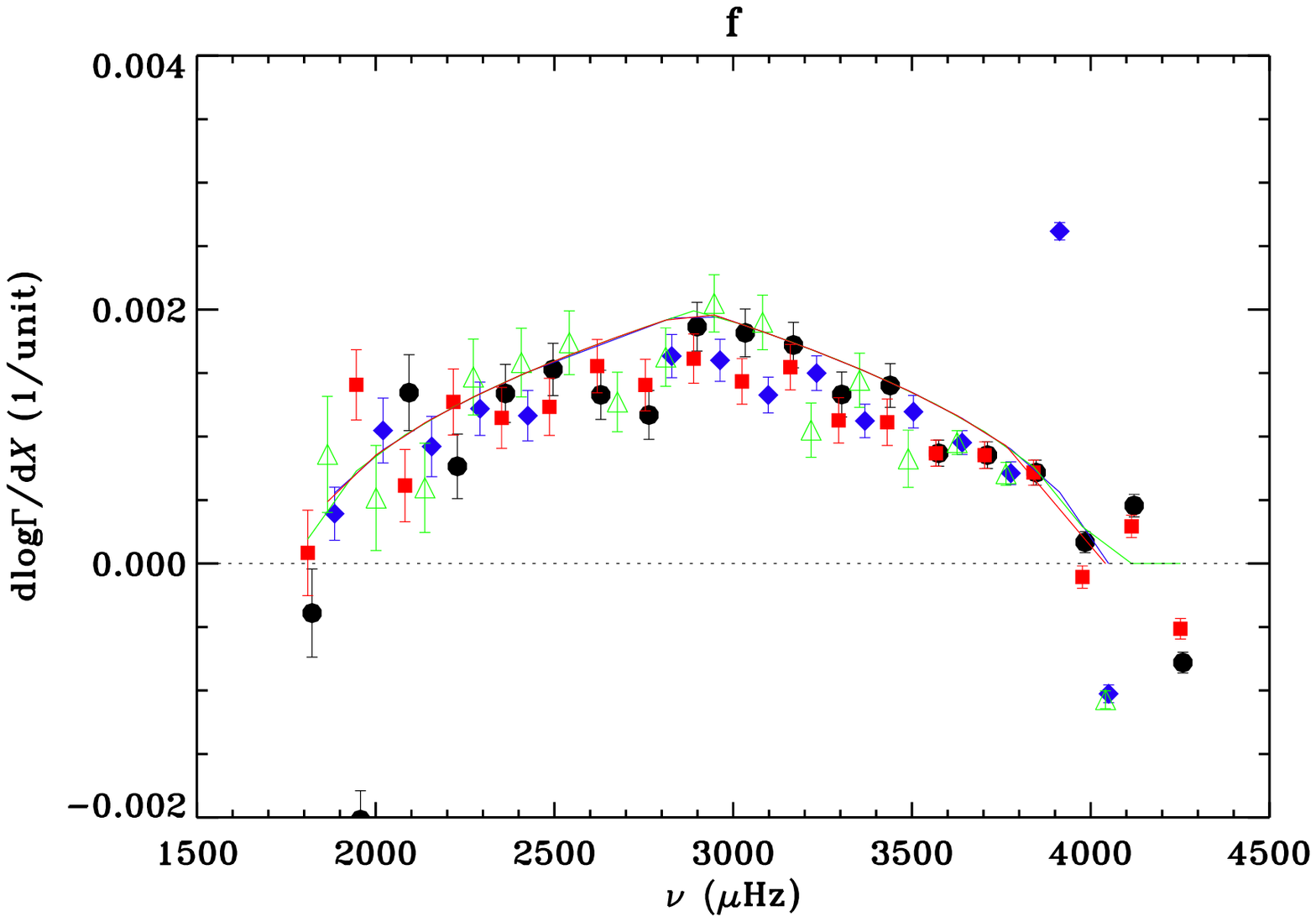}

\epsfxsize=0.4\linewidth\epsfbox{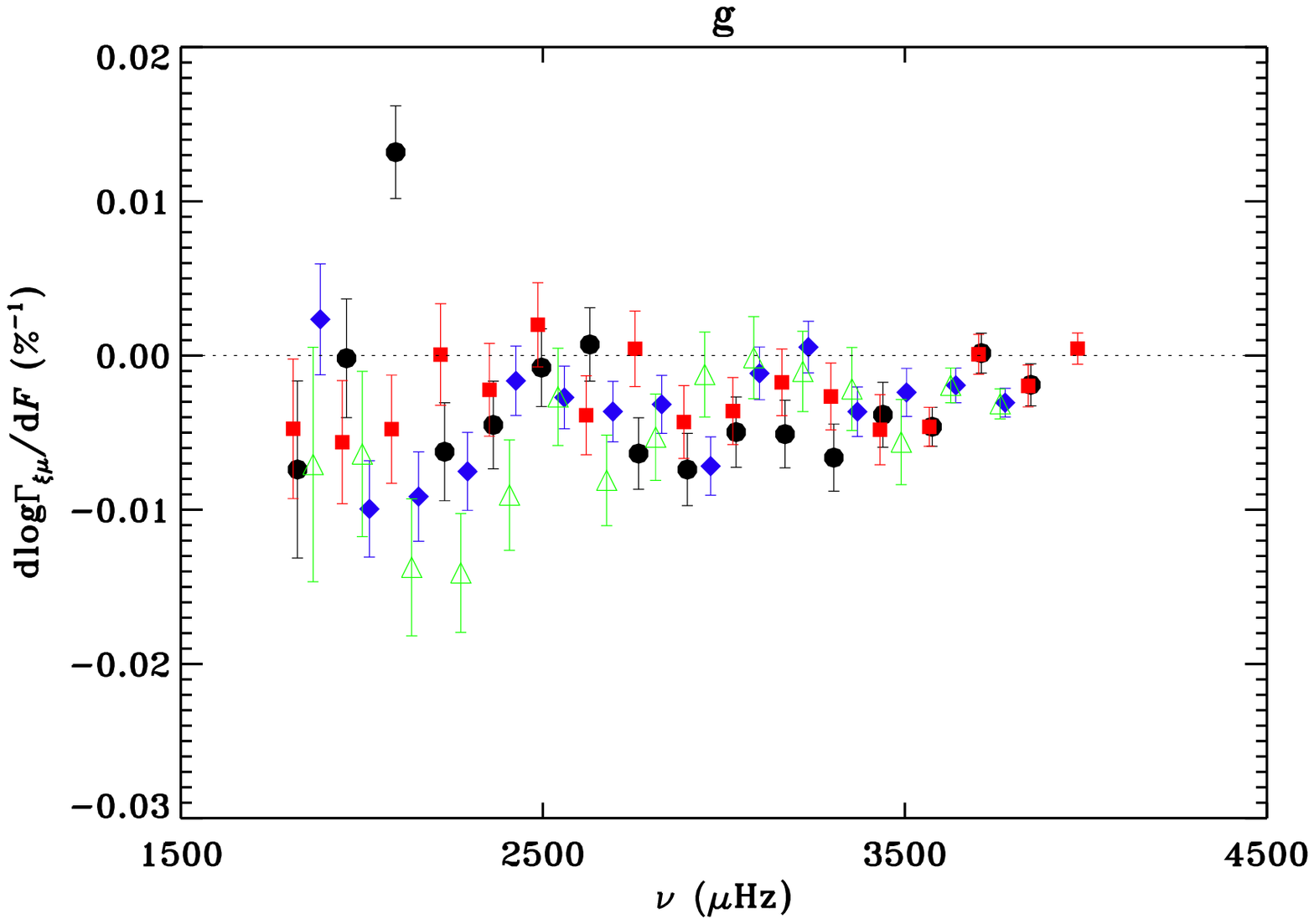}
\epsfxsize=0.4\linewidth\epsfbox{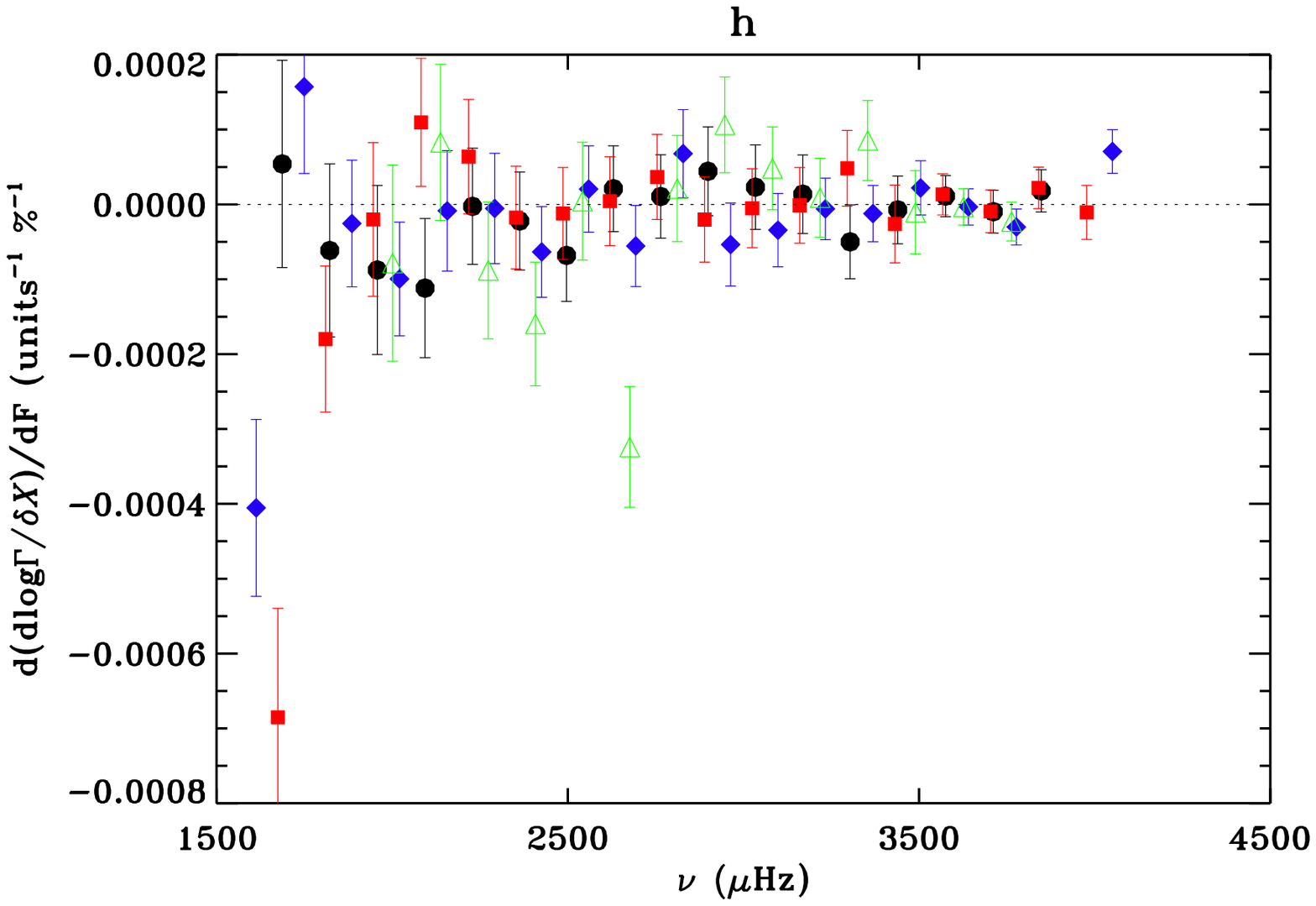}

\caption
{Sensitivity of fitted line width to activity index and fill, for artificial data. Panels (a)--(d) show fitted (symbols) and input (lines) $d\log\Gamma_{nl}/dX$
for MLE (a, c) and MCMC (b,d) fits to 11 one-year sets of artificial data at different activity levels, mimicking the solar cycle.  The results are for 100 per cent (a, b) and realistic (c, d) duty cycle. The bottom two rows, for MLE only, show the results of the tests with multiple realizations of the artificial data. Panels (e) and (f) show the sensitivity of the linewidth to activity index, $X$, with the results in panel (f) being for the test with no input frequency vrition. Panel (g) shows the sensitivity to duty cycle, $F$, for the test with 25 realizations of the artificial data. Panel~(h) shows the rate of change with duty cycle $F$ of the sensitivity $d\Gamma_{nl}/dX$ of mode width to activity index, for the  test where each of the 21 one-year BiSON window functions was applied to each of the 11 years of the SolarFLAG time series. 
Black circles represent $l=0$, blue diamonds $l=1$, red squares $l=2$, and
green open triangles $l=3$; for MCMC black circles represent $l=0/2$ pairs and blue diamonds $l=1/3$.}
\label{fig:flag6}
\end{figure*}

\subsubsection{Frequencies}

Figure~\ref{fig:flag2} (a--d) shows the input and fitted values of $d\nu_{nl}/dX$, the frequency shift per unit of the activity index $X$ for each fitting method. Note that the differential notation we use throughout is a convenient shorthand for the regression slope. In the case of the artificial data the linear relationship between activity and mode parameters is designed to be exact; the real data may exhibit small deviations from this but it is still a useful measure of the sensitivity to first order.
Except for a few modes at the higher end of the frequency range, the results from both of the algorithms agree within errors with the input values; the mean ratio between the fitted and input $d\nu_{nl}/dX$ for $0 \leq l  \leq 2$ and $2000\mu{\rm Hz} \leq \nu_{nl} \leq 3500 \mu{\rm Hz}$ is consistent with unity in each of the four cases.

Panels (e) and (f) of Figure~\ref{fig:flag2} illustrate the sensitivity of the frequency to activity 
for  the tests with 25 realizations of the artificial data, with panel (f) showing the results for the test with no input frequency variation. Apart from a few modes at frequencies above 3.7 mHz where the s/n ratio is low, we see no frequency change in the dataset where none was explicitly introduced. Figure~\ref{fig:flag2}(g) shows the sensitivity of the frequency to the duty cycle.  
These results were obtained by performing  for each mode a multiple linear regression in which the independent variables were the activity index $X$ and the duty cycle $F$ and the dependent variable was the mode frequency.
The results in Figure~\ref{fig:flag2}(e) clearly show the slightly lower $d\nu/dX$ for the $l=0$ modes -- a difference that is lost in the noise for the single-realization test.

In another test of the MLE fitting we applied each of the 21 one-year window functions from the BiSON network for 1993--2013 to each of the 11 years of the SolarFLAG time series. These results were used to check for any bias in the sensitivity of the frequency to activity level due to the different window function, as follows. The results were divided into 21 sets in which the 11 one-year spectra all had the same duty cycle but different activity indices. For each set a regression was performed for each mode, with the frequency shift as the dependent and the activity index as the independent variable, to obtain a set of slopes $d\nu_{nl}/dX$ for each value of $F$. Finally, a regression was performed for each mode with the duty cycle $F$ as the independent variable and the $d\nu/dX$ value as the dependent one. The results are shown in Figure~\ref{fig:flag2}(h). The sensitivity of the frequencies to activity index shows no systematic bias due to the window function.

\subsubsection{Amplitudes}

For mode amplitude $A$ (defined as in Equation~\ref{eq:rh2} above) and width $\Gamma$ we work with the natural logarithm of the parameter; a shift in the logarithmic parameter is equivalent to a fractional shift in the raw parameter.

Figure~\ref{fig:flag4}(a-d) shows the sensitivity, $d\log A_{nl}/dX$, of the natural logarithm of the amplitude parameter $A_{nl}$ to
activity index, obtained using the two methods for artificial data with 100 per cent and realistic duty cycles, compared with the input values.  The individual
$d\log A_{nl}/dX$ for the two methods are well correlated, with correlation coefficients of 0.77 for the 100 per cent duty cycle and 0.70 for the realistic one over the 47 modes common to both sets; the threshhold for 0.1 per cent probability of the same result arising for two unrelated populations of 47 samples is 0.47.

The mean ratios between fitted and input $dA_{nl}/dX$ values are $0.84\pm 0.08$ for MCMC and $0.85\pm 0.07$ for MLE in the 100 per cent duty cycle case and $0.83\pm 0.09$ for MCMC and $0.79\pm 0.10$ for MLE in the realistic duty cycle case. 
This result suggests that the sensitivity of the amplitude to the
activity index is being systematically underestimated by both methods. However, as we shall see below,
when we average over all realizations of the artificial data we find that the average bias is not as severe as for this case.

In panels (e) and (f) of Figure~\ref{fig:flag4}
we show the sensitivity of the
amplitude measurements to activity index 
for the
multiple-realization
tests with MLE fitting; panel (f) shows the results for the artificial data with no input frequency variation.  The lack of frequency variation does not significantly affect the amplitude variation. The results with multiple realizations clearly reveal the frequency-dependence of the shifts, but also confirm the systematic underestimation of the sensitivity. 
The sensitivity to duty cycle (panel (g)) has a small non-zero value that does not depend strongly on frequency; this is most likely due to the power in higher-order daily sidelobes that is not accounted for in the fitting. 

Figure~\ref{fig:flag4}(h) shows the variation with duty cycle of the amplitude shift per unit activity, from the test with MLE fitting where the window function was varied. Apart from a few low-frequency modes and the $l=3$ modes, the results show no systematic bias of the sensitivity measurement due to the window function.

The apparent underestimate of the amplitude shifts is of some concern. However, when we consider the MLE fits to the larger cohort of artificial time series, we obtain a more favourable result. Figure~\ref{fig:flagx}(a) shows the fitted amplitude shifts for data with realistic fill, averaged over 25 realizations of the artificial data for each mode and plotted as a function of the expected shift. In this case we find, for modes between 2.0 and 3.5 mHz, that the slope of a linear fit of fitted vs. input shifts is $0.92\pm 0.03$, $0.95\pm 0.02$, and 
$0.89\pm 0.02$ for $l=0$, $l=1$, and $l=2$, respectively, which suggests that the realization chosen for the main plots was a particularly `unfortunate' example and the real systematic underestimate of the sensitivity is likely to be less than ten per cent. We emphasise that the size of the fractional shifts is not affected by any systematic errors in the underlying values as long as these are not activity-dependent.

\subsubsection{Line width}

Figure~\ref{fig:flag6}(a--d) shows the input and fitted values of $d\log \Gamma_{nl}/dX$, the rate of change of line width with the activity index $X$, for the two methods. 
 As for the amplitude measurements, the individual-mode sensitivity results for the two methods are well correlated with one another, with a correlation coefficient of 0.59 in the 100 per cent duty cycle case  and 0.71 for the realistic duty cycle, over 27 values. The slope of a fit of MCMC vs MLE values is $0.96\pm 0.48$ for 100 per cent duty cycle and $1.08\pm 0.70$ for realistic duty cycle, which indicates that the values are consistent; furthermore, for each method the $d\Gamma/dX$ values for the 100 per cent and realistic duty cycle cases are consistent. The results of the tests with multiple realizations are shown in Fig.~\ref{fig:flag6}(e,f). Again, the frequency variation of the sensitivity is clearly seen in Fig.~\ref{fig:flag6}(e) and (f), with panel (f) being for the case with no input frequency variation. Again, the lack of frequency variation has no significant effect on the width variations. Fig.~\ref{fig:flag6}(g) shows that there is a small, more or less frequency-independent, effect of the duty cycle on the fitted line width, which is to be expected for the MLE algorithm due to the crude handling of the window function effects. Fig.~\ref{fig:flag6}(h) shows that the results of the MLE test with multiple duty cycles applied to the same realization of the artificial data; the duty cycle has no significant effect on the sensitivity of the linewidth to activity-related variations, so as long as the two factors are not correlated we can treat the effects as independent of one another.

As in the case of the amplitude, $d\Gamma/dX$ appears systematically underestimated, with a mean ratio between fitted and input values of $0.81\pm 0.09$ for 100 per cent and $0.79\pm 0.10$ for realistic duty cycle for MLE fits and  $0.84\pm 0.08$ for 100 per cent fill and $0.83\pm 0.09$ for realistic duty cycle from MCMC. For the test with MLE fits to 25 realizations of artificial data with realistic duty cycle (Fig.~\ref{fig:flagx}(b)), we obtain slopes of $0.87\pm 0.04$, $0.88\pm 0.03$, and $0.77\pm 0.03$ for $l=0$, $l=1$ and $l=2$ respectively; as with the amplitude, this suggests that our main sample realization was particularly unfavourable, but we might still be failing to recover the $l=2$ shifts completely.

\subsubsection{Energy supply rate}
We recall from Equation~\ref{eq:rh3} that the energy of a mode is proportional to the square of its amplitude, and the rate of energy supply to the mode $dE_{nl}/dt$ is proportional to $h_{nl}\Gamma_{nl}^2$
so for the logarithmic quantity we have

\begin{equation}
\log dE_{nl}/dt = 2\log \Gamma_{nl} + \log A_{nl} + {\mathrm {const}},
\end{equation}
where const includes a contribution from the mode mass and visibility function.

The energy supply rate is believed not to vary with the activity level, and the SolarFLAG data were designed to reflect this. In other words, we expect $2dA/dX+d\Gamma/dX=0$. As the $d\log A_{nl}/dX$ and $d\log\Gamma_{nl}/dX$ values are noisy, we check for this by performing a weighted least-squares fit with the individual $\delta\log A_{nl}$ and $\delta\log\Gamma_{nl}$ as $x$- and $y$-values. For the $l=0,1$ values between 2.0 and 3.5 mHz we obtain slopes of $-1.8\pm 0.58$ and $-1.47 \pm 0.90$ for the MLE fitting with 100 per centand realistic duty cycle respectively, and $-1.53\pm 0.48$ and $-1.34\pm 0.58$ for MCMC. Three of these values are within $1\sigma$ of the expected value of -2, and the other is only just outside the $1\sigma$ range. A fit with $d\log A_{nl}/dX$ as the abscissa and $d\log\Gamma_{nl}/dX$ as the ordinate gives agreement within the (large) error in all cases, and the mean value of $2d\log A_{nl}/dX+d\log\Gamma_{nl}/dX$ is consistent with zero in all four cases. We can therefore say that the fitting is not significantly biasing the variation of the energy supply rate, in spite of the small bias on the $d\log A_{nl}/dX$ and $d\log\Gamma_{nl}/dX$ values.

\subsubsection{Asymmetry}

\begin{figure*}

\centerline{\epsfxsize=0.45\linewidth\epsfbox{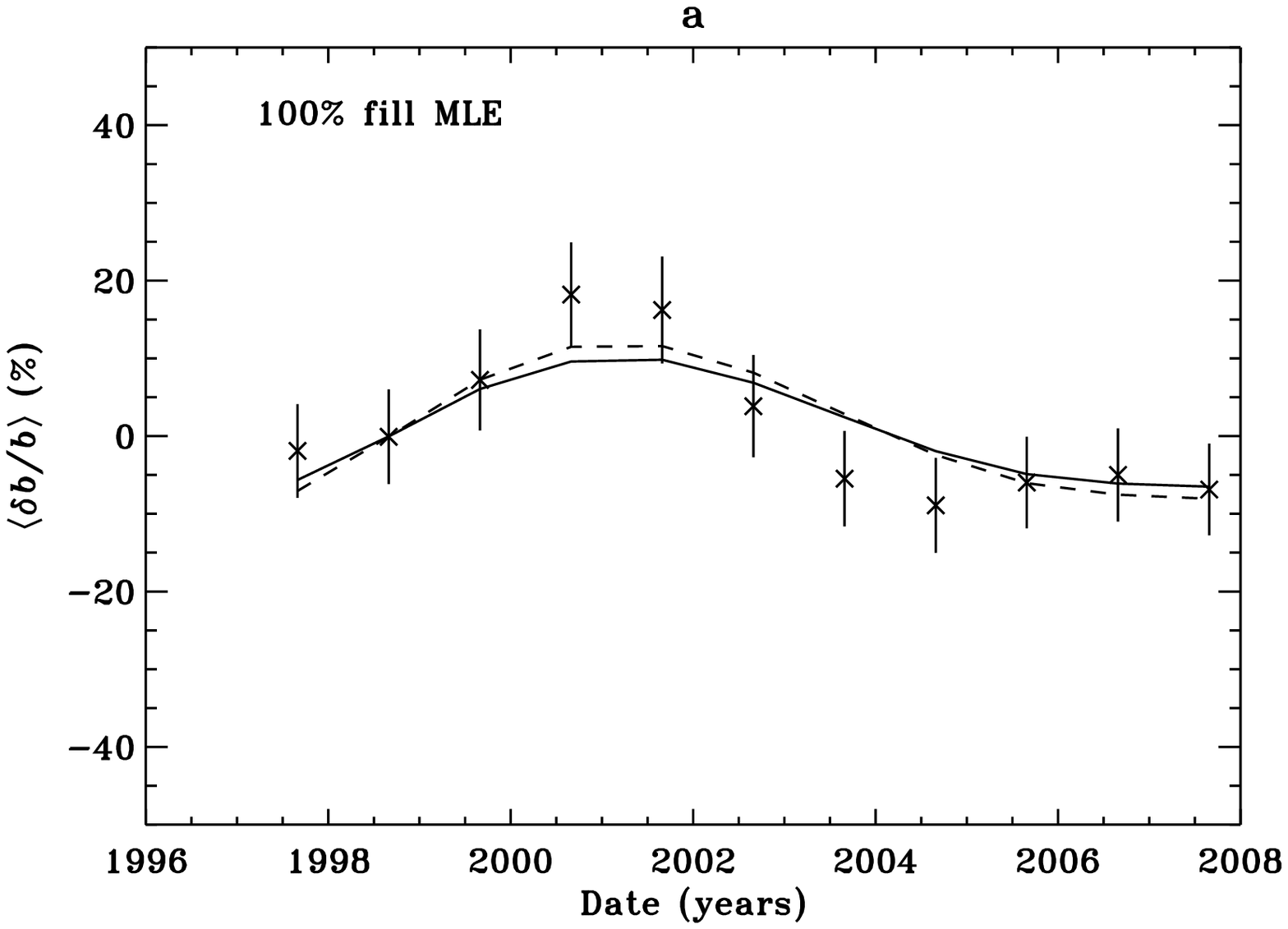}
\epsfxsize=0.45\linewidth\epsfbox{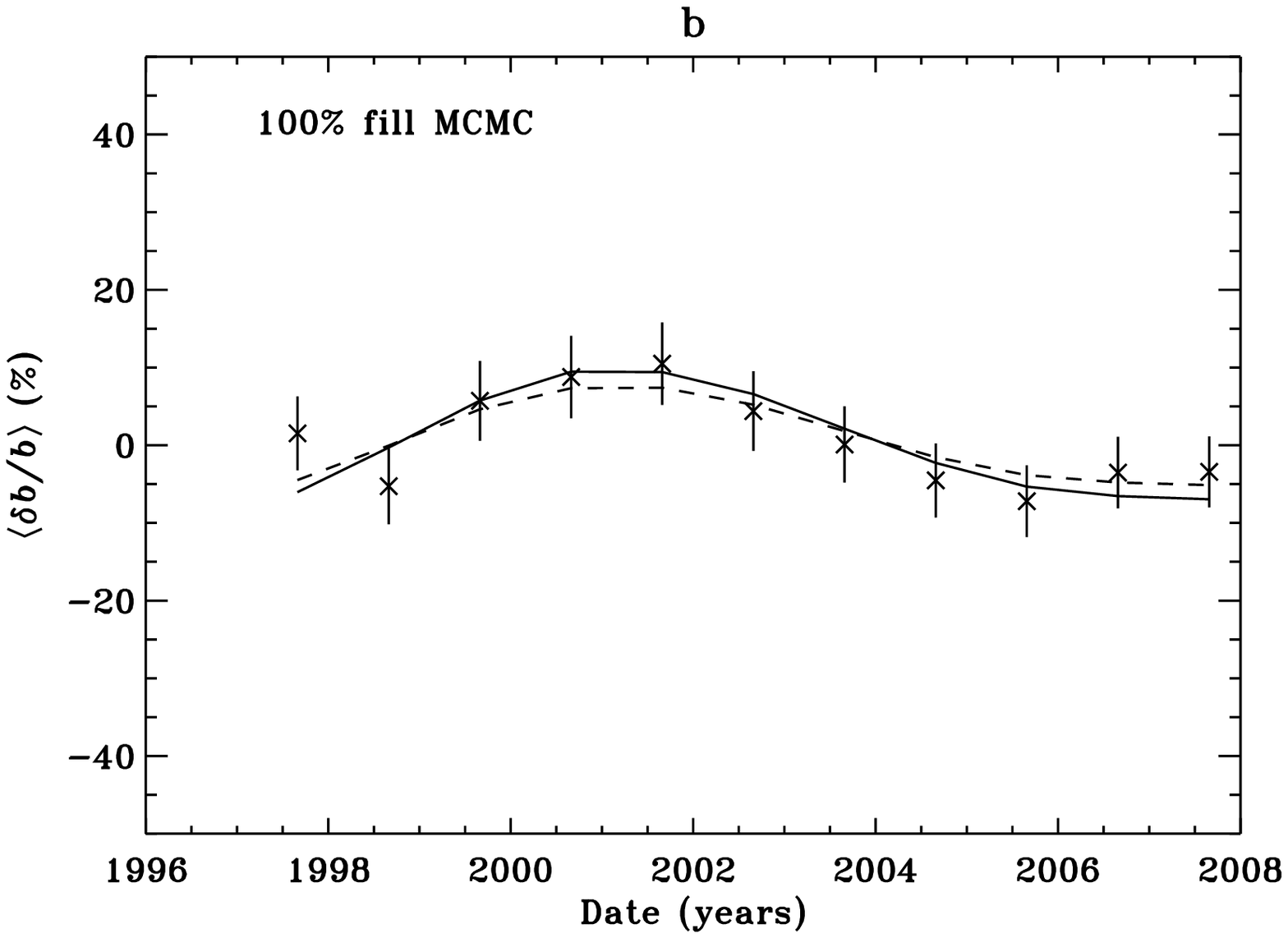}}
\centerline{\epsfxsize=0.45\linewidth\epsfbox{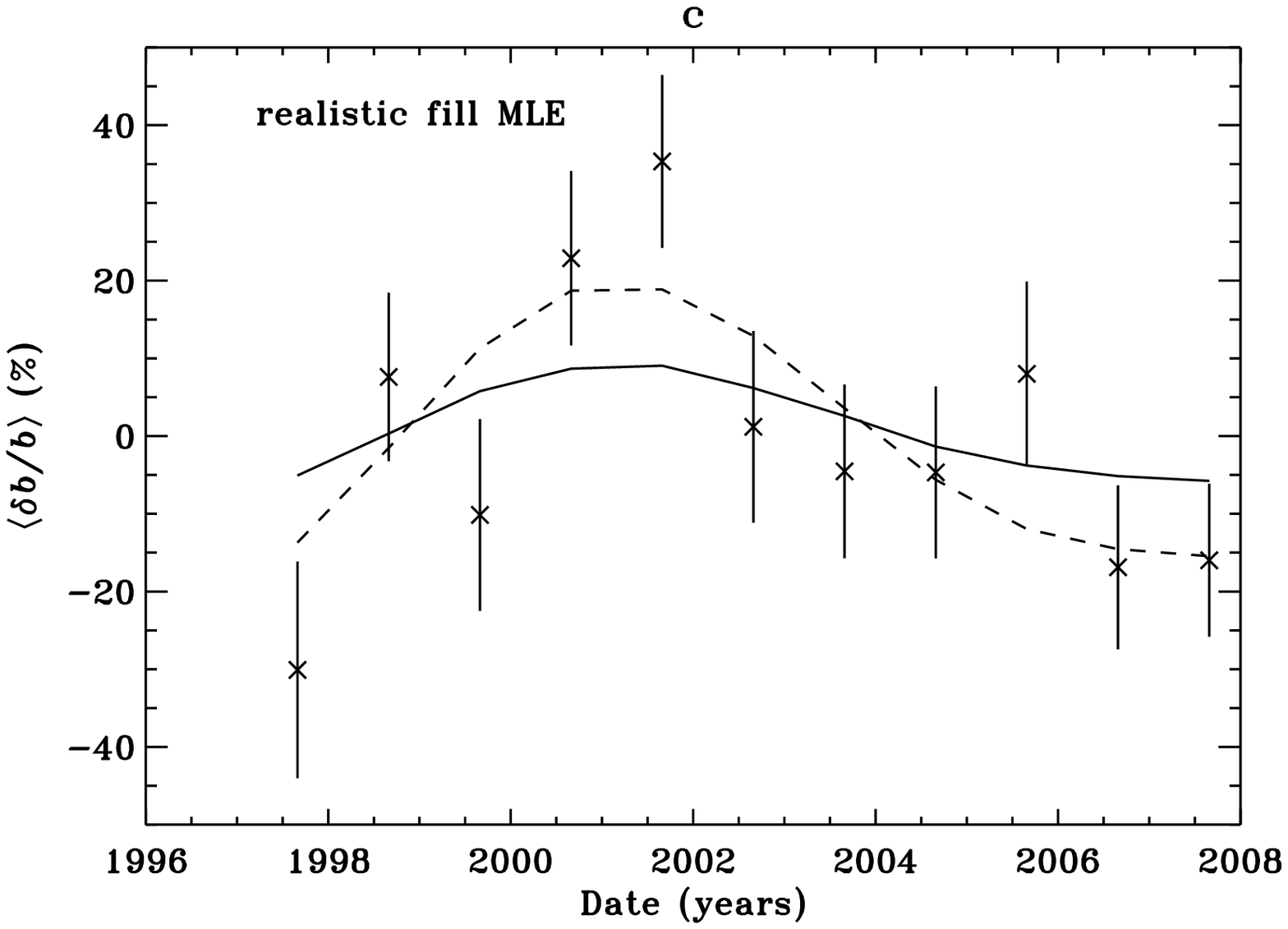}
\epsfxsize=0.45\linewidth\epsfbox{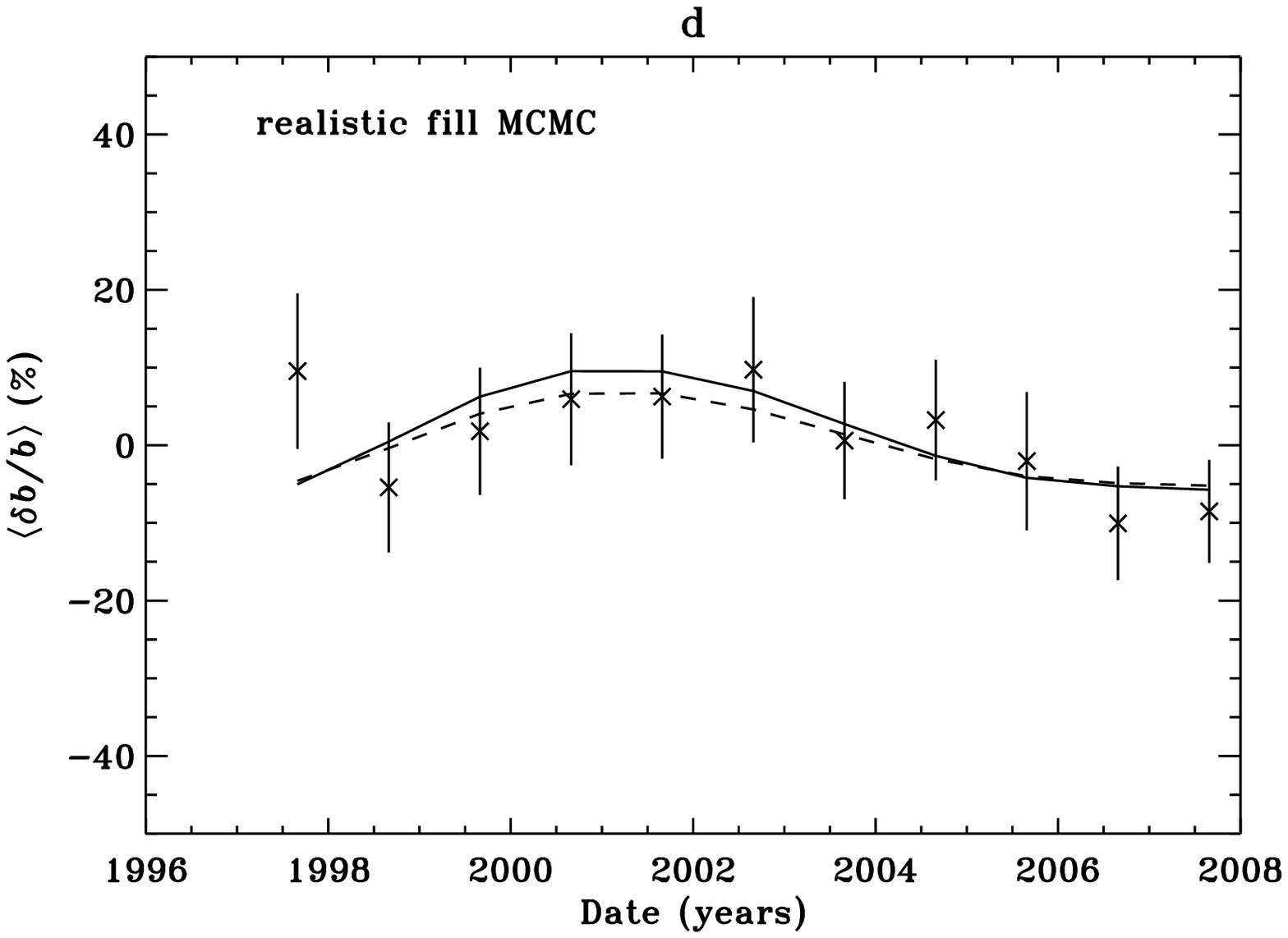}}

\caption{Fitted (symbols) and input (lines) mean fractional asymmetry shift $\delta b/b$ as a function of time
for MLE (a, c) and MCMC (b,d) fits to 11 one-year sets of artificial data, for 100 per cent (a, b) and realistic (c, d) duty cycle. The dashed lines represent the activity index scaled to the best linear least-squares fit to the asymmetry-shift data.}
\label{fig:flagasym}
\end{figure*}

The design of the SolarFLAG data is such that the fractional shift in the asymmetry is expected to be the same as that in the line width. Because the asymmetry determinations are noisier, we do not show the results per mode. Instead, we look at the error-weighted average fractional shift relative to the (also error-weighted) temporal mean value for each mode. Figure~\ref{fig:flagasym} shows the temporal variation of the mean fractional asymmetry shift, averaged over all modes between 2 and 3.5 mHz, compared with that which would have been obtained if the fitting returned exactly the input values with the same uncertainties. The correlation coefficients between the expected and measured mean asymmetry shifts are as follows: 0.79 and 0.65 for MLE with 100 per cent and realistic duty cycle, and 0.78 and 0.75 for MCMC. The slopes of linear least-squares fits with the input average fractional shift as the independent variable and the measured values as the dependent one are $1.15\pm 0.33$, $2.35\pm 0.68$ for MLE with 100 per cent and realistic duty cycle, and $0.82\pm 0.25$, $1.0\pm 0.49$ for MCMC. These results suggest that we can (marginally) detect a shift in the asymmetry of about the right magnitude and sign, but even when averaging over many mode pairs the random errors are too large to make a statement about any systematic over- or under-estimate of its size, particularly in the realistic-duty-cycle case.
We can also infer, with caution, that the MCMC fitting is performing somewhat better than the MLE in this respect. This is also true in the case (not shown) where the tests were repeated using MLE with the window function handled by convolution rather than sidelobe fitting, which suggests that the advantage is specifically in the MCMC approach rather than the window function convolution.

\subsection{Discussion of the test results}
\label{sec:testdisc}
The results of the simulated-data exercise give us confidence that with 11 years of observations we can determine the activity-related shifts in the frequencies without significant bias with either fitting method. For the amplitudes and line widths both methods may underestimate the shifts by around 10 per cent (or more for the $l=2$ widths). This may be the result of a peak profile model that does not fully capture the subtleties of the mode asymmetry. We also have marginal sensitivity to changes in the asymmetry, even with realistic duty cycle. As we are fitting data with asymmetric peaks using an asymmetric peak profile, we would not expect in this case to see any bias in the frequency shifts due to neglect of the asymmetry, and indeed, even though we only make a weak detection of the asymmetry variation, we are very successful in recovering the input frequency shift -- and also in recovering a zero input frequency shift. A brief consideration of the numbers makes this understandable. A frequency shift due to neglect of the asymmetry could not exceed the product of the line width and the absolute asymmetry parameter. At 3\,mHz the line width is about 1~$\mu{\rm Hz}$ and the maximum low-degree frequency variation is around 0.31~$\mu{\rm Hz}$ or one-third of the line width, which would require around a 30 per cent change in the absolute value of the line width -- far in excess of what is observed, and so large that the asymmetry would be obvious to the eye rather than requiring subtle analysis.

Given the good agreement between the two techniques it seems unlikely that the use of MCMC fitting will invalidate the major results from the previous two decades or more of MLE fitting of Sun-as-a-star data. 

A comparison of the error estimates returned by the two algorithms shows that the uncertainties for the frequency are very similar, as are those for the asymmetry parameter. For amplitude, we find that the uncertainties from MLE are slightly (about 10 per cent) higher than those for MCMC for the $l=0,1$ modes and about 15 per cent smaller for $l=2,3$ where the signal-to-noise is lower and the modes are made up of more components that are handled differently by the two algorithms, with the MLE fit having fewer independent parameters. For linewidth, the $l=0$ and $l=1$ uncertainties from MLE are, not surprisingly, somewhat (20 to 50 per cent) larger than those for the $l=0/2$ and $l=1/3$ pairs provided by the MCMC algorithm. 

These findings will guide our interpretation of the results of the fitting of observational data.

\section{BiSON data}
\label{sec:fitresults}

\subsection{Data}

\begin{figure}
\centerline{\epsfxsize=\linewidth\epsfbox{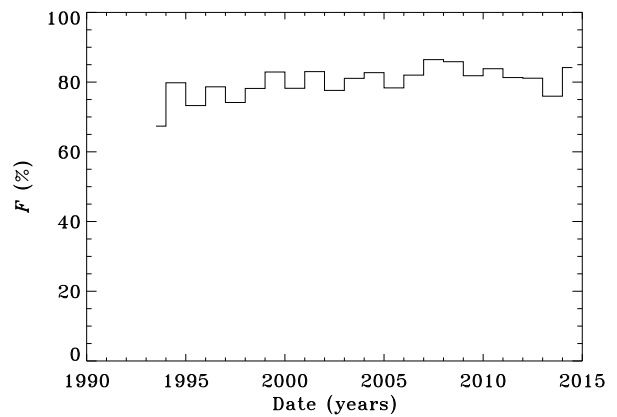}}
\caption{Yearly duty cycle for the BiSON network over the period covered by our analysis.}
\label{fig:fill}
\end{figure}

Early observations by what was to become the BiSON group have been made since 1976, but until the early 1990s the observations were relatively sparse. The final BiSON network was deployed in 1990--1992 and has been operating ever since, although there have been some changes to the instruments over time. 
Fig.~\ref{fig:fill} shows the yearly duty cycle of the network over time. For the purposes of this analysis, we use 22 non-overlapping spectra corresponding to the calendar years 1993--2014. We are fortunate in that over this period the BiSON duty cycle is essentially uncorrelated with the activity level. The data for 1992 and earlier years have much lower duty cycle and were excluded from this analysis.

\subsection{Analysis}

\subsubsection{Data preparation}
The BiSON data were prepared as described in \cite{2014MNRAS.441.3009D} and divided into one-year, non-overlapping time series.

\subsubsection{Choice of activity proxy}
It is common to express activity changes in terms of a linear relationship to an activity measure, although sometimes additional terms may be used, especially for higher-degree modes. Such analyses have been carried out using, for example, the sunspot number, the 10.7 cm radio flux (RF), and the global magnetic flux. \citet{1993ApJ...411L..45B} found that, for observations over the period 1984--1990, the RF and the MgII index gave better correlation with medium-degree frequency shifts than the global Kitt Peak magnetic field strength, and \citet{2007ApJ...659.1749C} also found better agreement with proxies such as the RF flux that have a greater sensitivity to weak flux. On the other hand, for low-degree modes \citet{2001MNRAS.322...22C} found that the correlations were essentially the same for all six proxies they examined, although there was a slight difference between the sensitivity to the Kitt Peak magnetic index in the falling phase of Cycle 22 and the rising phase of Cycle 23.  For local and latitudinally resolved measurements a spatially-resolved proxy is needed, and for this purpose a localized measure of the  photospheric magnetic field strength (often called a magnetic activity index, or MAI), tends to be preferred because the irradiance-based measures are not spatially resolved. The relationships between these proxies are not necessarily linear, and may not be consistent over time as instruments are upgraded or observing protocols change. Even magnetic measurements from different instruments may show the effects of small differences in calibration; also, many magnetic indices are only available as Carrington maps that cannot easily be translated into daily indices.  These considerations need to be borne in mind when using the measures to compare with long sequences of helioseismic data. For the current work, as we are dealing with Sun-as-a-Star data, we have chosen to use the daily RF index averaged over the exact period covered by each spectrum.

\subsection{Results}

For all variables, we first find the modes that are common to all the data sets being compared and then calculate the shifts for each mode relative to an error-weighted mean over all the data sets. These individual-mode shifts can then be combined in an error-weighted mean to show the overall variation. 

\subsubsection{Frequency}

\begin{figure*}
\centerline{\epsfxsize=0.45\linewidth\epsfbox{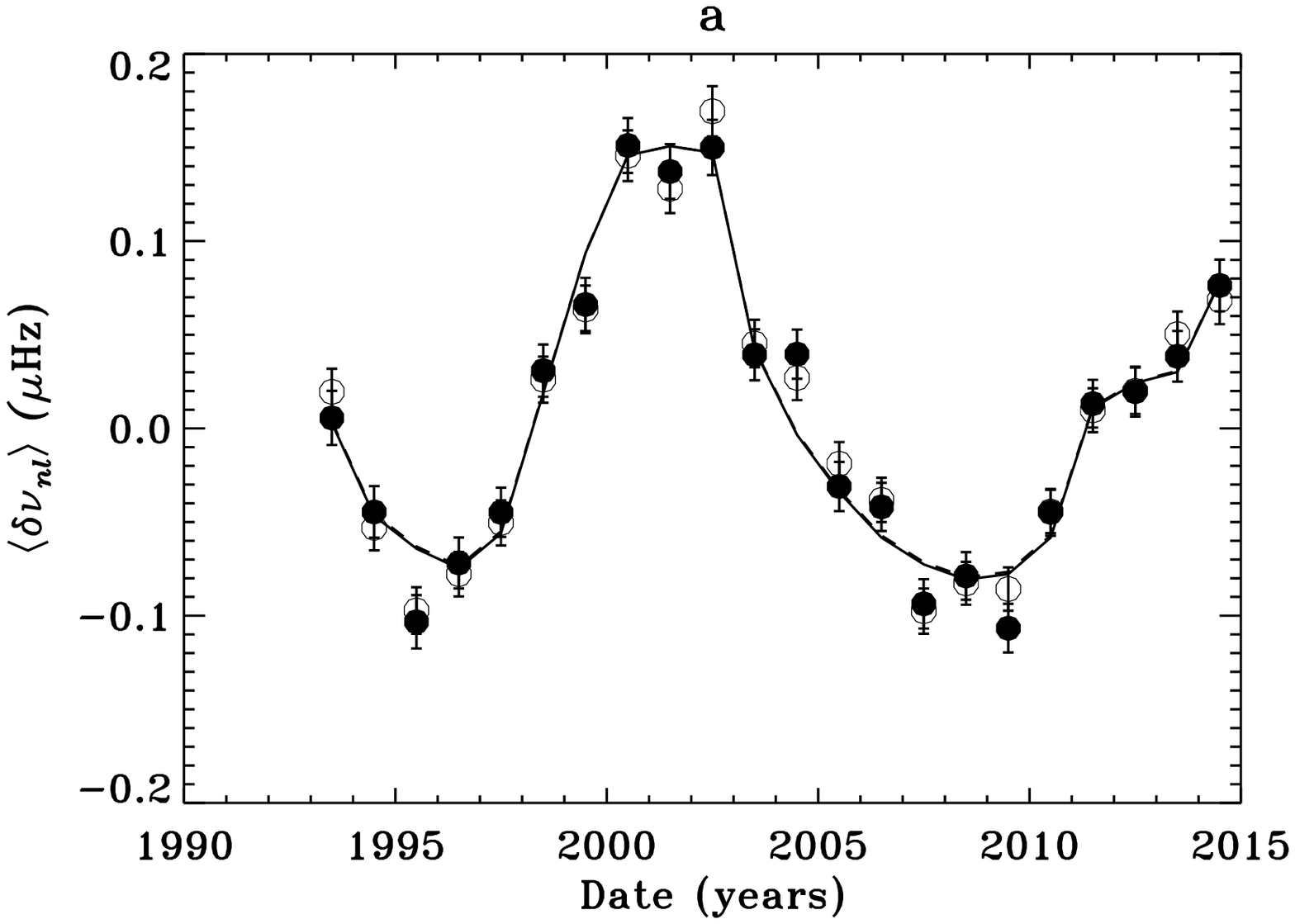}\epsfxsize=0.45\linewidth\epsfbox{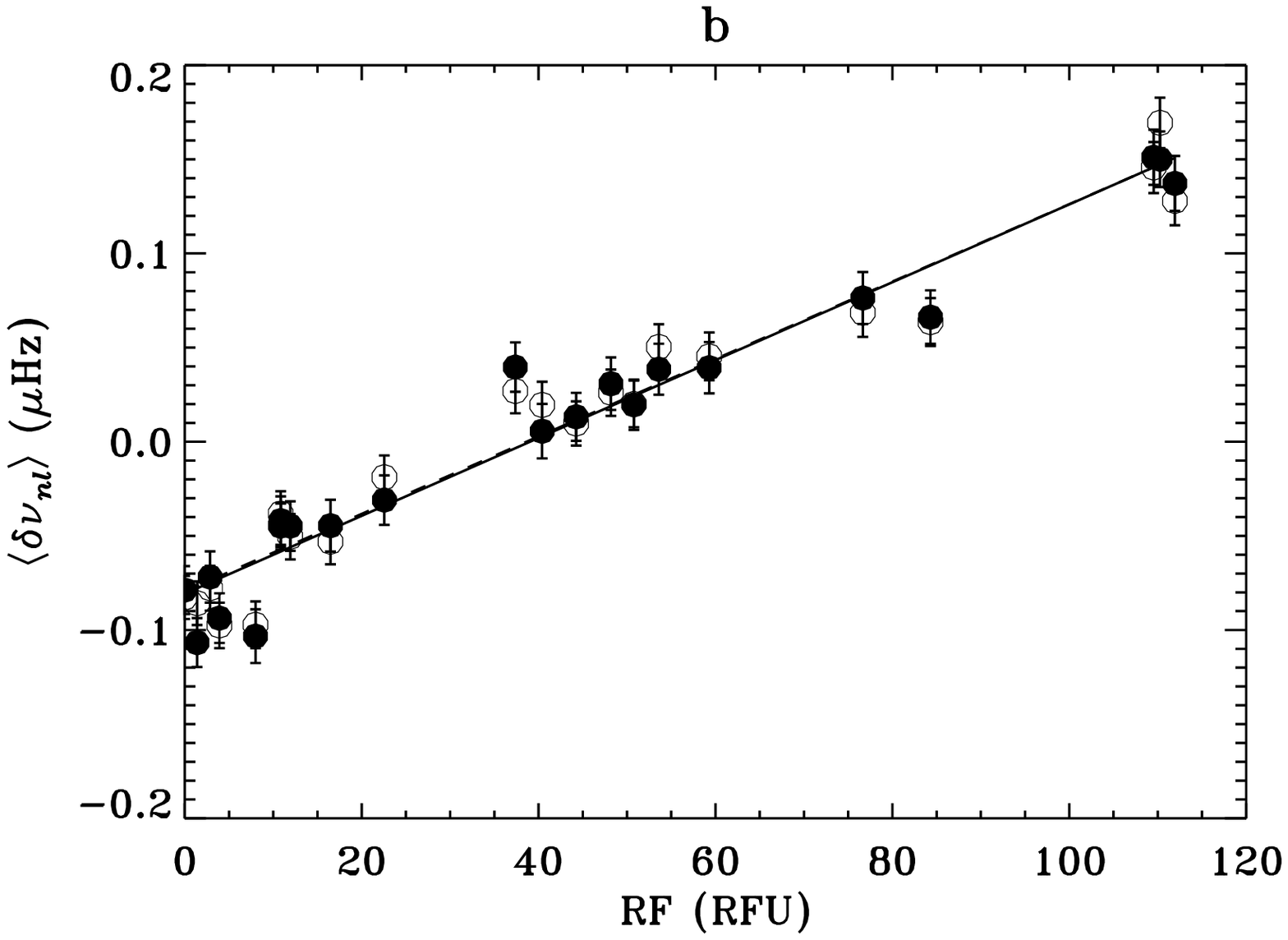}}

\centerline{
\epsfxsize=0.45\linewidth\epsfbox{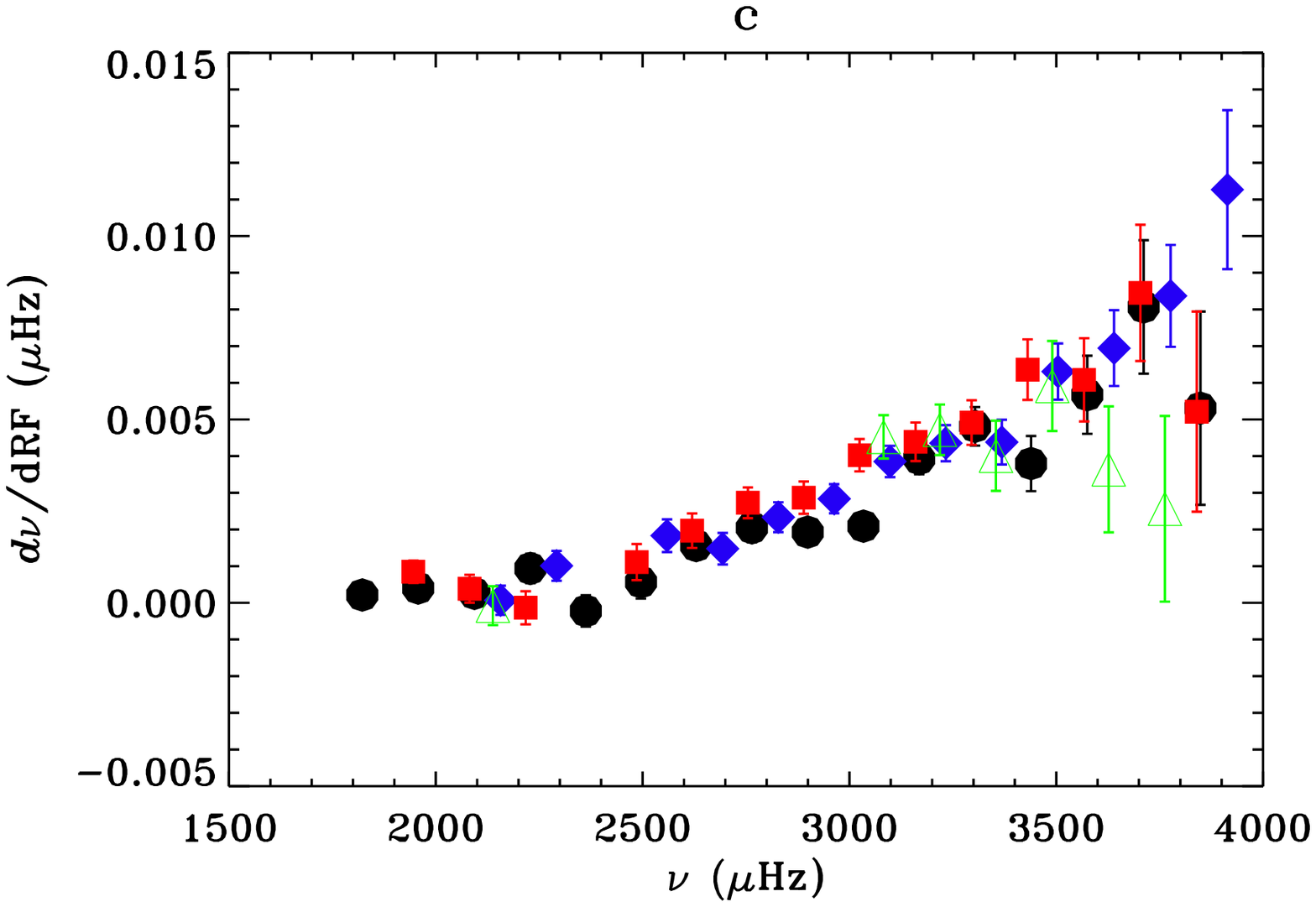}
\epsfxsize=0.45\linewidth\epsfbox{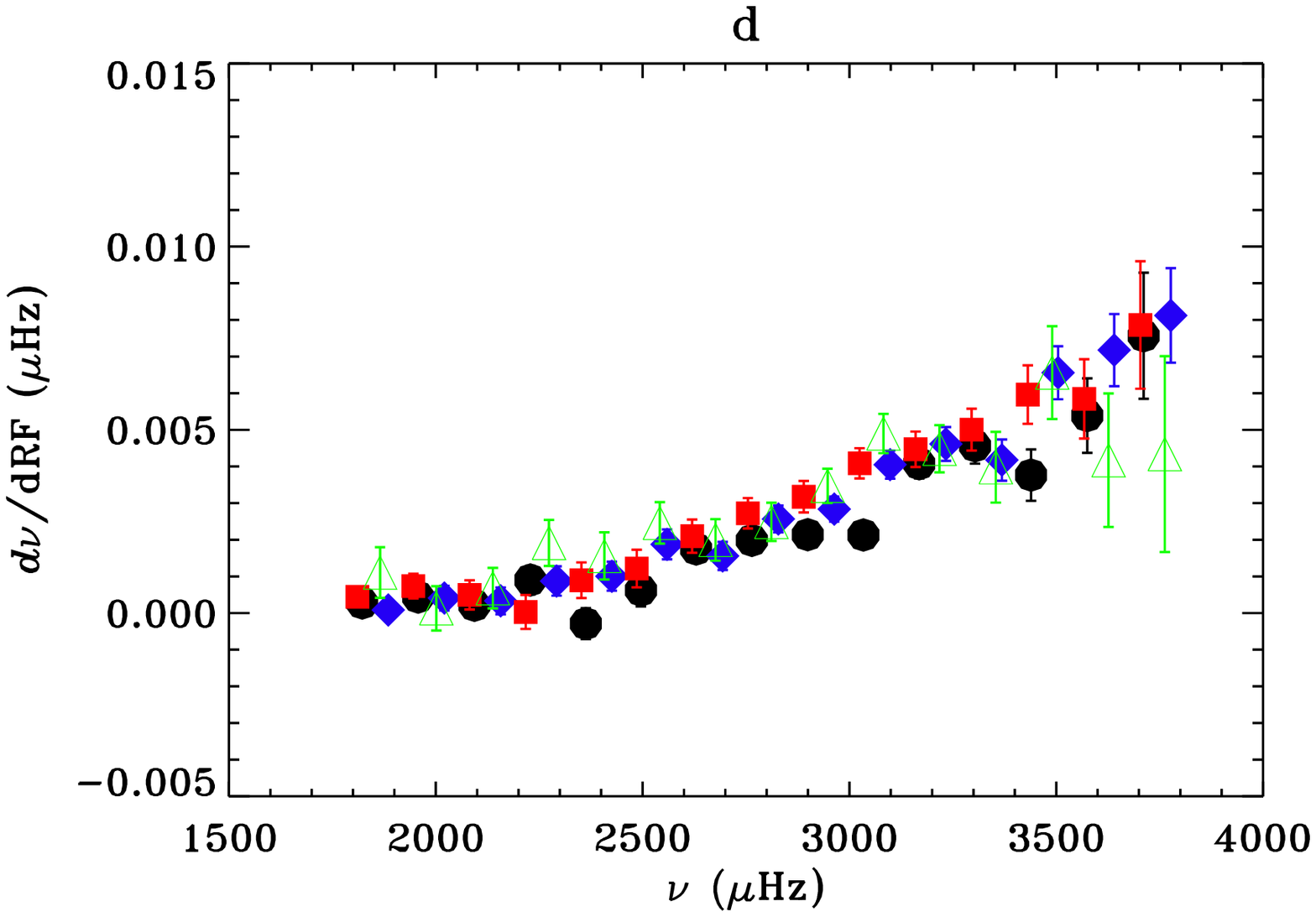}
}
\caption{Solar-cycle frequency changes from BiSON data 1993--2014. The top row shows frequency  variation from MLE (filled symbols) and MCMC (open symbols) fits, as an error-weighted mean over all modes with $l \leq 2$ and frequencies between 2 and 3.5 mHz, as a function of time (a) and RF index (b). The solid and dashed lines represent linear fits to the RF index for MLE (solid) and MCMC (dashed). The bottom row shows $d\nu_{nl}/dRF$ as a function of frequency for MLE (c) and
MCMC (d) fits, for  $l=0$ (black circles), $l=1$ (blue diamonds),
$l=2$ (red squares) and $l=3$ (green open triangles).}
\label{fig:freq1}
\end{figure*}

Figure~\ref{fig:freq1}(a,b) shows the mean frequency variation with time and with RF index for the two algorithms, and Figure~\ref{fig:freq1}(c,d) shows $d\nu_{nl}/dRF$ for each mode as a function of frequency for each method. The methods agree well, even in the small deviations from the linear fit; in fact, the correlation coefficient between the MCMC and MLE mean frequency shifts is 0.992, better than the correlation between the shifts and the RF index, which is 0.975 in each case. The threshold for significance at the 0.1 per cent level is  0.65 for 22 data points, so all of these are highly significant correlations. It is noticeable in Figure~\ref{fig:freq1}(a) that in both solar minima the frequency points tend to lie below the activity trend-line. The frequency dependence of the shifts is very clearly seen, and there is a hint that the $l=0$ shifts are lower than those for $l>0$, consistent with the expected behaviour due to the latitudinal distribution of the activity bands.

\subsubsection{Amplitude}

\begin{figure*}
\centerline{\epsfxsize=0.45\linewidth\epsfbox{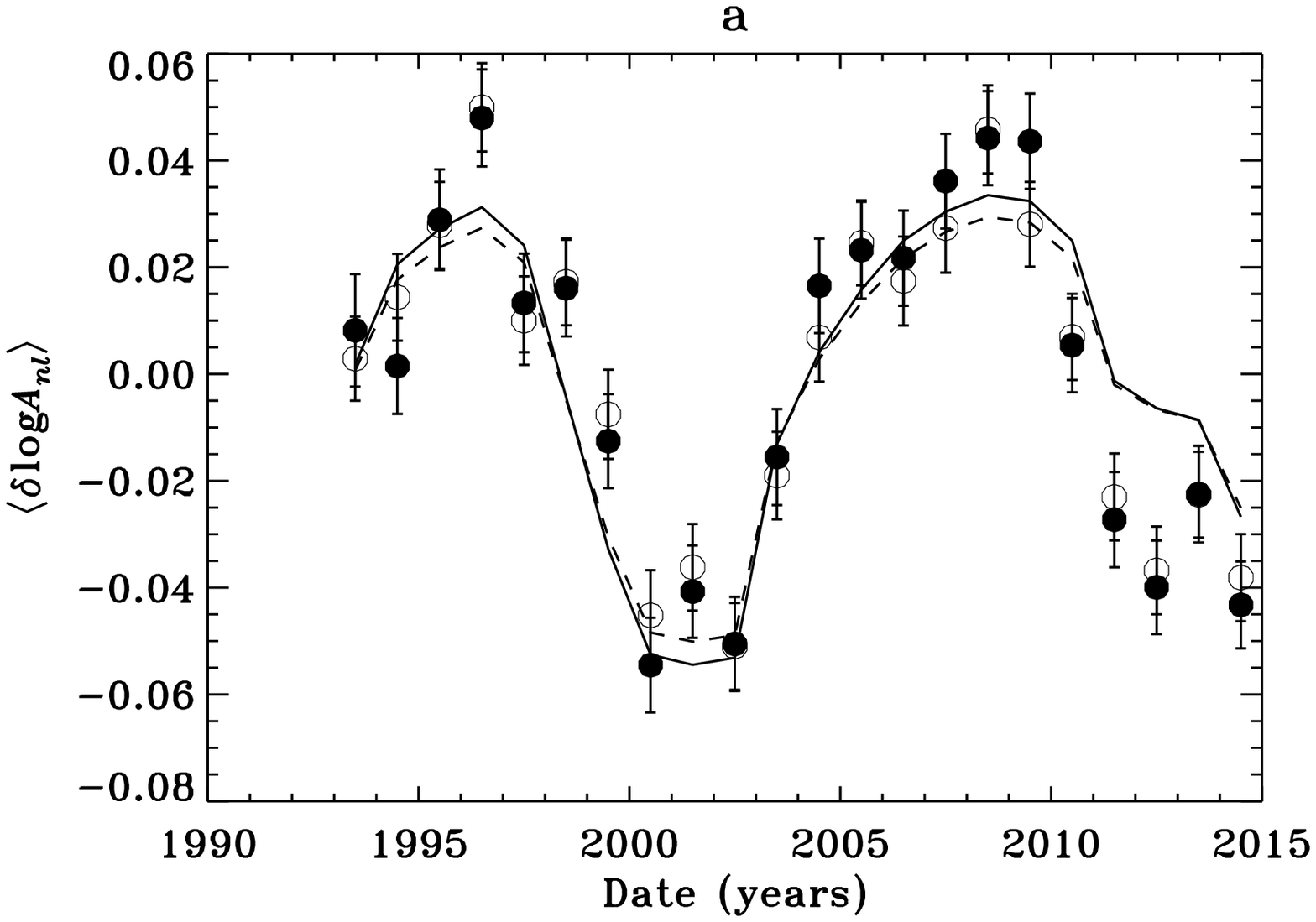}
\epsfxsize=0.45\linewidth\epsfbox{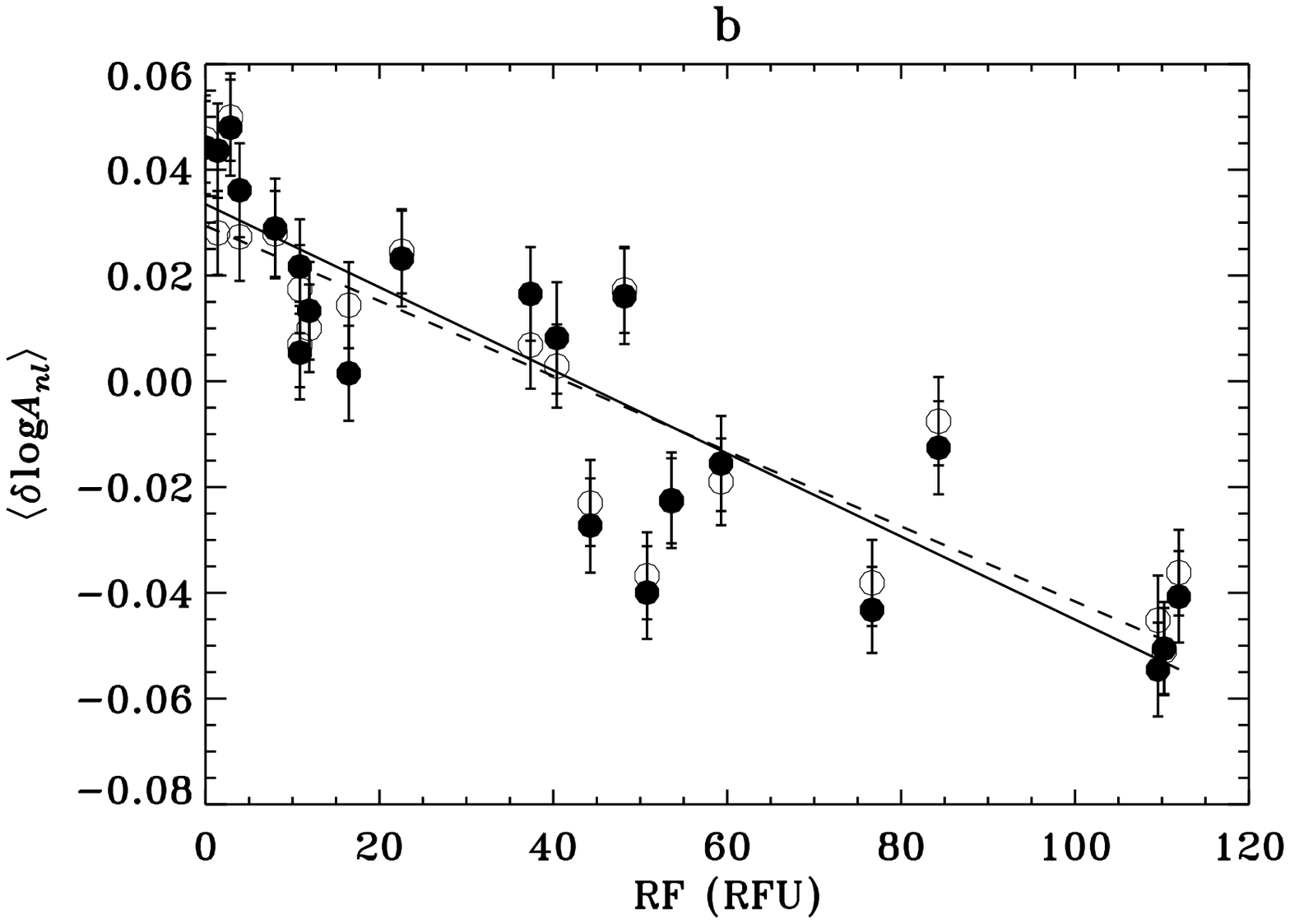}}

\centerline{
\epsfxsize=0.45\linewidth\epsfbox{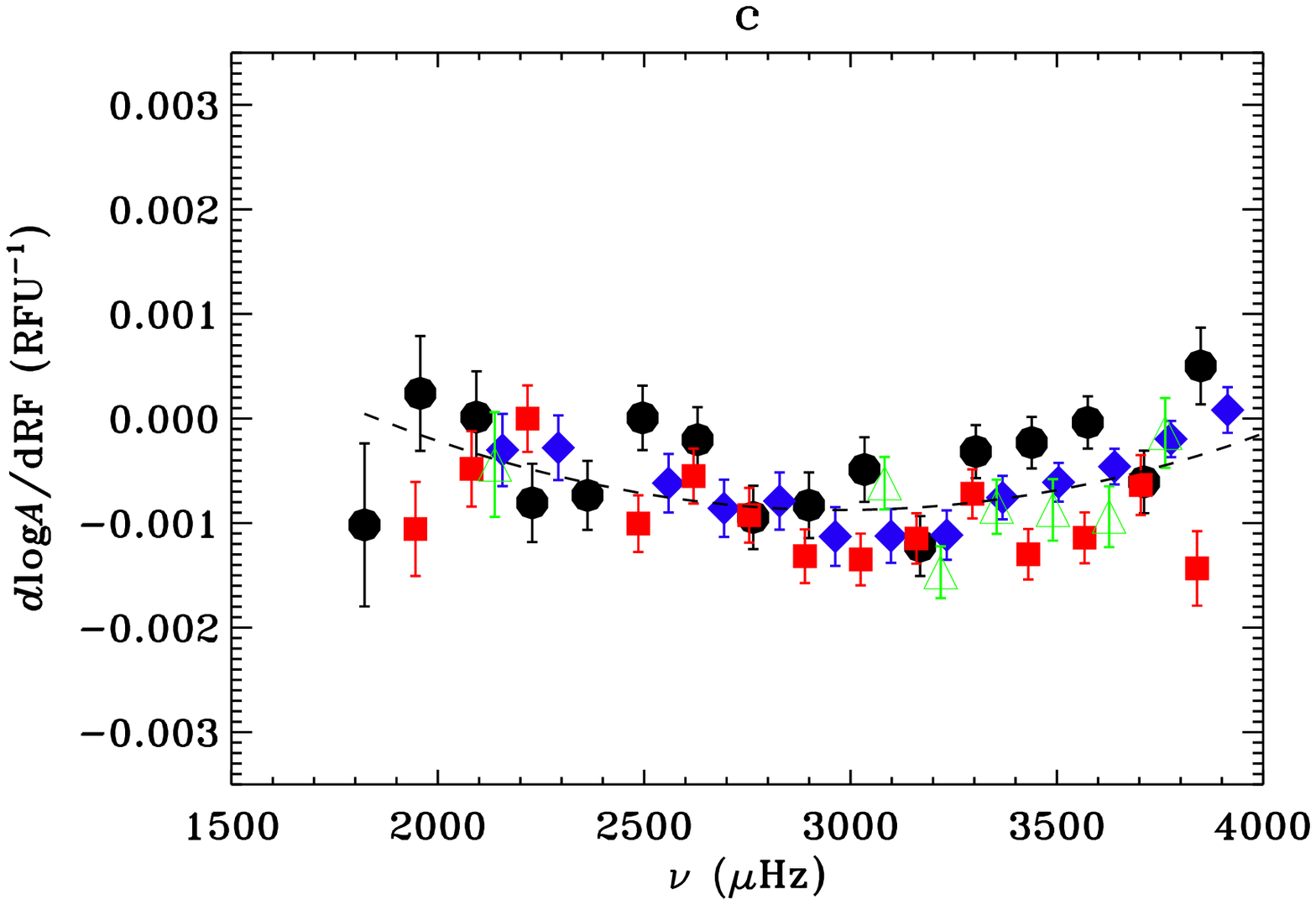}
\epsfxsize=0.45\linewidth\epsfbox{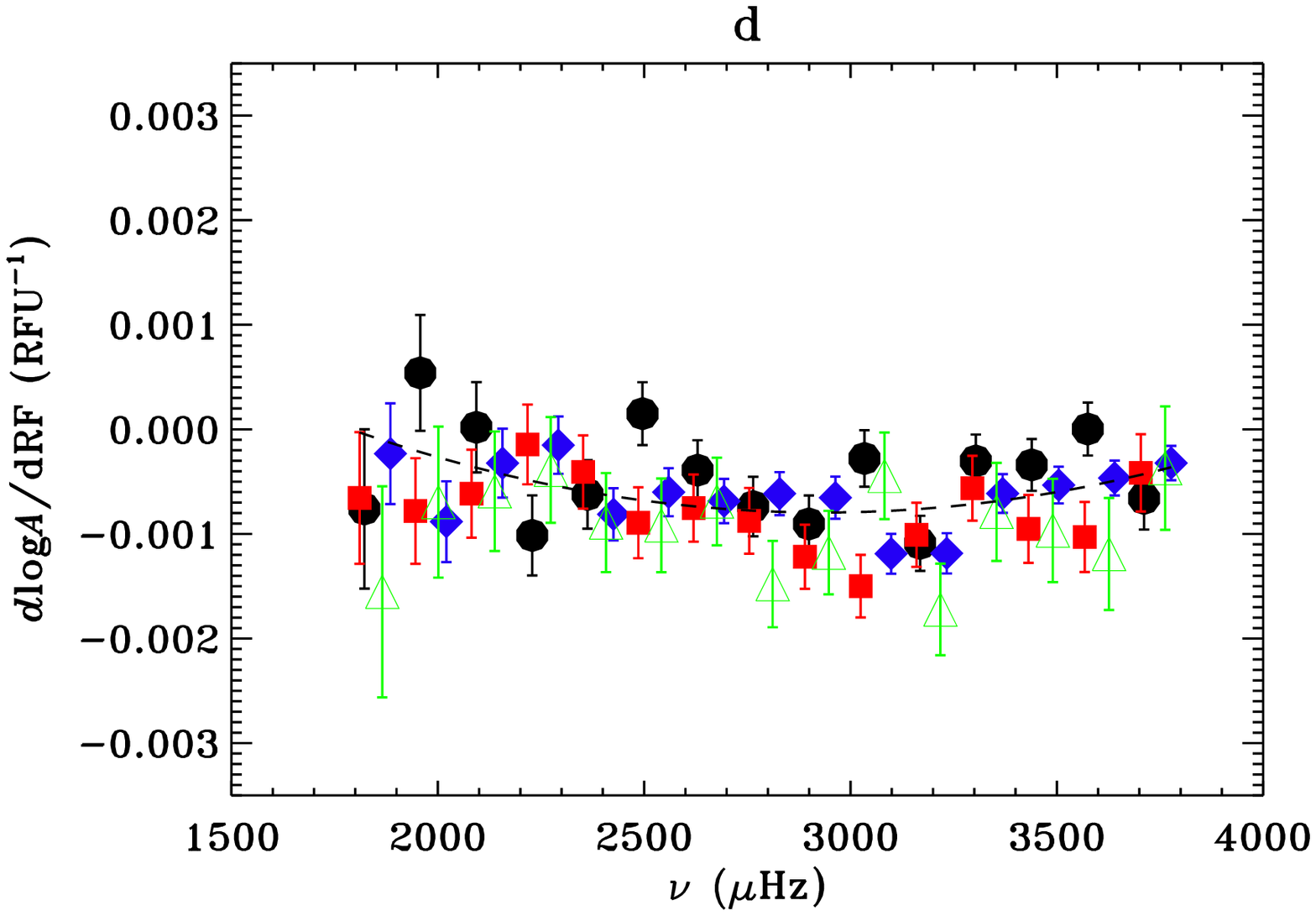}
}
\caption{Solar-cycle mode amplitude changes from fits to 1-year BiSON spectra from 1993--2013. The top row shows amplitude variation from  MLE (filled symbols) and MCMC (open symbols) fits, averaged over all modes with $l \leq 2$ and frequencies between 2 and 3.5 mHz, as a function of time (a) and RF index (b). The solid and dashed lines represent linear fits to the RF index for  MLE (solid) and MCMC (dashed). The bottom row shows shows $d\log A_{nl}/dB$ as a function of  frequency for MLE (c) and MCMC (d), for  $l=0$ (black circles), $l=1$ (blue diamonds),
$l=2$ (red squares) and $l=3$ (green open triangles). The dashed curve represents a quadratic function in frequency fitted to the modes with $l\leq 2$.}\label{fig:amp1}
\end{figure*}

Figure~\ref{fig:amp1}(a,b) shows the amplitude variation for the two algorithms as a function of time and as a function of RF index, and Figure~\ref{fig:amp1}(c,d) shows $d\log A_{nl}/dRF$ as a function of frequency for each algorithm Again, the two methods show good agreement, with a correlation coefficient of 0.98 between the two sets of mean shifts, while the correlation with the RF index is -0.88 in each case. 

The data plotted in Fig.~\ref{fig:amp1}(a,b) have been corrected for the window-function effect on the amplitude, but this makes very little difference to the result.  The mean shift changes by
$(-0.078\pm 0.005)$ per cent per RFU from MLE and $(-0.071\pm 0.005)$ per cent per RFU from MCMC, 
giving a change of about 
 $(-8.8\pm 0.6)$ per cent for MLE or $-8.0\pm 0.5$ per cent for MCMC  
from the highest to the lowest RF value.
The scatter of the points around the linear fit, together with the agreement between the methods, suggests that effects other than a simple linear dependence on activity may be involved. Curiously, the points for the years 2011--2013 in Figure~\ref{fig:amp1}(a) all fall below the fit to the RF index, suggesting that the modes may have been more strongly suppressed in the rising phase of Cycle 24.

The uncertainties are still too large to allow us to usefully quantify the frequency dependence of the individual-mode shifts, but there is a visible tendency for the modes in the middle of the five-minute band to be more strongly suppressed by activity. The reduction in $\chi^2$ obtained by fitting a quadratic function of frequency to the values with $l \leq 2$, rather than a constant, corresponds to a probability of 12 per cent for MLE and 23 per cent for MCMC that the result could have been obtained from a random distribution -- better than a $1\sigma$ result but not at the $2\sigma$ level.  The shape of the variation, with the greatest
sensitivity around 3\,mHz where the modes are strongest, is consistent with
the findings from resolved helioseismology.

\subsubsection{Line width}

\begin{figure*}

\centerline{\epsfxsize=0.45\linewidth\epsfbox{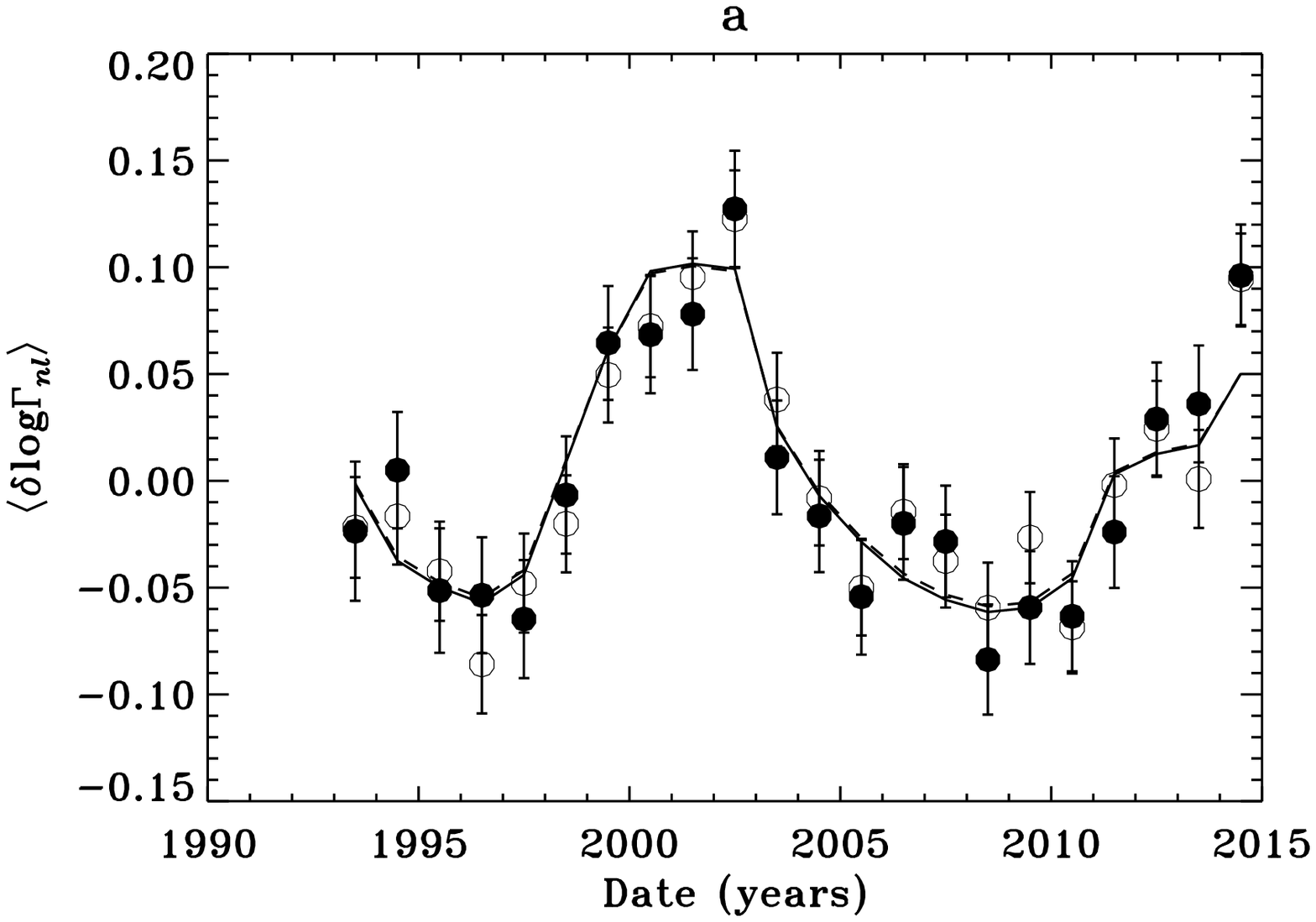}\epsfxsize=0.45\linewidth\epsfbox{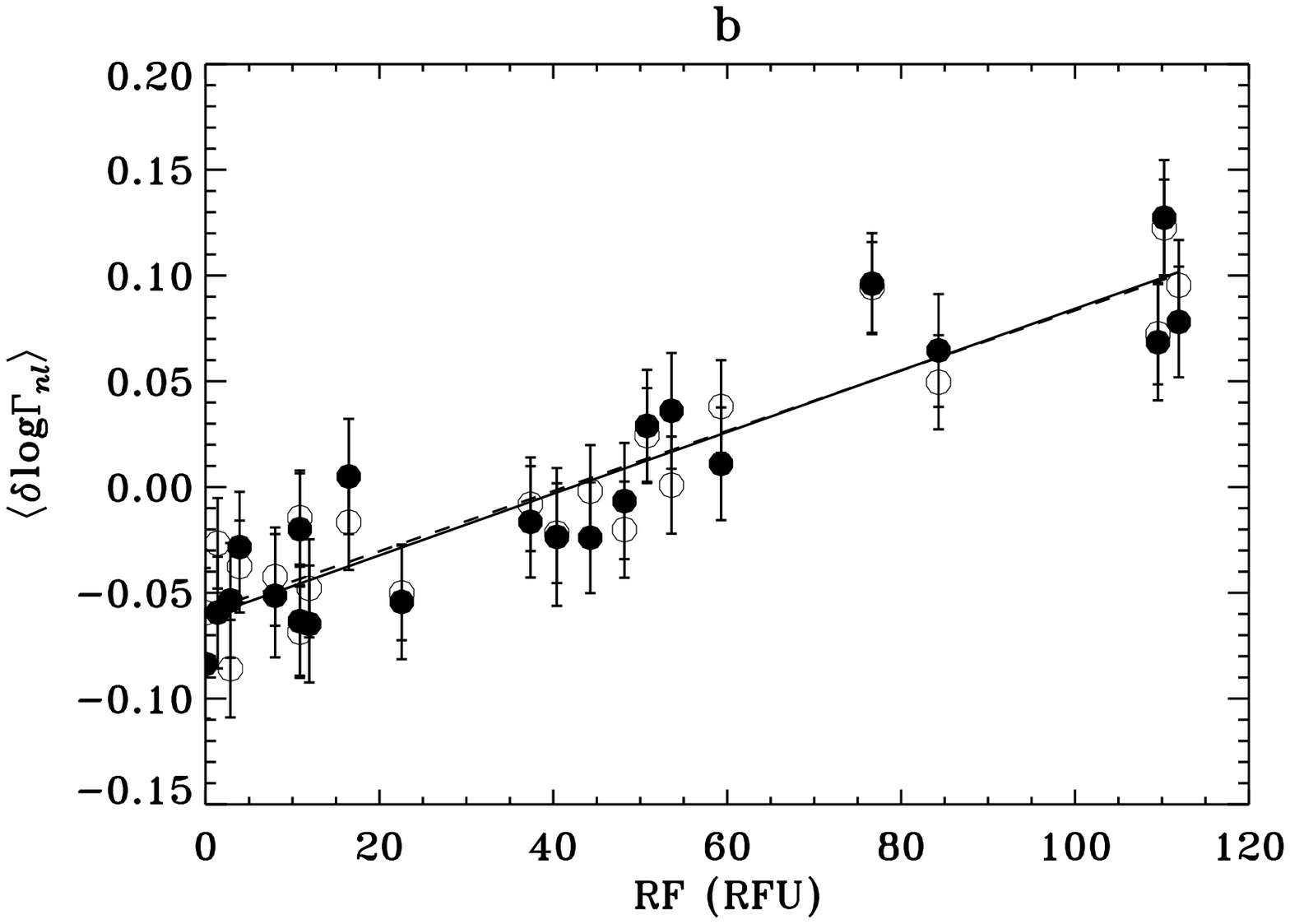}}

\centerline{
\epsfxsize=0.45\linewidth\epsfbox{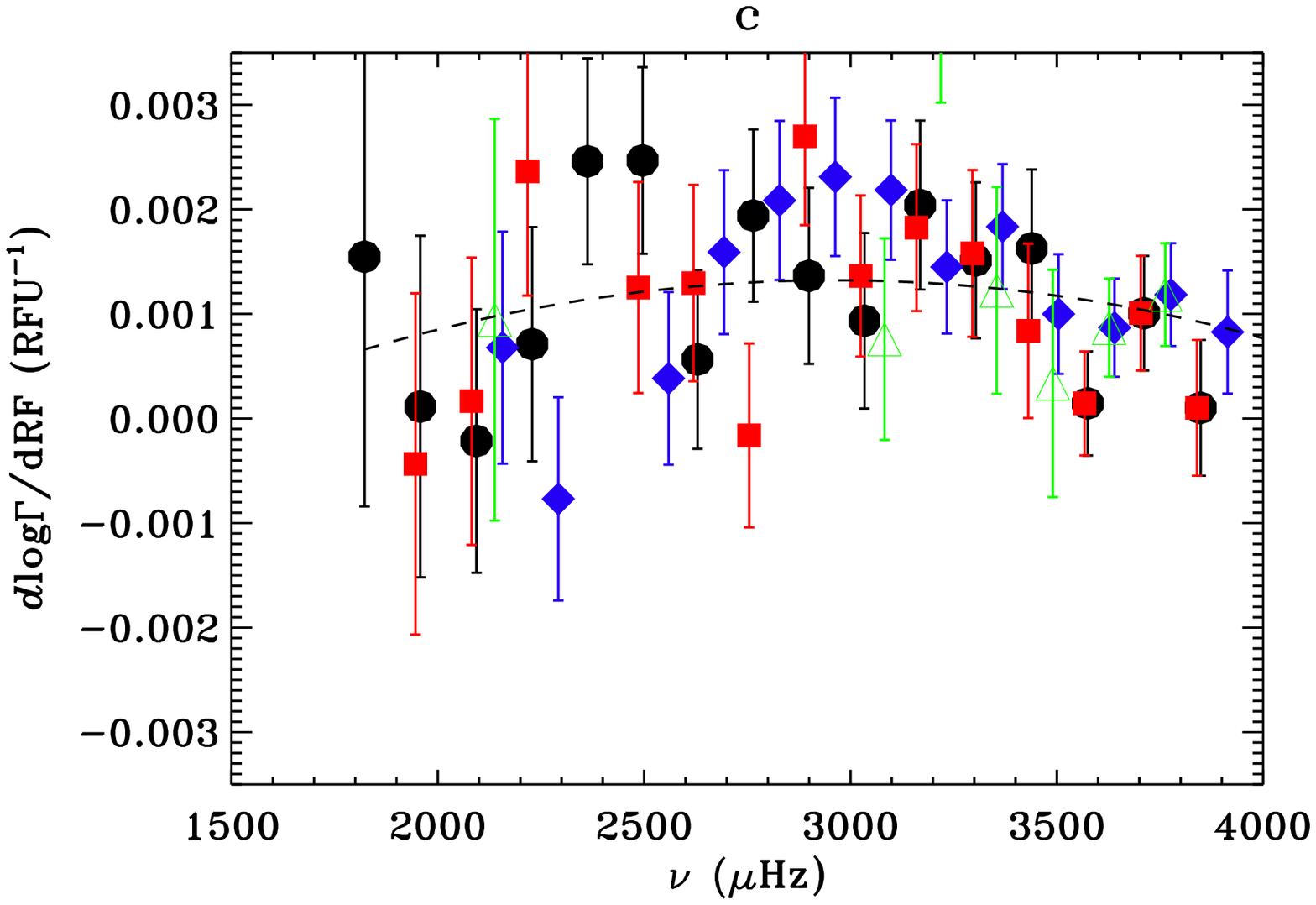}
\epsfxsize=0.45\linewidth\epsfbox{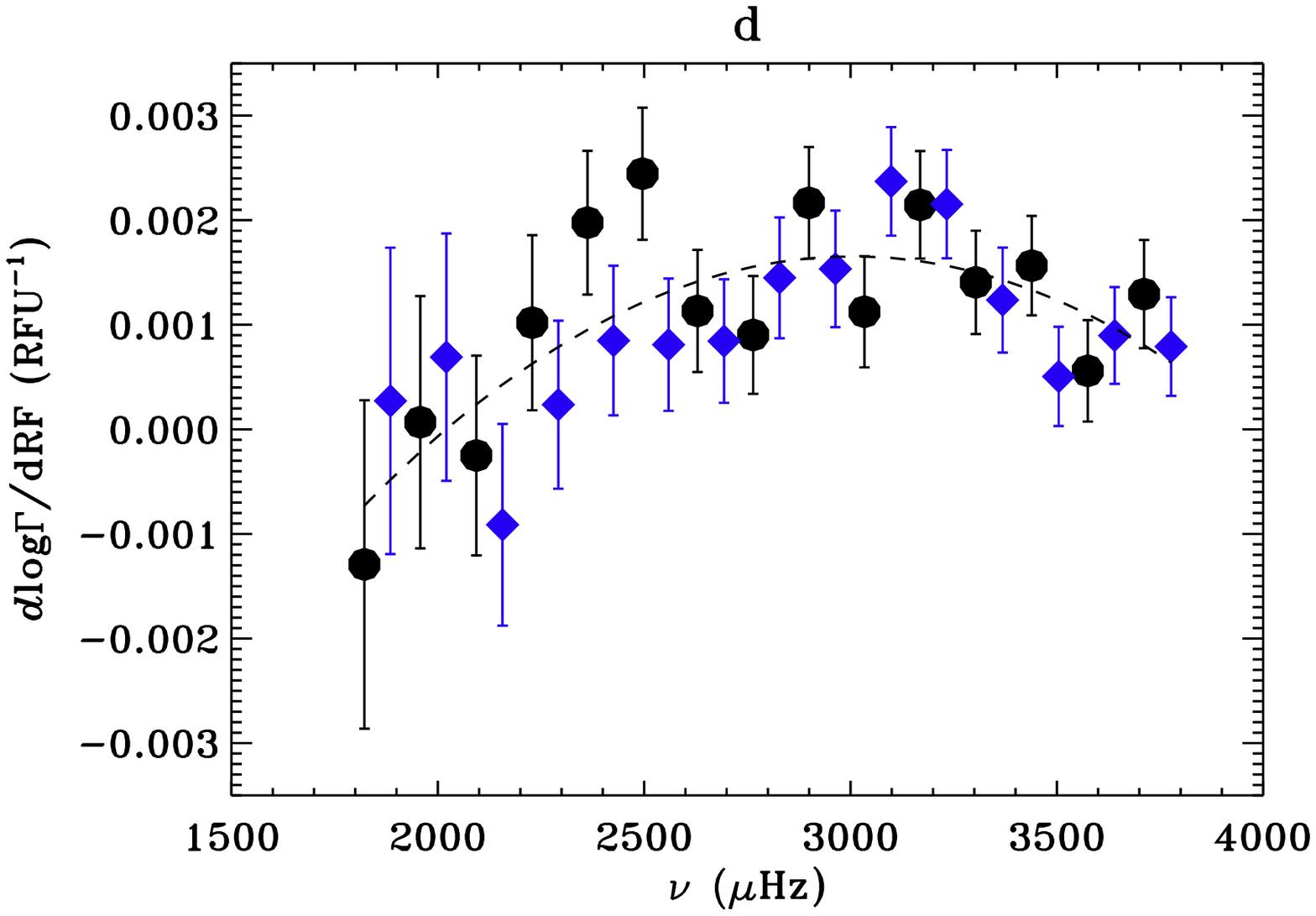}
}

\caption{Width variation with activity. Top row shows mean shift in $\log\Gamma_{nl}$ from  MLE (filled symbols) and MCMC (open symbols) fits, averaged over all modes with $l \leq 2$ and frequencies between 2 and 3.5 mHz, as a function of time (a) and RF index (b). The solid and dashed lines represent linear fits to the RF index for MLE (solid) and MCMC (dashed).
The bottom row shows $d\log\Gamma_{nl}/dRF$ as a function of  frequency for MLE (c) and MCMC (d) for $l=0$ (black circles), $l=1$ (blue diamonds), $l=2$ (red squares), and $l=3$ (green open triangles). For MCMC fits the $l=0$ value is for the $l=0/2$ pair and the $l=1$ for the $l=1/3$ pair. The dashed curve shows a quadratic fit to the $l\leq 2$ values with frequency as the independent variable.}
\label{fig:width1}
\end{figure*}

Figure~\ref{fig:width1}(a,b) shows the mean shift in $\log\Gamma$ for the two algorithms as a function of time and RF, averaged over the modes between 2 and 3.5\,mHz. 
Figure~\ref{fig:width1}(c,d) shows $d\log \Gamma/dRF$ as a function of frequency for each method. A direct comparison between the two data sets in this case is more difficult because the MCMC algorithm returns only one width per mode pair while the MLE gives two; also the different treatment of the window function affects the line width estimates. However, the two sets of mean shifts have a correlation coefficient of 0.946, while the correlation coefficient with the RF index is 0.922 for MLE and  0.951 for MCMC . Although the width shifts are larger than those in amplitude they also have larger uncertainties.  Given the uncertainties, we do not see any significant deviation from the linear trend of the mean shifts with RF index. The mean shift is $(0.146\pm 0.016)$ per cent per RFU for MLE and $(0.142 \pm 0.013)$ per cent per RFU for MCMC, corresponding to a change of $(16.3\pm 1.8)$ per cent and $(15.9 \pm 1.5)$ per cent respectively between minimum and maximum RF values.

With the available data, the statistical significance of any frequency dependence of $d\log\Gamma_{nl}/dRF$ is not high. The decrease in $\chi^2$ obtained by fitting a quadratic function of frequency to the values, rather than a constant value, corresponds to a 14 per cent probability for MLE and 10 per cent for MCMC of this result being obtained from random data -- again better than a $1\sigma$ result but not at the $2\sigma$ level.

\subsubsection{Energy supply rate}
\begin{figure*}

\centerline{\epsfxsize=0.45\linewidth\epsfbox{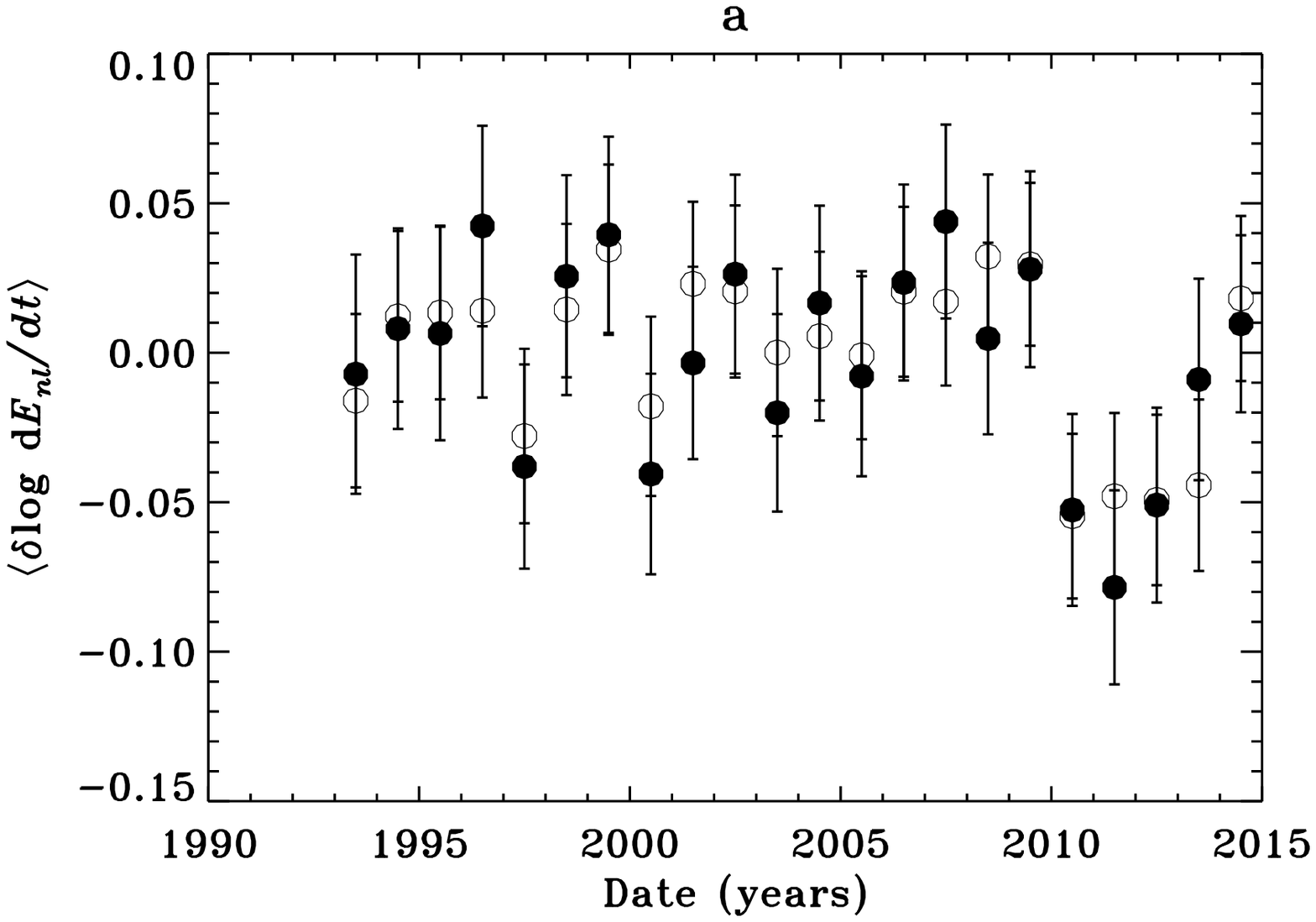}\epsfxsize=0.45\linewidth\epsfbox{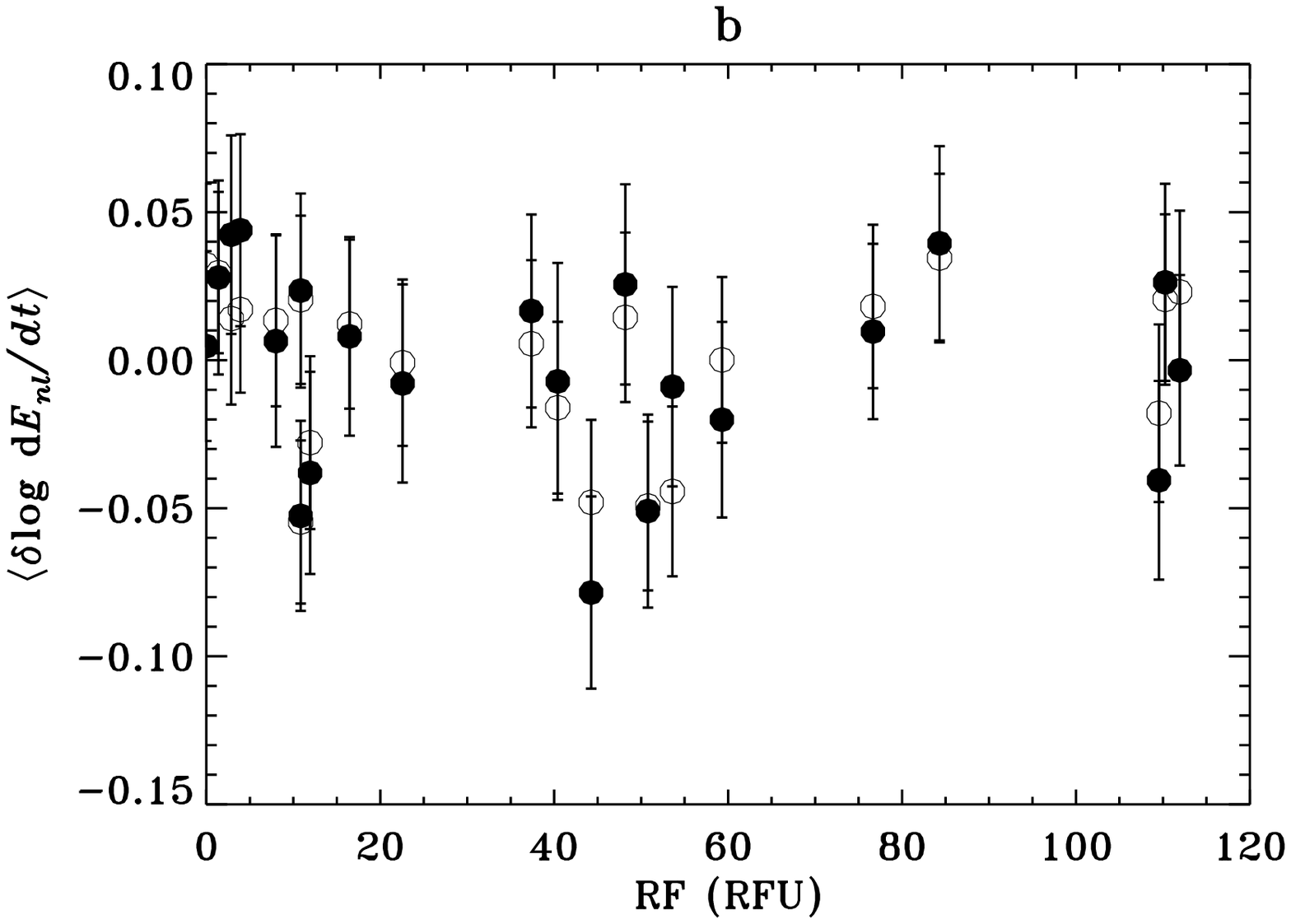}}

\caption{Mean change in log energy supply rate from MCMC (open symbols) and MLE (filled symbols) fits, averaged over all modes with $l \leq 2$ and frequencies between 2 and 3.5 mHz, as a function of time (left) and RF index (right). }
\label{fig:rate}
\end{figure*}

We plot in Figure~\ref{fig:rate} the variation with date and RF index of the
mean change in the natural log of the energy supply rate $E_{nl}$, formed by taking the sum of the mean line width shift and twice the mean amplitude shift. Following \citet{2000MNRAS.313...32C}, the $\approx 90$ per cent correlation between the $A$ and $\Gamma$ errors has been taken into account in the error propagation.
 The results imply a fractional change in the energy supply rate between minimum and maximum RF of $(-1.1\pm 2.2)$ per cent from MLE and $(0.17 \pm 1.9)$ per cent from MCMC. The previous result that there is no significant change in the energy supply rate with activity level is thus confirmed. On the other hand, there is a hint 
of a decrease in the energy supply rate from 2011--2013, which we see as an unexpectedly large decrease in the mode amplitude in this period.

\subsubsection{Asymmetry}

\begin{figure*}

\centerline{\epsfxsize=0.45\linewidth\epsfbox{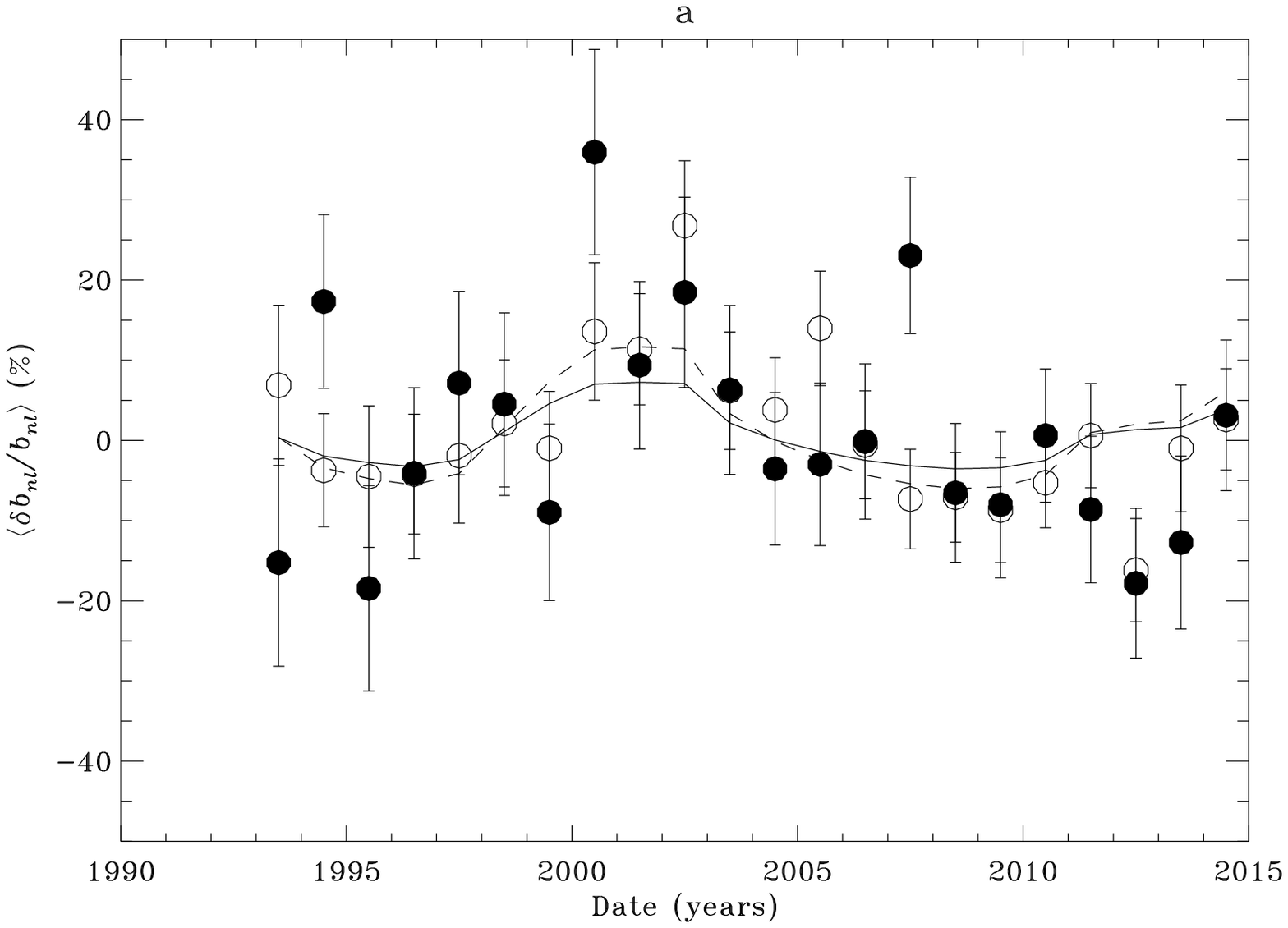}\epsfxsize=0.45\linewidth\epsfbox{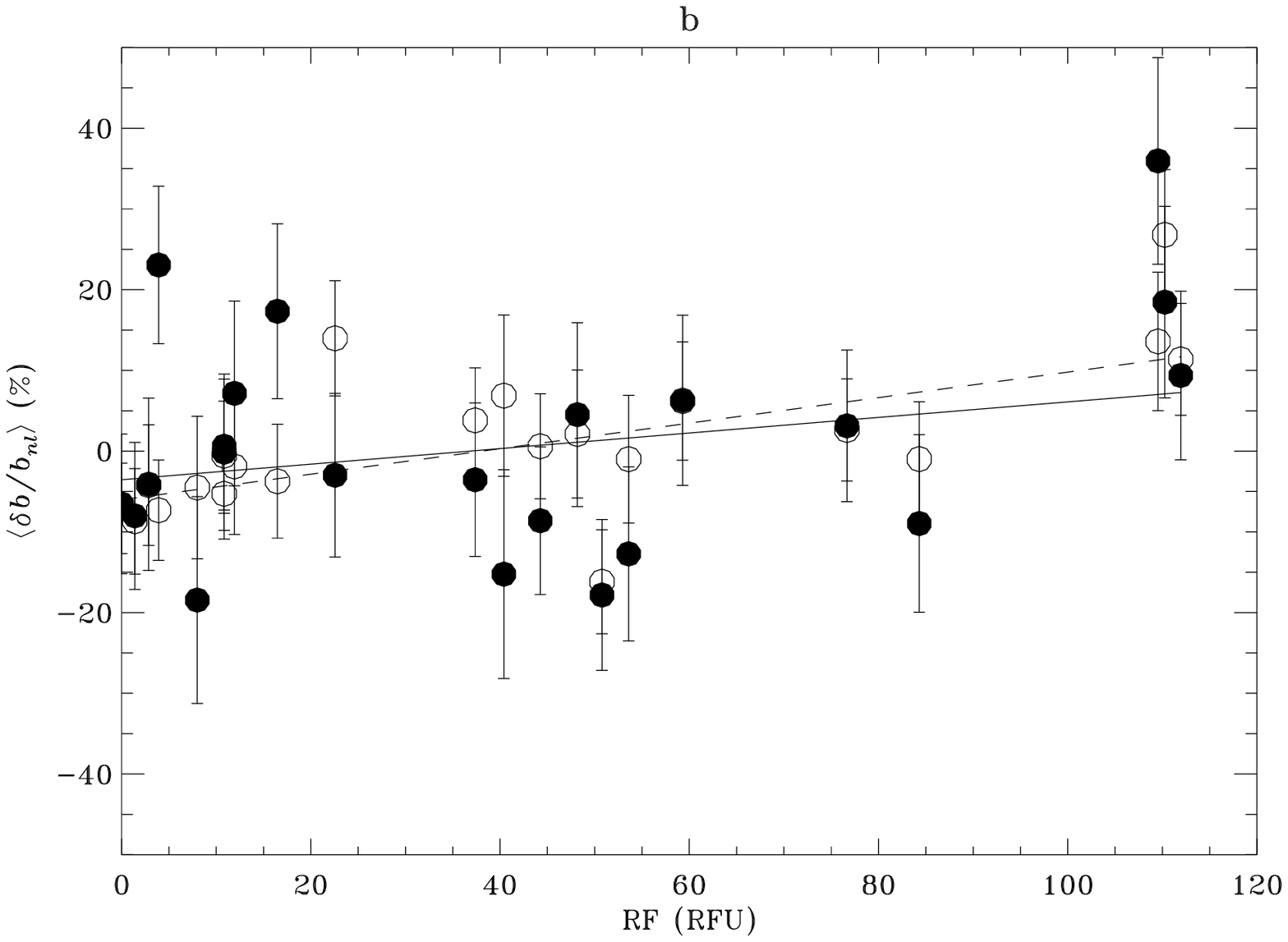}}

\caption{Asymmetry variation (mean fractional shift) from MLE (filled symbols) and MCMC (open symbols) fits, averaged over all modes with $l \leq 2$ and frequencies between 2 and 3.5 mHz, as a function of time (left) and RF index (right). The solid and dashed lines represent linear fits to the RF index for MLE (dashed)and MCMC (solid).}
\label{fig:asym1}
\end{figure*}

Figure~\ref{fig:asym1} shows the mean fractional shift in the asymmetry parameter $b$ as a function of time and of RF index, for the two algorithms. As the uncertainties are large, we do not show the shifts for individual mode pairs.
The coefficient of correlation between  mean shift and the RF index is 0.36 for MLE (just short of being significant at the 10 per cent level) and 0.66
for MCMC (significant at the 0.5 per cent level), and the correlation between the two sets of mean shifts is 0.46. The results of the artificial data tests lead us to place more reliance on the MCMC estimate.
Even though the result is marginally significant, it is consistent with the results in \citet{2007ApJ...654.1135J}, which only included data on the declining phase of the last cycle, where the asymmetry change is most pronounced.

\section{Discussion and Conclusions}
\label{sec:conclusion}

We have analysed 22 years of BiSON data for activity-related changes in the mode parameters using two different algorithms. 
The analysis was validated by the use of an improved process that allowed for the incorporation of the effects of line asymmetry.
The newer MCMC algorithm represents a small improvement over the older MLE, particularly in the search for changes in the asymmetry, but in general there is good agreement between the two algorithms. Both methods accurately reproduce the input frequency shifts in artificial data and slightly underestimate the shifts in mode amplitude and width, which is perhaps not surprising given the limitations of the model peak profile.

One of the objectives of the present work was to check that the previous reports of frequency shifts were not an artefact of underlying changes in the mode asymmetry, as suggested by \citet{2013ASPC..478..137K}. Contrary to this suggestion, we find that our fitting accurately reproduces the input frequency shifts in artificial data and finds shifts in the real data of the same magnitude as those reported in earlier work. 
Furthermore, the MCMC parameter estimation explicitly favours a frequency
change over a simple change in underlying asymmetry (with the model we
apply) and in any case the asymmetry change required to account for the frequency shift would be unrealistically large.

We have not attempted in this work to reproduce the kind of analysis of frequency shifts used by for example \citet{2012ApJ...758...43B}, in which spectra from heavily overlapped time periods were used and the shifts were averaged in frequency bands; our objective was instead to look for systematic biases in long-term trends. Having eliminated the fear of such biases in the frequency shifts, we will be able to proceed with confidence to a more detailed analysis of the short-term variations in further work. We do see deviations from the linear relationship between frequency and activity that are more evident at low activity and appear consistent with those reported by  \citet{2010ApJ...718L..19F, 2012ApJ...758...43B} and others.

Our amplitude and width changes also agree well with earlier work. \cite{1993MNRAS.265..888E} (who fitted using a symmetric Lorentzian peak profile) report a change of -35 per cent in mode area (height times width) between the 1986 minimum and the 1990 maximum, but based on their linear fit to sunspot number a change of -20 per cent would give a more realistic comparison with our result. As our amplitude is proportional to the square root of their mode area, that would translate to a roughly -10 per cent change in amplitude, which compares reasonably well with our estimate of around -8 per cent. \cite{2000MNRAS.313...32C}, in an analysis covering the period 1991--1997
and averaging over modes between 2.6 and 3.6 mHz, report changes of $(24\pm 3)$ per cent in line width and $-(22\pm 3)$ per cent in power between maximum and minimum, (corresponding to $(-11\pm 2)$ per cent in  amplitude). The 100-day periods analysed by
\cite{2000MNRAS.313...32C} correspond to a range in RF index of about 150 RFU, compared with the range of 110 in the 365-day average values in our analysis, which would make their 24 per cent change in line width equivalent to 18 per cent, within a $1\sigma$ difference from our 15.5 per cent value. For amplitude, their change of -11 per cent would be equivalent to $(8.1\pm 1.1)$ per cent, in good agreement with our analysis. It is not clear whether the $\approx 10$ per cent underestimate uncovered by our simulations applies to both results; the biases for symmetric fits may well be different.

We do not reproduce the finding of \citet{2007A&A...463.1181S} that the $l=0$ mode width is less sensitive to global activity than the higher-degree modes. It is not possible to check this using the MCMC fitting, as the $l=0$ and $l=2$ mode components are assumed to have the same width.

Our results on the asymmetry shifts are somewhat disappointing. However, we should note that the work of \cite{2007ApJ...654.1135J} was more narrowly focused on the asymmetry measurement and they carried out a more elaborate analysis to obtain the best possible results. Also, if we look only at the eight years of BiSON data that they used in the frequency range that they considered, we do reproduce their `marginally' significant result for the BiSON data. The much weaker correlation that we find for the whole dataset suggests that their result for BiSON may have been a statistical fluke; measuring the asymmetry in ground-based data with a duty cycle in the 70 per cent range is still challenging. In particular,  the intimate relationship between background and asymmetry may make it difficult to improve very much on these results using our current approach of pairwise fitting with a flat background for each pair. We note that the asymmetry shifts -- and the absolute asymmetry values -- observed for low-degree modes are too small to have a significant influence on the frequency shifts.

In general, both MLE and MCMC methods perform well and give consistent results for the solar-cycle variations of the mode parameters, as well as very similar uncertainties.  This has implications for future peakfinding strategies in both solar and stellar applications. For high-quality data the considerable extra computational expense of the Bayesian approach may not be justified unless we are specifically looking for asymmetry changes. We have not addressed the issue of the advantages that might be gained by using global or pseudo-global instead of pairwise fitting for solar data. Such methods are, again, more computationally intensive than the conventional approach, and when combined with the Bayesian method this expense becomes nearly prohibitive for large data sets. Our results suggest that it might be worthwhile instead to concentrate on developing MLE-based global or pseudo-global methods for use with solar data; such methods are already in use for asteroseismology of solar-like oscillators but may require refinement to deal with longer and higher-quality solar data sets. Any such methods would need to be cross-checked against earlier results. Our findings confirm that solar-cycle variations of mode frequency, lifetime, and amplitude from earlier Sun-as-a-star work are reliable.  If, as in the case of \citet{2013ASPC..478..137K}, the results obtained by novel methods differ substantially from those obtained by more conventional means, they should be treated with caution.

The data we have analysed cover the period up to the end of 2014, well into the weak but drawn out and double- or possibly multiple-humped maximum of Solar Cycle 24; we note that the RF value for 2014 is higher than that for 2013. Some of the issues around inter-cycle differences may become clearer once Cycle 24 definitely enters its declining phase. We await the next few years of measurements with interest.

\section*{Acknowledgements}
We would like to thank all those who are, or have been, associated with BiSON. BiSON is funded by the Science and Technology Facilities Council (STFC).  WJC thanks Sarbani~Basu
and Aldo~Serenelli for providing the model BS05(OP) frequencies. 

RH thanks 
 the National Solar Observatory for computing support.

\bibliography{bshifts}

\onecolumn

\appendix

\section{The solarFLAG complex frequency amplitude, and frequency
  power, spectrum}
\label{sec:appfull}

In the following description -- which is based on the detailed
discussions in
\citet{2006MNRAS.371.1731T}
\citet{2008AN....329..440C}
-- we model the $p$-modes as forced,
damped oscillators having a high $Q$. The frequency response of a
given mode (with $n$, $l$ and $m$) is then just a Lorentzian, which
may be written in complex amplitude form as:
 \begin{equation}
 L_{nlm}(\nu) = \Re{[A_{nlm}(\nu)]} + i\Im{[A_{nlm}(\nu)]},
 \label{eq:lor1}
 \end{equation}
where:
 \[
 \Re{[A_{nlm}(\nu)]} = \frac{\xi_{nlm} H_{nlm}^{1/2}}{1+\xi_{nlm}^2},
 \]
and
 \begin{equation}
 \Im{[A_{nlm}(\nu)]} =  \frac{H_{nlm}^{1/2}}{1+\xi_{nlm}^2}
 \label{eq:lor2}
 \end{equation}
are the real and imaginary parts, respectively. 
In the above,
$H_{nlm}$ is the height of the peak, and
 \[
 x_{nlm}=(\nu-\nu_{nlm})/(\Gamma_{nlm}/2),
 \]
where $\nu_{nlm}$ and $\Gamma_{nlm}$ are the central frequency and
line width of the mode, respectively.

The observed response in the frequency domain also depends on the
spectral response of the excitation function, which we call
$E_{nlm}(\nu)$. The complex amplitude, $A_{nlm}(\nu)$, of the full
response is therefore actually:
 \begin{equation}
 A_{nlm}(\nu) = L_{nlm}(\nu) E_{nlm}(\nu).
 \end{equation}
We excite the modes with the granulation-like noise, which has a
response that is white only locally in the vicinity of the
resonance. The power spectral density of this granulation-like noise
follows the
\citet{1985ESASP.235..199H}
power-law model given in
Equation~\ref{eq:harbasic}. To obtain the excitation response,
$E_{nlm}(\nu)$, we must take the square root of the power spectral
density of the granulation-like noise; and we must also normalize to
the value of the granulation-like noise at the frequency of the
resonance. This gives:
 \begin{equation}
 E_{nlm}(\nu) = \left( \frac{1+(2 \pi \nu_{nlm} \tau)^2}
                {1+(2 \pi \nu \tau)^2} \right)^{1/2}.
 \label{eq:E}
 \end{equation}
The power spectral density of this mode is then:
 \begin{equation}
 {\rm PSD}_{nlm} = |A_{nlm}(\nu)|^2 = [L_{nlm}(\nu) E_{nlm}(\nu)][L_{nlm}^*(\nu) E_{nlm}^*(\nu)].
 \label{eq:psdnlm}
 \end{equation}

Now let us write down equations to describe the complete solarFLAG
spectrum. This spectrum is comprised of many overtones, whose
excitation is correlated; and also correlated and uncorrelated
background noise. We begin with a description of the power spectral
density of the overtones of a given ($l$,\,$m$). The excitation of the
overtones is correlated, with the coefficient $\rho$ describing this
correlation. The modes are also correlated with the granulation-noise
background, and this correlation is also 
set equal to $\rho$.

First, consider the equations which describe the real and imaginary
parts of the complex amplitude given by the overtones. We must sum
over all the radial orders $n$ of the chosen ($l$,\,$m$). We therefore
have:
 \begin{equation}
 \Re{[A_{lm}(\nu)]} = \displaystyle\sum_{{\rm over}~n}
                     \left(
                    \frac{H_{nlm} E_{nlm} \xi_{nlm}}{1+\xi_{nlm}^2}
                    \right)^{1/2},
 \label{eq:imr}
 \end{equation}
and
 \begin{equation}
 \Im{[A_{lm}(\nu)]} = \displaystyle\sum_{{\rm over}~n}
                     \left(
                    \frac{H_{nlm} E_{nlm}}{1+\xi_{nlm}^2}
                    \right)^{1/2}.
 \label{eq:ima}
 \end{equation}
Note that to keep things tidy, we have dropped the
explicit dependence of $E_{nlm}$ and $\xi_{nlm}$ on $\nu$ in the equations
above, and in what follows. The observed power spectral density of the
overtones \emph{and} the granulation-like noise for this ($l$,\,$m$)
is then given by:
 \begin{equation}
  {\rm PSD}_{lm}(\nu) = |\rho|\, \psi_{lm}(\nu) \psi_{lm}^{*}(\nu)
  + (1-|\rho|)\, \left( \displaystyle\sum_{{\rm over}~n}
              \left(\frac{H_{nlm}E_{nlm}}{1+\xi_{nlm}^2}\right) + n_{lm}
              \right),
 \label{eq:psdlm}
 \end{equation}
where
 \begin{equation}
 \psi_{lm}(\nu) = \Re{[A_{lm}(\nu)]} + i\,\Im{[A_{lm}(\nu)] - n_{lm}^{1/2}}.
 \label{eq:powlm}
 \end{equation}
There are two important things to notice about
Equations~\ref{eq:psdlm} and~\ref{eq:powlm}. First the description in
Equation~\ref{eq:psdlm} has two parts: one part, prefixed by the
factor $|\rho|$, describes the correlated contributions to the
spectrum; while the other part, prefixed by the factor
$(1-|\rho|)$, describes the uncorrelated contributions. Second,
notice how the background factor $n_{lm}^{1/2}$ in
Equation~\ref{eq:powlm} is prefixed by a minus sign. This makes the
correlation of the modes and background \emph{negative}, which in turn
means we get negative asymmetry as in real Doppler velocity
observations.

Finally, the total power spectrum density is given by summing,
incoherently, the power spectral densities of all ($l$,\,$m$) in the
spectrum, and the uncorrelated photon shot-noise background, $n_{\rm
psn}$ (see Section~\ref{sec:beta}), to give:
 \begin{equation}
  {\rm PSD}(\nu) = \displaystyle\sum_{{\rm all}~(l,m)} {\rm PSD}_{lm}(\nu) +
  n_{\rm psn}.
 \label{eq:psdtot}
 \end{equation}

\label{lastpage}

\end{document}